\documentclass[twocolumn,showpacs,preprintnumbers,showkeys,superscriptaddress]{revtex4}
\usepackage{amssymb}
\usepackage{graphicx}
\usepackage{dcolumn}
\usepackage{bm}
\usepackage{appendix}

\def\eqref#1{Eq.~(\ref{eq:#1})}

\begin{document}

\title{Generalized Density Matrix Revisited: Microscopic Approach
to Collective Dynamics in Soft Spherical Nuclei}
\author{L. Y. Jia}  \email{jial@nscl.msu.edu}
\affiliation{National Superconducting Cyclotron Laboratory and
Department of Physics and Astronomy, Michigan State University, East
Lansing, Michigan 48824, USA }

\date{\today}

\begin{abstract}

The generalized density matrix (GDM) method is used to calculate
microscopically the parameters of the collective Hamiltonian. Higher
order anharmonicities are obtained consistently with the lowest
order results, the mean field [Hartree-Fock-Bogoliubov (HFB)
equation] and the harmonic potential [quasiparticle random phase
approximation (QRPA)]. The method is applied to soft spherical
nuclei, where the anharmonicities are essential for restoring the
stability of the system, as the harmonic potential becomes small or
negative. The approach is tested in three models of increasing
complexity: the Lipkin model, model with factorizable forces, and
the quadrupole plus pairing model.

\end{abstract}


\vspace{0.4in}

\maketitle

\newpage

\section{Introduction \label{sec_intro}}

A long-standing question of microscopic description of nuclear
collective motion belongs to the class of problems which are left
behind by the advancing army that currently is mostly interested in
new frontiers, in our case, in drip line physics. Meanwhile, we
still lack a systematic theory based on first principles and
inter-nucleon interactions that would allow us to fully understand
numerous collective phenomena in the low-energy region of medium and
heavy nuclei and satisfactorily describe the data. In relatively
light nuclei, the shell model (what is nowadays called configuration
interaction) with effective nucleon-nucleon forces usually works
well although even here the abundant numerical results sometimes
require some kind of model interpretation. In heavier nuclei, the
necessary orbital space is too large for direct numerical
diagonalization.

Phenomenological models frequently work well, first of all the
geometric Bohr Hamiltonian \cite{Bohr,Bohr_Hamil} and the
interacting boson model (IBM) \cite{IBM}. However, the relation
between their parameters and the underlying microscopic structure
remains uncertain. Moreover, some assumptions of such models turn
out to be unreliable. For example, the identification in the IBM of
the prescribed boson number with the number of valence fermionic
pairs breaks down in the attempt to explain very long
``quasivibrational" bands extended, without considerable changes in
spacing, up to spin values much greater than the finite boson number
would allow, see for example the ground state band in $^{110}$Cd
close to the equidistant ladder up to $J^{\pi}=28^{+}$.

The microscopic theory is relatively successful in well deformed
nuclei. Various mean-field methods, including the modern energy
density functional approach \cite{DFT1,DFT2} with pairing, indicate
regions of nuclei with clearly pronounced deformed energy minima.
With the microscopic definition of shape, one can calculate the
moment of inertia by the cranking model and the generator coordinate
method, construct rotational bands built on different intrinsic
configurations and explain back-bending and similar phenomena
\cite{Back_Bending}.

In our opinion, the status of microscopic theory is still
underdeveloped with respect to spherical nuclei, especially in the
case of the presence of a low-lying collective mode. The standard
way of defining such modes is based on the quasiparticle random
phase approximation (QRPA). This is essentially the harmonic
approximation that determines the frequency and two-quasiparticle
structure of the collective phonons. If the multipole coupling is
strong, the collective mode has a large amplitude, the frequency
falls down, and the QRPA reveals the instability. In reality, this
is not necessarily a point of phase transition. Rather, this is the
region of strong anharmonicities outside of the reach of the QRPA.
Phenomenologically, this can be described by a special choice of
potential and rotational parameters in the Bohr Hamiltonian which
are close to the ${\cal O}(6)$ limit of the IBM with a
gamma-unstable potential. Currently we do not have a reliable
microscopic approach to quantify collective behavior of this type.
Another practically important question related to anharmonicities is
the mode-mode coupling. The coexistence and interaction of soft
quadrupole and octupole modes are relevant, for example, to the
search of mechanisms for many-body enhancement of the nuclear Schiff
moment and the atomic electric dipole moment \cite{Mul_Mode}.

Instead of the direct diagonalization of the primary nucleon
Hamiltonian, it seems reasonable to work out a procedure for the
microscopic derivation of the effective collective Hamiltonian.
Typical collective states can usually be identified by their quantum
numbers, low energies and large transition probabilities. Being
interconnected by large matrix elements of corresponding collective
operators they form a collective subspace of the total Hilbert space
of the system. In the case of a soft multipole mode, it is often
possible to label the empirical levels by the phonon quantum
numbers, even if their energies and transition rates noticeably
differ from the predictions of the harmonic approximation. This
difference results from anharmonic effects which still keep the
geometric nature of the mode. Therefore our approach will be to
develop the road to a consistent mapping of the underlying nucleonic
dynamics onto that inside the collective subspace.

The idea of this approach goes back to the boson expansion technique
suggested long ago \cite{first_anh}; a detailed review of work in
this direction can be found in \cite{rmp_boson}. The formalism of
the generalized density matrix (GDM) reformulating earlier work
\cite{GDM_Klein} by Kerman and Klein seems to be the most
appropriate for our goal \cite{BZ1,BZ2,BZ3,Zele_ptps}. This
formalism was applied to collective rotation
\cite{BZ1,BZ3,BZ4,Zele5} and large amplitude collective motion
\cite{LA1,LA2,LA3,LA4} generalizing the time-dependent mean-field
method \cite{Negele}. Here we apply the GDM approach to collective
vibrations in soft spherical nuclei.

The generalized density matrix $R_{12}=a^{\dagger}_{2}a_{1}$ is the
set of operators defined originally in the entire Hilbert space
[{\sl 1} and {\sl 2} here represent a complete set of
single-particle (s.p.) quantum numbers]. The microscopic Hamiltonian
provides exact operator equations of motion (e.o.m.) for this set.
Taking matrix elements of these equations between the states of the
collective family we map the equations onto the dynamics of the
collective operators inside this family. The choice of the
collective Hamiltonian should be quite general dictated by the type
and symmetries of collective motion under study. Comparison with
microscopic dynamics determines the collective parameters. The
lowest orders give naturally the mean field [Hartree-Fock-Bogoliubov
(HFB) equation] and the harmonic part (QRPA). Next orders determine
anharmonicities. These higher order terms are not assumed to be
perturbative, they are separated only by their operator structure in
the collective space. Simple estimates \cite{Zele_estimate,
Zele_AIP} show that in many generic cases the quartic anharmonicity
with respect to the quadrupole coordinate plays an important role.
In fact, this was earlier confirmed by specific realistic
applications \cite{fitting} of the phenomenological anharmonic
Hamiltonian; $^{100}$Pd is probably the clean example of such
dynamics.

We start with the discussion, Sec. \ref{sec_GDM}, of the general
procedure of the GDM method. In Sec. \ref{sec_Soft} we consider
systems near the critical point (small RPA frequency $\omega^2$).
Sec. \ref{sec_Lipkin} and Sec. \ref{sec_qq} are devoted to the
Lipkin model and factorizable force model, respectively, which
traditionally serve as a testing ground for various theoretical
approaches. Sec. \ref{sec_real_nucl} discusses the GDM method
applied to realistic nuclei with pairing and rotational symmetry. In
Sec. \ref{sec_QQPP} we give the results for a quadrupole plus
pairing Hamiltonian, with a semi-realistic numerical example. Sec.
\ref{sec_conclusion} summarizes our method and discusses future
working directions. The details of calculations are given in the
Appendices.

\section{The Generalized Density Matrix Method  \label{sec_GDM}}

In this section we reveal the essence of the GDM method, in a simple
system without complications due to rotational symmetry and pairing
correlations. A single collective mode is assumed; the case of
multiple modes is discussed briefly in Appendix
\ref{app_mode_coupling}. The main result, beyond the well known HF
equation and RPA, is a relation (\ref{quartic_sol_con}) involving
cubic and quartic anharmonicities.

\subsection{Preparation}

The starting point is the effective microscopic fermionic
Hamiltonian
\begin{eqnarray}
H = \sum_{12} Z_{12} a_1^\dagger a_2 + \frac{1}{4} \sum_{1234}
V_{1234} a_1^\dagger a_2^\dagger a_3 a_4 .    \label{H_f}
\end{eqnarray}
We find it convenient for $H$, $Z_{12}$ and $V_{1234}$ to be
dimensionless; in other words $H$ is measured in some unit of
energy. We have assumed in eq. (\ref{H_f}) a two-body force,
inclusion of three-body forces is discussed in Appendix
\ref{app_3b}. In accordance with the discussion in Sec.
\ref{sec_intro}, we assume that $H$ has a band of collective states
$\{ |C_i\rangle \}$ characterized by low energies and large
transition amplitudes. We assume that there exists a reference state
$|\Phi\rangle$, a collective mode operator $A^\dagger = (\alpha - i
\pi) / \sqrt{2}$ ($\alpha$, $\pi$ are collective coordinate and
momentum), such that \emph{approximately}
\begin{eqnarray}
[\alpha , \pi] = i ,    \label{ap_com}
\end{eqnarray}
\begin{eqnarray}
|C_i\rangle = [~ c_{i0} + c_{i1} A^\dagger + c_{i2} (A^\dagger)^2 +
\ldots \nonumber \\
+ c_{i-1} A + c_{i-2} (A)^2 + \ldots ~]~ |\Phi\rangle ,
\label{col_sta}
\end{eqnarray}
\begin{eqnarray}
\langle C_1 | ~ H ~ | C_2 \rangle = \langle C_1 | ~ E_0 +
\frac{\omega^2}{2} ~ \alpha^2 + \frac{1}{2} ~ \pi^2 \nonumber \\
+ \frac{\Lambda^{(30)}}{3} ~\alpha^3 +
\frac{\Lambda^{(12)}}{4}~ \{\alpha, \pi^2\} \nonumber \\
+ \frac{\Lambda^{(40)}}{4} ~\alpha^4 + \frac{\Lambda^{(22)}}{8} ~\{
\alpha^2 , \pi^2 \} + \frac{\Lambda^{(04)}}{4} ~\pi^4 + \ldots ~ |
C_2 \rangle .   \label{H_b}
\end{eqnarray}
Eq. (\ref{ap_com}) says that $A^\dagger$ is effectively a boson
operator. Eq. (\ref{col_sta}) says that the collective band $\{
|C_i\rangle \}$ can be built by repeated action of $A^\dagger$ or
$A$ on the reference state $|\Phi\rangle$. Later $|\Phi\rangle$ will
be identified as the HF ground state. Eq. (\ref{H_b}) says that
within the band, the effect of the fermionic Hamiltonian can be
approximated by an expansion over the bosonic operators, where we
keep all time-even terms up to quartic anharmonicities ($\alpha$ is
time-even, $\pi$ is time-odd).

Now our goal is to map the exact e.o.m. in the full Hilbert space
onto collective dynamics inside the band subspace. We will use
contractions and normal ordering of operators. They are defined as:
\begin{eqnarray}
A^{\bullet} B^\bullet \equiv \langle \Phi | A B | \Phi \rangle ,  \\
N[A B] \equiv A B - \langle \Phi | A B | \Phi \rangle .
\end{eqnarray}
Without paring, the reference state $|\Phi\rangle$ has a definite
particle number,
\begin{eqnarray}
\langle \Phi | a_1^\dagger a_2 | \Phi \rangle \equiv \rho_{21}, ~~~
\langle \Phi | a_1 a_2 | \Phi \rangle = \langle \Phi | a_1^\dagger
a_2^\dagger | \Phi \rangle = 0 .    \label{dm_def}
\end{eqnarray}
$\rho$ is the usual single-particle density matrix. Normal ordering
of more than two operators is defined by the Wick theorem:
\begin{eqnarray}
a_1^\dagger a_2^\dagger a_3 a_4 = N[a_1^\dagger a_2^\dagger a_3 a_4]
+ \rho_{41} N[a_2^\dagger a_3]
- \rho_{31} N[a_2^\dagger a_4] \nonumber \\
- \rho_{42} N[a_1^\dagger a_3] + \rho_{32} N[a_1^\dagger a_4] +
\rho_{41} \rho_{32} - \rho_{31} \rho_{42} .
\end{eqnarray}
Equivalently, normal ordering puts quasiparticle creation operators
to the left of annihilation operators.

The generalized density matrix operator is defined in the full space
as
\begin{eqnarray}
R_{12} \equiv a_2^\dagger a_1 = \rho_{12} + N[a_2^\dagger a_1]
\equiv \rho_{12} + R^N_{12} .
\end{eqnarray}
The Hamiltonian (\ref{H_f}) in the normal ordering form is
\begin{eqnarray}
H = \langle \Phi | H | \Phi \rangle + \sum_{12} f_{12} N[a_1^\dagger
a_2] \nonumber \\
+ \frac{1}{4} \sum_{1234} V_{1234} N[a_1^\dagger
a_2^\dagger a_3 a_4] ,  \label{H_norm_order}
\end{eqnarray}
where we have introduced the self-consistent field operator
\begin{eqnarray}
W\{R\}_{12} \equiv \sum_{34} V_{1432} R_{34} , ~~~ f_{12} = Z_{12} +
W\{\rho\}_{12} ,   \label{W_def}
\end{eqnarray}
and $\langle \Phi | H | \Phi \rangle = \sum_{12} (Z_{12} +
\frac{1}{2} W\{\rho\}_{12}) \rho_{21}$ is the average energy of the
reference state.

The exact e.o.m. for the density matrix operator in the full
many-body Hilbert space is
\begin{eqnarray}
[ R_{12} , H ] = [ a_2^\dagger a_1 , H ] = \nonumber \\
{[}f, \rho]_{12} + [f, R^N]_{12} +
[W\{R^N\}, \rho]_{12} \nonumber \\
+ \frac{1}{2} \sum_{345} ( V_{1345} N[a_2^\dagger a_3^\dagger a_4
a_5]-V_{5432} N[a_5^\dagger a_4^\dagger a_3 a_1]) .   \label{e.o.m.}
\end{eqnarray}
Since we are only interested in the band subspace, we take matrix
elements of eq. (\ref{e.o.m.}) between two collective states:
\begin{eqnarray}
\langle C_i | [ R_{12} , H ] | C_j \rangle = \nonumber \\
\langle C_i |~ [f, \rho]_{12} + [f, R^N]_{12} +
[W\{R^N\}, \rho]_{12} \nonumber \\
+ \frac{1}{2} \sum_{345} ( V_{1345} N[a_2^\dagger a_3^\dagger a_4
a_5]-V_{5432} N[a_5^\dagger a_4^\dagger a_3 a_1])  ~| C_j \rangle .
\label{bk_e.o.m.}
\end{eqnarray}
We assume that within the band the effect of $R_{12}$ can be
approximated by a boson expansion:
\begin{eqnarray}
\langle C_i | R_{12} | C_j \rangle = \langle C_i |~ \rho_{12} +
r^{(10)}_{12} \alpha + r^{(01)}_{12} \pi \nonumber \\
+ r^{(20)}_{12} \frac{\alpha^2}{2} + r^{(02)}_{12} \frac{\pi^2}{2} +
r^{(11)}_{12} \frac{\{\alpha,\pi\}}{2} + r^{(30)}_{12}
\frac{\alpha^3}{3} \nonumber \\
+ r^{(03)}_{12} \frac{\pi^3}{3} + r^{(21)}_{12} \frac{\{\alpha^2,
\pi\}}{4} + r^{(12)}_{12} \frac{\{\alpha, \pi^2\}}{4} + \ldots ~|
C_j \rangle ,  \label{R_exp}
\end{eqnarray}
where we keep explicitly terms up to quartic anharmonicities. A
convenient normalization is: a term with $m$ of $\alpha$ and $n$ of
$\pi$ has a factor of $1/(m n)$; each anti-commutator gives an
additional $1/2$. Similarly, for $N[a_4^\dagger a_3^\dagger a_2
a_1]$ we have
\begin{eqnarray}
\langle C_i | N[a_4^\dagger a_3^\dagger a_2 a_1] | C_j \rangle =
\langle C_i |~ \frac{1}{2} r^{(20)}_{1234} \alpha^2 + \frac{1}{2}
r^{(02)}_{1234} \pi^2 \nonumber \\
+ \frac{1}{2} r^{(11)}_{1234} \{\alpha,\pi\} + \frac{1}{3}
r^{(30)}_{1234} \alpha^3 + \frac{1}{3} r^{(03)}_{1234} \pi^3
\nonumber \\
+ \frac{1}{4} r^{(21)}_{1234} \{\alpha^2, \pi\} + \frac{1}{4}
r^{(12)}_{1234} \{\alpha, \pi^2\} + \ldots ~ | C_j \rangle ,
\label{RR_exp}
\end{eqnarray}
where we have assumed that the expansion starts from $\alpha^2$,
$\pi^2$, $\{\alpha,\pi\}$, as explained in Appendix \ref{app_ap_mb}.
Now the r.h.s. of eq. (\ref{bk_e.o.m.}) is written as an expansion
over boson operators.

The l.h.s. of eq. (\ref{bk_e.o.m.}) is approximately given by:
\begin{eqnarray}
\langle C_i | [ R_{12} , H ] | C_j \rangle \approx \nonumber
\\
\langle C_i | [ \rho_{12} + r^{(10)}_{12} \alpha + \ldots ~,~ E_0
+ \frac{\omega^2}{2} ~ \alpha^2 + \ldots ] | C_j \rangle ,
\nonumber
\end{eqnarray}
where we have restricted the intermediate states (between $R_{12}$
and $H$) by those of the collective subspace $\{ |C_i\rangle \}$,
since $H$ is a collective operator: the matrix elements of $H$
connecting the collective band with states of a different nature are
small. This is the main approximation of the method; influence of
the neglected ``environment'' states can be later accounted for with
the use of statistical assumptions \cite{chaotic}. After calculating
commutators like $[\pi , \alpha^2] = - 2 i \alpha$, the l.h.s. is
written as a boson operator expansion. Then we equate in eq.
(\ref{bk_e.o.m.}) l.h.s. and r.h.s. coefficients of the same phonon
structure: $1$, $\alpha$, $\pi$, $\alpha^2/2$ ... The resultant
equations are examined below.

\subsection{Zero Order: Mean Field (Hartree-Fock)  \label{sec_GDM_mf}}

Terms without $\alpha$ or $\pi$ in eq. (\ref{bk_e.o.m.}) give
\begin{eqnarray}
[f, \rho]_{12} = 0 .   \label{HF}
\end{eqnarray}
Thus $f$ and $\rho$ can be diagonalized simultaneously in some s.p.
basis:
\begin{eqnarray}
f_{12} = \delta_{12} e_1 , ~~~ \rho_{12} = \delta_{12} n_1 ,
\end{eqnarray}
providing mean-field s.p. energies and occupation numbers. We will
always use this s.p. basis. If we restrict the reference state
$|\Phi\rangle$ to be a Slater determinant, then the occupation
numbers $n_1$ can be only $0$ or $1$; in this case eq. (\ref{HF}) is
the usual HF equation, $|\Phi\rangle$ is the HF ground state. More
general choices, such as the thermal ensemble, are also possible.
For future convenience we define
\begin{eqnarray}
e_{12} \equiv e_1 - e_2 , ~~~ n_{12} \equiv n_1 - n_2 .
\end{eqnarray}
We assume that degenerate s.p. levels have the same occupancies,
\begin{eqnarray}
e_1 = e_2  ~\Rightarrow~  n_1 = n_2 ,
\end{eqnarray}
but the reverse is not necessarily true.

\subsection{First Order: Random Phase Approximation  \label{sec_GDM_harm}}

Terms linear in $\alpha$ and $\pi$ in eq. (\ref{bk_e.o.m.}) give
\begin{eqnarray}
\underline{\pi} : ~~~~~~~~~~~~ i r^{(10)} = [f, r^{(01)}] + [w^{(01)}, \rho] ,   \label{RPA_a}  \\
\underline{\alpha} : ~~~~~ - i \omega^2 r^{(01)} = [f, r^{(10)}] +
[w^{(10)}, \rho] ,   \label{RPA_p}
\end{eqnarray}
where $w^{(10)} = W\{r^{(10)}\}$, and $w^{(01)} = W\{r^{(01)}\}$ are
the corresponding components of the mean field. This is the set of
RPA equations. The formal solution is
\begin{eqnarray}
r^{(10)}_{12} = \frac{n_{12}}{(e_{12})^2 - \omega^2}
[-i \omega^2 ~ w^{(01)}_{12} + e_{12} w^{(10)}_{12}] ,  \label{RPA_r10} \\
r^{(01)}_{12} = \frac{n_{12}}{(e_{12})^2 - \omega^2} [i ~
w^{(10)}_{12} + e_{12} w^{(01)}_{12}] .  \label{RPA_r01}
\end{eqnarray}
Note that $r^{(10)}$ and $r^{(01)}$ have only $n_1 \ne n_2$ matrix
elements. From eqs. (\ref{W_def}), (\ref{RPA_r10}) and
(\ref{RPA_r01}) we obtain a linear homogenous set of equations for
$w^{(10)}$ and $w^{(01)}$:
\begin{eqnarray}
w^{(10)}_{34} = \sum_{12} V_{3214} \frac{n_{12}}{(e_{12})^2 -
\omega^2} [- i \omega^2
w^{(01)}_{12} + e_{12} w^{(10)}_{12}] ,   \label{RPA_w10} \\
w^{(01)}_{34} = \sum_{12} V_{3214} \frac{n_{12}}{(e_{12})^2 -
\omega^2} [ i w^{(10)}_{12} + e_{12} w^{(01)}_{12}] .~
\label{RPA_w01}
\end{eqnarray}
Introduce the matrix $M$:
\begin{eqnarray}
M = \left(
  \begin{array}{cc}
    M^{a} & M^{b} \\
    M^{c} & M^{d} \\
  \end{array}
\right) ,  \label{M_def}
\end{eqnarray}
where ($e_3 \ne e_4$ , $e_1 \ne e_2$)
\begin{eqnarray}
M^a_{(34),(12)} = M^d_{(34),(12)} = \delta_{(12),(34)} - V_{3214}
\frac{n_{12} e_{12}}{(e_{12})^2 - \omega^2} ,~  \label{Mad_def} \\
M^b_{(34),(12)} = \omega^2 \cdot M^c_{(34),(12)} = V_{3214}
\frac{\omega^2 ~ n_{12}}{(e_{12})^2 - \omega^2} ,~  \label{Mbc_def}
\end{eqnarray}
in which $\delta_{(12),(34)} = \delta_{13} \delta_{24}$. Then the
$e_3 \ne e_4$ part of eqs. (\ref{RPA_w10}) and (\ref{RPA_w01}) is
written as
\begin{eqnarray}
M \cdot \left(
          \begin{array}{c}
            w^{(10)} \\
            i \cdot w^{(01)} \\
          \end{array}
        \right) = 0 .   \label{RPA_set}
\end{eqnarray}
Non-zero solution requires a zero determinant:
\begin{eqnarray}
{\rm{Det}} [ M ] = 0 .   \label{RPA}
\end{eqnarray}
Eq. (\ref{RPA}) is the RPA secular equation determining the harmonic
frequency $\omega^2$.

By eq. (\ref{RPA}) the transpose matrix of $M$, $M^T$, has a zero
eigenvalue. Assume the corresponding eigenvector is $(\lambda_{34} ,
\chi_{34})$:
\begin{eqnarray}
M^T \cdot \left(
        \begin{array}{c}
          \lambda \\
          \chi \\
        \end{array}
      \right)
= 0 \cdot \left(
        \begin{array}{c}
          \lambda \\
          \chi \\
        \end{array}
      \right)
= 0 ~\Rightarrow~\left(
                       \begin{array}{cc}
                         \lambda^T & \chi^T \\
                       \end{array}
                     \right)
\cdot M = 0 .~~    \label{lk_def}
\end{eqnarray}
In other words, the row vectors of $M$ are linearly dependent.
$\lambda$ and $\chi$ are used later.

The normalization of $r^{(10)}$, $r^{(01)}$ can be fixed by the
so-called saturation principle as explained in Appendix
\ref{app_sat}:
\begin{eqnarray}
1 = \sum_{12} \frac{- n_{12}}{[(e_{12})^2 - \omega^2]^2} \nonumber
\\
\cdot (i w^{(10)}_{12} + e_{12} w^{(01)}_{12})( \omega^2
w^{(01)}_{21} + i e_{21} w^{(10)}_{21}) .  \label{norm}
\end{eqnarray}

Starting from the next (second) order, $r^{(mn)}_{1234}$ of eq.
(\ref{RR_exp}) begin to appear in the e.o.m., the saturation
principle is also used to express them in terms of $r^{(mn)}_{12}$
of eq. (\ref{R_exp}), as explained in Appendix \ref{app_sat}.

\subsection{Second Order: Cubic Anharmonicity  \label{sec_GDM_cubic_anharm}}

Using eqs. (\ref{r20_1234}-\ref{r11_1234}), the second order terms,
$\alpha^2/2$, $\pi^2/2$, $\{\alpha,\pi\}/2$, in eq.
(\ref{bk_e.o.m.}) give
\begin{eqnarray}
- 2 i \omega^2 r^{(11)} - 2 i \Lambda^{(30)} r^{(01)} = \nonumber \\
{[}f, r^{(20)}] + [w^{(20)}, \rho] + 2 [w^{(10)}, r^{(10)}] ,  \label{Cubic_1} \\
- i \omega^2 r^{(02)} + i r^{(20)} + i \Lambda^{(12)} r^{(10)} = \nonumber \\
{[}f, r^{(11)}] + [w^{(11)}, \rho]
+ [w^{(10)}, r^{(01)}] + [w^{(01)}, r^{(10)}] ,  \label{Cubic_2} \\
2 i r^{(11)} - i \Lambda^{(12)} r^{(01)} = \nonumber \\
{[}f, r^{(02)}] + [w^{(02)}, \rho] + 2 [w^{(01)}, r^{(01)}] .
\label{Cubic_3}
\end{eqnarray}

First we consider the $e_1 = e_{1'}$ matrix elements
$r^{(20)}_{11'}$, $r^{(11)}_{11'}$ and $r^{(02)}_{11'}$. As one can
check, eqs. (\ref{Cubic_1}) and (\ref{Cubic_3}) give the same
$r^{(11)}_{11'}$. But $r^{(02)}_{11'}$ and $r^{(20)}_{11'}$ are not
fully determined since eq. (\ref{Cubic_2}) determines only the
difference $\omega^2 r^{(02)}_{11'} - r^{(20)}_{11'}$. We fix them
by the saturation principle, the $e_1 = e_{1'}$ matrix elements of
eqs. (\ref{x_higher}) and (\ref{p_higher}):
\begin{eqnarray}
r^{(20)}_{11'} = - i {[ p, r^{(10)} ]}_{11'} ,~~ r^{(02)}_{11'} = i
{[ x, r^{(01)} ]}_{11'} , \nonumber \\
r^{(11)}_{11'} = i {[x,
r^{(10)}] }_{11'} = - i {[p, r^{(01)}] }_{11'} .  \label{sat_11p}
\end{eqnarray}
It is straightforward to show that eq. (\ref{sat_11p}) is consistent
with eqs. (\ref{Cubic_1}-\ref{Cubic_3}).

Next the $e_1 \ne e_2$ matrix elements $r^{(20)}_{12}$,
$r^{(11)}_{12}$ and $r^{(02)}_{12}$ are solved formally in terms of
$w^{(20/02/11)}$ and $\Lambda^{(30/12)}$ from eqs.
(\ref{Cubic_1}-\ref{Cubic_3}). Then by eq. (\ref{W_def}) we obtain
the linear set of equations for $w^{(20)}$, $w^{(02)}$ and
$w^{(11)}$.

\subsection{Third Order: Quartic Anharmonicity  \label{sec_GDM_quartic_anharm}}

Using eqs. (\ref{r30_1234}-\ref{r12_1234}), the third order terms,
$\alpha^3/3$, $\{\alpha^2,\pi\}/4$, $\{\alpha,\pi^2\}/4$, $\pi^3/3$,
in eq. (\ref{bk_e.o.m.}) give
\begin{eqnarray}
- \frac{3i}{2} \omega^2 r^{(21)} - 3 i \Lambda^{(30)} r^{(11)} - 3 i
\Lambda^{(40)} r^{(01)} = \nonumber \\
{[}f, r^{(30)}] + [w^{(30)}, \rho] + \frac{3}{2} [w^{(20)},
r^{(10)}] + \frac{3}{2} [w^{(10)},
r^{(20)}] ,~~ \label{quartic_1} \\ \nonumber \\
- 2 i \omega^2 r^{(12)} + 2 i r^{(30)} - 2 i \Lambda^{(30)} r^{(02)}
+ 2 i \Lambda^{(12)} r^{(20)} \nonumber \\
+ i \Lambda^{(22)} r^{(10)} = [f, r^{(21)}] + [w^{(21)}, \rho] +
[w^{(20)},r^{(01)}] \nonumber \\
+ [w^{(01)},r^{(20)}] + 2 [w^{(11)},r^{(10)}]
+ 2 [w^{(10)},r^{(11)}] ,~~  \label{quartic_2} \\ \nonumber \\
- 2 i \omega^2 r^{(03)} + 2 i r^{(21)} + i \Lambda^{(12)} r^{(11)} -
i \Lambda^{(22)} r^{(01)} = \nonumber \\
{[}f, r^{(12)}] + [w^{(12)}, \rho] + [w^{(10)},r^{(02)}] \nonumber
\\
+ [w^{(02)},r^{(10)}] + 2 [w^{(11)},r^{(01)}] +
2 [w^{(01)},r^{(11)}] ,~~    \label{quartic_3} \\ \nonumber \\
\frac{3 i}{2} r^{(12)} - \frac{3 i}{2} \Lambda^{(12)} r^{(02)} + 3
i \Lambda^{(04)} r^{(10)} = \nonumber \\
{[}f, r^{(03)}] + [w^{(03)}, \rho] + \frac{3}{2} [w^{(02)},
r^{(01)}] + \frac{3}{2} [w^{(01)}, r^{(02)}] .~~  \label{quartic_4}
\end{eqnarray}

The $e_1 = e_{1'}$ matrix elements $r^{(30/21/12/03)}_{11'}$ can now
be found in terms of the lower order quantities from eqs.
(\ref{quartic_1}-\ref{quartic_4}). The $e_1 \ne e_2$ matrix elements
$r^{(30/21/12/03)}_{12}$ can be calculated formally in terms of
$w^{(30/21/12/03)}$ and $\Lambda^{(40/22/04)}$ from eqs.
(\ref{quartic_1}-\ref{quartic_4}). By eq. (\ref{W_def}) we obtain a
linear set of equations for $w^{(30)}$, $w^{(21)}$, $w^{(12)}$ and
$w^{(03)}$. However, this set is not linearly independent. Thus we
have a solvability condition. To see this, keeping only
$(r/w)^{(30/21/12/03)}$ and $\Lambda^{(40/22/04)}$ terms, eq.
(\ref{quartic_1}) + $\frac{1}{2} \omega^2 \times$eq.
(\ref{quartic_3}) gives
\begin{eqnarray}
- i \omega^2 [\frac{1}{2} r^{(21)} + \omega^2 r^{(03)}] = [f,
r^{(30)} + \frac{1}{2} \omega^2 r^{(12)}] \nonumber \\
+ [w^{(30)} + \frac{1}{2} \omega^2 w^{(12)}, \rho] + 3 i
\Lambda^{(40)} r^{(01)} \nonumber \\
+ i \frac{1}{2} \omega^2
\Lambda^{(22)} r^{(01)} + \ldots  \label{like_RPA1}
\end{eqnarray}
$\frac{1}{2} \times $eq. (\ref{quartic_2}) + $\omega^2 \times$eq.
(\ref{quartic_4}) gives
\begin{eqnarray}
+ i [r^{(30)} + \frac{1}{2} \omega^2 r^{(12)}] = [f, \frac{1}{2}
r^{(21)} + \omega^2 r^{(03)}] \nonumber \\ + [\frac{1}{2} w^{(21)} +
\omega^2 w^{(03)}, \rho] - i \frac{1}{2} \Lambda^{(22)} r^{(10)}
\nonumber \\
- 3 i \omega^2 \Lambda^{(04)} r^{(10)} + \ldots \label{like_RPA2}
\end{eqnarray}
The variable parts of eqs. (\ref{like_RPA1}) and (\ref{like_RPA2})
have the same structure as the RPA equations (\ref{RPA_a}) and
(\ref{RPA_p}). Introducing temporarily $x = r^{(30)} + \frac{1}{2}
\omega^2 r^{(12)}$ , $y = \frac{1}{2} r^{(21)} + \omega^2 r^{(03)}$,
we can solve $x$, $y$ in terms of $W\{x\}$, $W\{y\}$. Using eq.
(\ref{W_def}) to obtain linear equations for $W\{x\}$, $W\{y\}$, the
$e_3 \ne e_4$ part is written as
\begin{eqnarray}
M \cdot \left(
          \begin{array}{c}
            W\{x\} \\
            i ~ W\{y\} \\
          \end{array}
        \right) =
\left(
  \begin{array}{c}
    A \\
    B \\
  \end{array}
\right) + \ldots ,     \label{quartic_RPA}
\end{eqnarray}
where the matrix $M$ is defined in eqs. (\ref{M_def}-\ref{Mbc_def});
$A$, $B$ consist of $\Lambda^{(40)}$, $\Lambda^{(22)}$ and
$\Lambda^{(04)}$ terms. Multiplying eq. (\ref{quartic_RPA}) from
left by $(\lambda^T ~ \chi^T)$ and using eq. (\ref{lk_def}) we come
to the solvability condition:
\begin{small}
\begin{eqnarray}
3 \Lambda^{(40)} \sum_{12} \sum_{e_3 \ne e_4} [\chi_{34} - e_{12}
\lambda_{34}] \cdot V_{3214} \frac{i ~ r^{(01)}_{12}}{(e_{12})^2 -
\omega^2}  \nonumber \\
+ \frac{1}{2} \Lambda^{(22)} \sum_{12} \sum_{e_3 \ne e_4} V_{3214}
\nonumber \\
\cdot \frac{\omega^2 \lambda_{34} \cdot [r^{(10)}_{12} - i e_{12}
r^{(01)}_{12}] - \chi_{34} \cdot [ e_{12} r^{(10)}_{12} - i \omega^2
r^{(01)}_{12}
]}{(e_{12})^2 - \omega^2} \nonumber \\
+ 3 \omega^2 \Lambda^{(04)} \sum_{12} \sum_{e_3 \ne e_4} [\omega^2
\lambda_{34} - e_{12} \chi_{34}] \cdot V_{3214}
\frac{r^{(10)}_{12}}{(e_{12})^2 - \omega^2}  = \ldots ~
\label{quartic_sol_con}
\end{eqnarray}
\end{small}
The r.h.s. ``$\ldots$'' contains only the lower order quantities,
including $\Lambda^{(30)}$ and $\Lambda^{(12)}$. On the l.h.s. the
coefficients of the $\Lambda^{(40)}$, $\Lambda^{(22)}$,
$\Lambda^{(04)}$ terms are of order $1$, $\omega^2$, $\omega^4$,
respectively, although this might not be obvious from eq.
(\ref{quartic_sol_con}). It follows from examining the expressions
of $M$, $A$ and $B$ in eq. (\ref{quartic_RPA}). This point will be
important for the discussion in Sec. \ref{sec_Soft}.

At the current stage, the cubic and quartic anharmonicities are not
completely fixed, we find only one relation (\ref{quartic_sol_con})
constraining them. However, we are able to obtain $\Lambda^{(30)}$
and $\Lambda^{(40)}$ near the critical point $\omega^2 \approx 0$,
with important applications, as will be explained in Sec.
\ref{sec_Soft}. Even with this limitation, eq.
(\ref{quartic_sol_con}) is useful. One could fit the ratios of
$\Lambda^{(mn)}$ with the experimental data, then use eq.
(\ref{quartic_sol_con}) to determine their magnitudes. This is
especially interesting for the cases with certain symmetries, where
the ratios are known. Results in this direction will be discussed
elsewhere.

\subsection{Self-consistent Hamiltonian Conditions}

If the approach is self-consistent, substituting the solutions of
eqs. (\ref{R_exp}) and (\ref{RR_exp}) into eq. (\ref{H_f}) should
provide eq. (\ref{H_b}). Namely,
\begin{eqnarray}
E_0 = \langle \Phi | H | \Phi \rangle = \sum_{12} (Z_{12} + \frac{1}{2} W\{\rho\}_{12}) \rho_{21} ,  \\
\Lambda^{(mn)} = \sum_{1} e_{1} r^{(mn)}_{11} + \frac{1}{4}
\sum_{1234} V_{1234} r^{(mn)}_{4321} .   \label{H_self}
\end{eqnarray}
We checked eq. (\ref{H_self}) explicitly up to the cubic
anharmonicities. Concerning the quartic anharmonicity, we checked
the combinations $\Lambda^{(40)} - \frac{1}{2} \omega^2
\Lambda^{(22)}$ and $\omega^2 \Lambda^{(04)} - \frac{1}{2}
\Lambda^{(22)}$; since from the fourth order e.o.m. (not listed)
only $r^{(40)}_{11} - \frac{1}{2} \omega^2 r^{(22)}_{11}$ and
$\omega^2 r^{(04)}_{11} - \frac{1}{2} r^{(22)}_{11}$ are determined,
similar to the situation in the cubic order.\\

In summary, this section discusses the general procedure of the GDM
method. The exact e.o.m. for the density matrix operators are mapped
onto the collective subspace by taking matrix elements between
states of this family. Comparing terms with the same phonon operator
structure, order by order, we get equations for the GDM. In each
order, the GDM is solved from a set of coupled linear equations in
terms of lower order quantities. The bosonic Hamiltonian
coefficients $\Lambda^{(mn)}$ appear as parameters in the solution.

At the current stage the anharmonicities are not completely fixed;
we find only one relation (\ref{quartic_sol_con}) involving cubic
and quartic anharmonicities, appearing in the third order as a
solvability condition. In the next section, we will show that the
cubic potential $\Lambda^{(30)}$ and quartic potential
$\Lambda^{(40)}$ can be determined in a special case -- around the
critical point $\omega^2 \approx 0$.

\section{Systems near the Critical Point  \label{sec_Soft}}

Anharmonicities become important when the harmonic potential
$\omega^2 \alpha^2 / 2$ becomes small or negative. This is the case
in many realistic medium and heavy nuclei away from magic numbers
\cite{rmp_QQPP}. The quartic potential $\Lambda^{(40)}$ and higher
terms restore the stability of the system. At the same time, the
system can be deformed by odd anharmonicities; the potential is flat
at the bottom, or gamma-unstable. Near the critical point $\omega^2
\approx 0$, we are able to determine the cubic potential term
$\Lambda^{(30)}$ and the quartic potential term $\Lambda^{(40)}$.
Deformation due to $\Lambda^{(30)}$ will be studied separately. In
this work we concentrate on the case of small $\Lambda^{(30)}$,
consistent with the idea of soft \emph{spherical} nuclei.

We make an assumption in the spirit of Landau phase transition
theory: in eq. (\ref{H_b}), the leading potential term $\omega^2
\alpha^2 / 2$ vanishes at the critical point, while other higher
order terms $\Lambda^{(mn)}$ remain \emph{finite}. Taylor expanding
$\Lambda^{(mn)}$ over $\omega^2$,
\begin{eqnarray}
\Lambda^{(mn)} = \Lambda^{(mn)}_c + \Lambda^{(mn)}_1 \omega^2 +
\Lambda^{(mn)}_2 \omega^4 + \ldots ,  \label{Lmn_exp}
\end{eqnarray}
the leading constant term $\Lambda^{(mn)}_c$ is finite.

Near the critical point the stability of the system is restored by
higher order anharmonicities, e.g. $\Lambda^{(40)} \alpha^4 / 4$.
Thus $\langle C_i | \alpha | C_j \rangle$, $\langle C_i | \pi | C_j
\rangle$ ... are finite. Consequently $r^{(mn)}_{12}$ in eq.
(\ref{R_exp}) is finite, since the l.h.s. $\langle C_i | a_2^\dagger
a_1 | C_j \rangle$ is finite. Again we call the finite leading
constant term in a Taylor expansion $r^{(mn)}_{c 12}$.

We can obtain $\Lambda^{(40)}_c$ from eq. (\ref{quartic_sol_con}) by
keeping only leading constant terms (neglecting terms with
$\omega^2$, $\omega^4$ ...), as explained below eq.
(\ref{quartic_sol_con}). Another approach is possible: neglecting
$\omega^2$ terms earlier, in each e.o.m. For convenience, we use
$\doteq$ instead of $=$ if an equation is correct in constant terms
but not in $\omega^2$ terms or higher. In this way we determine
$\Lambda^{(30)}_c$, eq. (\ref{L30_c}), as well as
$\Lambda^{(40)}_c$, eq. (\ref{L40_c}).

\subsection{RPA}

Keeping only the constant terms of eq. (\ref{RPA_w10}) we have
\begin{eqnarray}
w^{(10)}_{34} \doteq \sum_{12} V_{3214} \frac{n_{12}}{e_{12}} ~
w^{(10)}_{12} .  \label{RPA_w10_c}
\end{eqnarray}
Defining a square matrix
\begin{eqnarray}
D_{(34),(12)} \equiv \delta_{(12),(34)} - V_{3214}
\frac{n_{12}}{e_{12}}   ,~~  (e_3 \ne e_4 , e_1 \ne e_2) ,~
\label{D_def}
\end{eqnarray}
the $e_3 \ne e_4$ part of eq. (\ref{RPA_w10_c}) is written as $D
w^{(10)} \doteq 0$. Since the quantities $w^{(10)}_{c 12}$ do not
vanish, we have ${\rm{Det}}[ D ] \doteq 0$. Thus $D^T$, the
transpose matrix of $D$, has a $0$ eigenvalue. More accurately,
$D^T$ has an eigenvalue of order $\omega^2$; because Det$[D]$, the
product of all eigenvalues of $D^T$, is of order $\omega^2$. Assume
that the eigenvector corresponding to this eigenvalue is
$\eta_{34}$:
\begin{eqnarray}
D^T \eta ~\doteq~ 0 ~ \eta ~=~ 0  ~~~\Rightarrow~~~  \eta^T D
~\doteq~ 0 .        \label{eta_def}
\end{eqnarray}

\subsection{Cubic Anharmonicity}

Keeping only the constant terms of eq. (\ref{Cubic_1}),
\begin{eqnarray}
- 2 i \Lambda^{(30)} r^{(01)}_{12} \doteq e_{12} r^{(20)}_{12} -
n_{12} w^{(20)}_{12} + 2 [w^{(10)} , r^{(10)}]_{12} ,~~
\label{Cubic_1_c}
\end{eqnarray}
and calculating $w^{(20)}_{34}$ from eq. (\ref{Cubic_1_c}), the $e_3
\ne e_4$ part is written as
\begin{eqnarray}
D \cdot w^{(20)} \doteq C ,   \label{w20_c}
\end{eqnarray}
where $D$ is defined in eq. (\ref{D_def}), $C$ contains
$\Lambda^{(30)}$ and lower order quantities. Multiplying eq.
(\ref{w20_c}) from left by $\eta^T$ and using eq. (\ref{eta_def}) we
obtain
\begin{eqnarray}
\Lambda^{(30)} \cdot \sum_{e_1 \ne e_2} \sum_{e_3 \ne e_4}
\eta_{34} V_{3214} \frac{r^{(01)}_{12}}{e_{12}} ~~~\doteq \nonumber \\
i \sum_{e_1 \ne e_2} \sum_{e_3 \ne e_4} \eta_{34} V_{3214}
\frac{[w^{(10)} , r^{(10)}]_{12}}{e_{12}} \nonumber \\
- \frac{1}{2} \sum_{e_1 = e_{1'}} \sum_{e_3 \ne e_4} \eta_{34}
V_{31'14} {[ p , r^{(10)} ]}_{11'} ,  \label{L30_c}
\end{eqnarray}
where $p$ is given in eq. (\ref{p_ele}). Eq. (\ref{L30_c}) gives
$\Lambda^{(30)}_c + O(\omega^2)$. Then $w^{(20)}_{12}$ is solved
from eq. (\ref{w20_c}) with an overall factor still undetermined.

Similarly, from eq. (\ref{Cubic_2}) we obtain an equation $D \cdot
w^{(11)} \doteq \ldots$ Multiplying it by $\eta^T$ we obtain
\begin{eqnarray}
\Lambda^{(12)} \sum_{e_1 \ne e_2} \sum_{e_3 \ne e_4} \eta_{34}
V_{3214} \frac{r^{(10)}_{12}}{e_{12}}  \nonumber \\
+ \sum_{e_1 \ne e_2} \sum_{e_3 \ne e_4} \eta_{34} V_{3214} \frac{n_{12}}{(e_{12})^2} w^{(20)}_{12} \doteq  \nonumber \\
2 i \Lambda^{(30)} \sum_{e_1 \ne e_2} \sum_{e_3 \ne e_4} \eta_{34}
V_{3214} \frac{r^{(01)}_{12}}{(e_{12})^2}  \nonumber \\
+ 2 \sum_{e_1
\ne e_2} \sum_{e_3 \ne e_4} \eta_{34} V_{3214} \frac{[w^{(10)} ,
r^{(10)}]_{12}}{(e_{12})^2}
\nonumber \\
- i \sum_{e_1 \ne e_2} \sum_{e_3 \ne e_4} \eta_{34} V_{3214}
\frac{[w^{(10)}, r^{(01)}]_{12} + [w^{(01)}, r^{(10)}]_{12}}{e_{12}}
\nonumber \\
+ \sum_{e_1 = e_{1'}} \sum_{e_3 \ne e_4} \eta_{34} V_{31'14}
[w^{(01)} , r^{(01)}]_{11'} . ~~  \label{fix_f_w20}
\end{eqnarray}
From eq. (\ref{fix_f_w20}) the undetermined overall factor in
$w^{(20)}_{12}$ is fixed as a function of $\Lambda^{(12)}$. Then
from the equation $D \cdot w^{(11)} \doteq \ldots$ we solve for
$w^{(11)}_{12}$ as a function of $\Lambda^{(12)}$, with an overall
factor still undetermined. After doing similar manipulation on eq.
(\ref{Cubic_3}), the undetermined overall factor in $w^{(11)}_{12}$
is fixed as a function of $\Lambda^{(12)}$. $w^{(02)}_{12}$ is
solved as a function of $\Lambda^{(12)}$, with an overall factor
still undetermined.

In summary, there remain two undetermined parameters in this order:
$\Lambda^{(12)}$ and an overall factor in $w^{(02)}_{12}$. We will
see them explicitly in the factorizable force model (Sec.
\ref{sec_qq}).

\subsection{Quartic Anharmonicity}

Similarly, we obtain from eqs. (\ref{quartic_1}) and
(\ref{quartic_2}):
\begin{small}
\begin{eqnarray}
\Lambda^{(40)} ~\cdot~ \sum_{e_1 \ne e_2} \sum_{e_3 \ne e_4}
\eta_{34} V_{3214} \frac{r^{(01)}_{12}}{e_{12}} \doteq  \nonumber
\\
\frac{i}{3} \Lambda^{(12)} \sum_{e_1 = e_{1'}} \sum_{e_3 \ne e_4}
\eta_{34}
V_{31'14} r^{(20)}_{11'} \nonumber \\
- \Lambda^{(30)} \sum_{e_1 \ne e_2} \sum_{e_3 \ne e_4} \eta_{34}
V_{3214} \frac{r^{(11)}_{12}}{e_{12}}  \nonumber \\
- \frac{i}{3}
\Lambda^{(30)} \sum_{e_1 =
e_{1'}} \sum_{e_3 \ne e_4} \eta_{34} V_{31'14} r^{(02)}_{11'} \nonumber \\
+ \frac{i}{2} \sum_{e_1 \ne e_2} \sum_{e_3 \ne e_4} \eta_{34}
V_{3214} \frac{[w^{(20)}, r^{(10)}]_{12} + [w^{(10)},
r^{(20)}]_{12}}{e_{12}}  \nonumber \\
- \frac{1}{6} \sum_{e_1 = e_{1'}} \sum_{e_3 \ne e_4} \eta_{34}
V_{31'14} ([w^{(20)},r^{(01)}]_{11'} + [w^{(01)},r^{(20)}]_{11'})
\nonumber
\\
- \frac{1}{3} \sum_{e_1 = e_{1'}} \sum_{e_3 \ne e_4} \eta_{34}
V_{31'14} ([w^{(11)},r^{(10)}]_{11'} + [w^{(10)},r^{(11)}]_{11'}) .
~~ \label{L40_c}
\end{eqnarray}
\end{small}
Eq. (\ref{L40_c}) gives $\Lambda^{(40)}_c + O(\omega^2)$. There is
one unknown parameter $\Lambda^{(12)}$; quantities $(r/w)^{(20)}$
and $(r/w)^{(11)}$ depend implicitly on $\Lambda^{(12)}$.\\

In summary, this section fixes the cubic potential $\Lambda^{(30)}$
(\ref{L30_c}) and the quartic potential $\Lambda^{(40)}$
(\ref{L40_c}) near the critical point $\omega^2 \approx 0$, by
considering the leading terms of the e.o.m. Deformation due to
$\Lambda^{(30)}$ will be studied elsewhere. Near the critical point,
the stability of the system should be restored by the quartic
potential $\Lambda^{(40)}$, if it is positive and large. In the
following we test this idea in three models of increasing
complexity: the Lipkin model (Sec. \ref{sec_Lipkin}), model with
factorizable forces (Sec. \ref{sec_qq}), and the quadrupole plus
pairing model (Sec. \ref{sec_QQPP}).

\section{Lipkin Model  \label{sec_Lipkin}}

We test the GDM method in the Lipkin model \cite{Lipkin} where the
analytical solution is available. As we will see, the agreement is
perfect (Sec. \ref{sec_Lipkin_comp}). Then we discuss some problems
inherent to the bosonic approach itself (Sec.
\ref{sec_Lipkin_numerical}).

\subsection{Exact Solution  \label{sec_Lipkin_exact}}

In this model, there are two s.p. levels with energies $\pm
\frac{1}{2}$ (the spacing is the energy unit), each with degeneracy
$\Omega + 1$. The model Hamiltonian contains only ``vertical''
transitions ($\sigma = \pm 1$; $l = 1 , 2 , ... , \Omega + 1$):
\begin{eqnarray}
H = \sum_{\sigma , l} \frac{\sigma}{2} a_{\sigma, l}^\dagger
a_{\sigma, l} + \frac{\kappa}{2} \sum_{\sigma , l , l'} a_{\sigma ,
l}^\dagger a_{\sigma , l'}^\dagger a_{- \sigma , l'} a_{- \sigma ,
l} . \label{H_Lipkin}
\end{eqnarray}

The quasi-spin operators,
\begin{eqnarray}
J_+ = J_-^\dagger = J_x + i J_y = \sum_l a_{+1,l}^\dagger a_{-1,l} ,
\nonumber \\
J_z = \frac{1}{2} \sum_{\sigma , l} \sigma a_{\sigma ,
l}^\dagger a_{\sigma , l} ,      \label{J_Lip}
\end{eqnarray}
satisfy the angular momentum algebra. Using eq. (\ref{J_Lip}) the
Hamiltonian (\ref{H_Lipkin}) is written as
\begin{eqnarray}
H = J_z + \frac{1}{2} \kappa (J_+^2 + J_-^2) , \label{H_J}
\end{eqnarray}
and the total quasi-spin $J$ is a good quantum number. With the
Holstein-Primakoff transformation (HPT),
\begin{eqnarray}
J_+ = J_-^\dagger = A^\dagger \sqrt{2 J - A^\dagger A} , ~~~ J_z = -
J + A^\dagger A ,   \label{HP_tran}
\end{eqnarray}
where $A^\dagger$ and $A$ are bosonic creation and annihilation
operators with commutation relation $[A , A^\dagger] = 1$, the
Hamiltonian (\ref{H_J}) is written as an expansion over $A^\dagger$
and $A$; or $\alpha$ and $\pi$ by the canonical transformation
\begin{eqnarray}
A = \frac{1}{\sqrt{2}} ( i u \alpha + v \pi ) ,~ A^\dagger =
\frac{1}{\sqrt{2}} ( - i u \alpha + v \pi ) ,~   u v = - 1 .~~
\label{can_tran}
\end{eqnarray}

Assuming $J \gg 1$, we keep only the leading order in $1/J$. Under
the choice
\begin{eqnarray}
u \approx \sqrt{1 + 2 \kappa J} , ~~~ v = - \frac{1}{u} ,
\label{u_v_choice}
\end{eqnarray}
the Hamiltonian becomes
\begin{eqnarray}
H = \frac{\omega^2}{2} ~ \alpha^2 + \frac{1}{2} ~ \pi^2 +
\frac{\Lambda^{(40)}}{4} ~\alpha^4 + \frac{\Lambda^{(04)}}{4} ~\pi^4
,   \label{H_b_Lipkin}
\end{eqnarray}
with
\begin{eqnarray}
\omega^2 \approx 1 - 4 \kappa^2 J^2 ,~~~ \Lambda^{(40)} \approx
\kappa u^4   ,~~~ \Lambda^{(04)} \approx - \kappa v^4 .
\label{H_b_Lip_Lmn}
\end{eqnarray}
Other $\Lambda^{(mn)}$ vanishes in their leading order of $1/J$.

Around the critical point $\omega^2 \approx 0$,
\begin{eqnarray}
\kappa \approx \frac{1}{2 J} ,~~~ u \approx \sqrt{2} ,~~~ v
\approx - \frac{1}{\sqrt{2}} ,  \label{omega_exact_Lip} \\
\Lambda^{(40)} \approx \frac{2}{J} ,~~~ \Lambda^{(04)} \approx -
\frac{1}{8 J} .  \label{L4004_c_Lip}
\end{eqnarray}

\subsection{The GDM Method}

Applying the GDM method to the Hamiltonian (\ref{H_Lipkin}), we have
solved for $r^{(mn)}_{12}$ explicitly in terms of $\Lambda^{(mn)}$
following Sec. \ref{sec_GDM}. Below we summarize the main results.
In the mean-field order, the HF s.p. levels are the same as the
original s.p. levels. Introducing $n \equiv n_{\sigma = - 1} -
n_{\sigma = 1}
> 0$, where $n_{\sigma}$ are occupation numbers of s.p. levels, in
the harmonic order the RPA secular equation (\ref{RPA}) becomes
\begin{eqnarray}
\omega^2 = 1 - (n \kappa \Omega)^2 .  \label{omega_Lipkin}
\end{eqnarray}
In the quartic order, the solvability condition
(\ref{quartic_sol_con}) becomes
\begin{eqnarray}
\frac{3}{\omega^2} \Lambda^{(40)} + \Lambda^{(22)} + 3 \omega^2
\Lambda^{(04)} =    \nonumber \\
\frac{10}{3} ~ \frac{1}{\omega^4} ~ (\Lambda^{(30)})^2 + 2
\frac{1}{\omega^2} ~ \Lambda^{(30)} \Lambda^{(12)} + \frac{3}{2}
(\Lambda^{(12)})^2  \nonumber \\
+ \frac{12}{n(\Omega + 1)} ~ \frac{1
- \omega^2}{\omega^2} .~~     \label{qua_sol_con_Lip}
\end{eqnarray}

\subsection{Comparison with Exact Solution  \label{sec_Lipkin_comp}}

The quantum number $J$ is found from eq. (\ref{J_Lip}):
\begin{eqnarray}
J = |J_z|_{\max} = \frac{\Omega + 1}{2} ~ |n_1 - n_{-1}| =
\frac{n(\Omega + 1)}{2} .  \label{J_Omega}
\end{eqnarray}

We assume $2 J = n(\Omega+1) \gg 1$. In the harmonic order, the RPA
secular equation (\ref{omega_Lipkin}) agrees with the HPT frequency
equation (\ref{H_b_Lip_Lmn}). In the quartic order, the HPT
solutions (\ref{H_b_Lip_Lmn}) satisfy our solvability condition
(\ref{qua_sol_con_Lip}).

The diverging behavior of eq. (\ref{qua_sol_con_Lip}) around the
critical point $\omega^2 \approx 0$ gives $\Lambda^{(40)}_c$. The
$\Lambda^{(30)}_c$ term must vanish as seen from the presence of the
term $\frac{10}{3} ~ \frac{1}{\omega^4} ~ (\Lambda^{(30)})^2$, which
is the only one divergent as $\omega^{-4}$. Equating the l.h.s. and
r.h.s. diverging terms $\sim \omega^{-2}$ we obtain
\begin{eqnarray}
\Lambda^{(40)}_c = \frac{4}{n(\Omega + 1)} .  \label{L40_c_Lip}
\end{eqnarray}
This agrees with the HPT solution (\ref{L4004_c_Lip}),
$\Lambda^{(40)} \approx \frac{2}{J} = \frac{4}{n(\Omega+1)}$. If we
follow the procedure in Sec. \ref{sec_Soft}, we obtain the same
result (\ref{L40_c_Lip}).

\subsection{Numerical Diagonalization and Discussion  \label{sec_Lipkin_numerical}}

Here we discuss some problems inherent to the bosonic approach
itself. The bosonic Hamiltonian (\ref{H_b}) is usually diagonalized
in the infinite phonon space; practically the space is enlarged
until convergence is reached. However, there exists a maximal phonon
number, close to the \emph{active} valence particle number in the
system. Applying the phonon creation operator $A^\dagger$ too many
times to the ground state, we run out of valence particles. We will
call this finite phonon space ``physical space''. Only if e.g. the
first excitation energy has reached convergence within the physical
space, it is valid to formally enlarge the Hilbert space to the
infinite space. This point is especially important for the soft
modes, where amplitudes of vibrations are large and may exceed the
range (maximal $\langle \alpha^2 \rangle$) of the physical space.

We illustrate this problem in the Lipkin model where we know the
physical space exactly. The HPT (\ref{HP_tran}) maps the angular
momentum space $\{ | J M \rangle \}$ onto the phonon space $\{ | n
\rangle \}$ (see Ref. \cite{rmp_boson}):
\begin{eqnarray}
| J M \rangle ~~~\rightarrow~~~ | n = M + J \rangle ,
\end{eqnarray}
where $| n \rangle$ is the eigenstate of $A^\dagger A$. Since $- J
\le M \le J$, we have $0 \le n \le 2J$. By eq. (\ref{J_Omega}), $2 J
= n(\Omega+1)$ is just the valence particle number.

Now we consider the possibility of diagonalizing eq.
(\ref{H_b_Lipkin}) in the infinite space. The negative
$\Lambda^{(04)} \pi^4 / 4$ term causes divergence. Thus we have two
steps of approximations: first, the $\Lambda^{(04)} \pi^4 / 4$ term
can be neglected when diagonalizing eq. (\ref{H_b_Lipkin}) in the
physical space $\{ | n \le 2 J \rangle \}$; second, the space can be
increased to the infinite space $\{ | n \le + \infty \rangle \}$.

The negative $\Lambda^{(04)} \pi^4 / 4$ term is smaller than the
$\pi^2 / 2$ term in the physical space (especially for the first few
excited states), on the $\omega^2
> 0$ side of the critical point. Eqs.
(\ref{u_v_choice}) and (\ref{H_b_Lip_Lmn}) give
\begin{eqnarray}
| \Lambda^{(04)} | = | \frac{1}{ 4 J + 4 \kappa J^2 + 1/\kappa} |
\le \frac{1}{ 8 J } .   \label{L04_ine}
\end{eqnarray}
The equality sign in eq. (\ref{L04_ine}) holds at the critical point
when $\kappa = \kappa_c = 1/(2J)$. On the $\omega^2 > 0$ side
\begin{eqnarray}
\frac{\langle n | \pi^4 | n \rangle}{\langle n | \pi^2 | n \rangle}
= \frac{ (6 n^2 + 6 n + 3) / (4 v^4) }{ (2 n + 1) / (2 v^2) } \le
\frac{ 6 n^2 + 6 n + 3 }{ 2 n + 1 } ,  \label{p4_o_p2}
\end{eqnarray}
where the equality sign holds at the critical point when $v^2 =
1/2$. Consequently
\begin{eqnarray}
\left| \frac{\langle\Lambda^{(04)} \pi^4 / 4\rangle}{\langle \pi^2/2
\rangle} \right| < \frac{1}{2} \cdot \frac{1}{ 8 J } \cdot 6 J =
\frac{3}{ 8 } . \label{quar_to_harm_ratio}
\end{eqnarray}
The upper limit of eq. (\ref{quar_to_harm_ratio}) is reached at the
critical point for the state with the maximal number of phonons. We
see that in the physical space the negative $\Lambda^{(04)} \pi^4 /
4$ term does not reverse the order of states. For the first few
excited states the upper limit in eq. (\ref{quar_to_harm_ratio}) is
actually \emph{much smaller}, of the order $1/J$, because the upper
limit in eq. (\ref{p4_o_p2}) is of the order $1$.

The space can be safely increased to the infinite space when $J$ is
large enough. The range of the physical space $\langle n = 2J |
\alpha^2 | n = 2J \rangle \sim J$ increases linearly with $J$. The
zero-point vibrations $\langle \alpha^2 \rangle$ in the first few
excited states also increase, but much slower. On the $\omega^2
> 0$ side, an upper limit is obtained when dropping the harmonic potential
$\omega^2 \alpha^2 / 2$ in eq. (\ref{H_b_Lipkin}), in which case
$\langle \alpha^2 \rangle \sim (\Lambda^{(40)})^{-\frac{1}{3}} \sim
J^{\frac{1}{3}}$. However, it is not justified when the collectivity
is not so large, or if $\Lambda^{(40)}$ is numerically small (thus
large zero-point vibrations, see Sec. \ref{sec_qq_example}).

We do a numerical example to illustrate the above two steps of
approximations. The results for the first excitation energy $E_1 -
E_0$, at the critical point $\omega^2 = 0$, are presented in Table
\ref{Table_1}. In the last two lines eq. (\ref{H_J}) is diagonalized
directly in the $\{ |JM\rangle \}$ space, where $\kappa$ takes the
critical value corresponding to $\omega^2 = 0$. In the last line the
critical $\kappa$ is calculated by the RPA secular equation
(\ref{omega_Lipkin}), with $n=1$. In the second last line the
critical $\kappa$ is calculated from
\begin{eqnarray}
\omega^2 \approx 1 - 4 \kappa^2 (J^2 + J) .
\label{better_Lipkin_RPA}
\end{eqnarray}
Eq. (\ref{better_Lipkin_RPA}) is better than eq. (\ref{H_b_Lip_Lmn})
because it is accurate not only in the leading order but also in the
next order of $1/J$.

The difference between line $1$ and line $4$ comes from neglecting
higher orders in $1/J$ of $\Lambda^{(mn)}$; between line $1$ and
line $2$ from neglecting the negative $- \frac{1}{8 J}
\frac{\pi^4}{4}$ term; between line $2$ and line $3$ from increasing
the space. We see that they agree quite well, and better for larger
$J$. The difference between line $4$ and line $5$ is because the RPA
secular equation is accurate in the leading order of $1/J$ but not
in the next order, which is the source of the biggest error in our
method.

In summary we argue that the existence of a \emph{finite} physical
boson space is general, in which the bosonic Hamiltonian should be
diagonalized. This Hamiltonian may have ``divergent-looking'' terms
[e.g. the negative $\Lambda^{(04)}$ term in eq. (\ref{H_b_Lipkin})],
which are indeed well-behaved in the \emph{finite} physical space.

However in general the exact physical space is unknown. Further
approximations are needed if the microscopically calculated (e.g. by
GDM) bosonic Hamiltonian is used to reproduce the spectrum of the
original fermionic Hamiltonian. First, the ``divergent'' terms must
be small and have little influence on the interested quantities,
thus they can be dropped. Second, the interested quantities must
have reached convergence within the physical space, thus formally
the bosonic Hamiltonian (without the ``divergent'' terms) can be
diagonalized in the infinite boson space. If the above two
conditions are not satisfied, the bosonic Hamiltonian encounters
serious difficulties or might be inapplicable in reproducing the
correct spectrum.

\section{Factorizable Force Model  \label{sec_qq}}

Here we consider the factorizable force model where the GDM method
provides approximate analytical results. They will be compared with
the exact results obtained by the shell model diagonalization. First
we introduce a Hermitian multipole operator
\begin{eqnarray}
Q = \sum_{12} q_{12} a_1^\dagger a_2 .     \label{Q_f}
\end{eqnarray}
For simplicity we assume $q_{12}$ is real; its hermiticity implies
$q_{12} = q_{21}$. Furthermore, we assume that $Q$ is time-even. The
model Hamiltonian is
\begin{eqnarray}
H = \sum_1 \epsilon_1 a_1^\dagger a_1 + \frac{1}{4} \sum_{1234} ( -
\kappa q_{14} q_{23} + \kappa q_{13} q_{24} ) a_1^\dagger
a_2^\dagger a_3 a_4 .~~
\end{eqnarray}
By definition of this model, the two-body part is different from
\begin{eqnarray}
- \frac{\kappa}{2} Q \cdot Q = - \frac{\kappa}{2} \sum_{12} (q
q)_{12} a_1^\dagger a_2  \nonumber \\
+ \frac{1}{4} \sum_{1234} ( -
\kappa q_{14} q_{23} + \kappa q_{13} q_{24}) a_1^\dagger a_2^\dagger
a_3 a_4
\end{eqnarray}
by a one-body term.

\subsection{The GDM Method  \label{sec_qq_GDM} }

The mapping of $Q$ is performed by substituting eq. (\ref{R_exp})
into eq. (\ref{Q_f}):
\begin{eqnarray}
Q\{R\} \equiv {\rm{Tr}}\{q R\} =  \nonumber \\
{\rm{Tr}}\{q \rho\} + {\rm{Tr}}\{q r^{(10)}\} \alpha + {\rm{Tr}}\{q
r^{(01)}\} \pi + {\rm{Tr}}\{q r^{(20)}\} \frac{\alpha^2}{2}  \nonumber \\
+ {\rm{Tr}}\{q r^{(02)}\} \frac{\pi^2}{2} + {\rm{Tr}}\{q r^{(11)}\} \frac{\{\alpha,\pi\}}{2} + \ldots \nonumber \\
= Q^{(00)} + Q^{(10)} \alpha + 0 \cdot \pi + Q^{(20)}
\frac{\alpha^2}{2}  \nonumber \\
+ Q^{(02)} \frac{\pi^2}{2} + 0 \cdot \frac{\{\alpha,\pi\}}{2} +
\ldots ,~   \label{Q_exp}
\end{eqnarray}
where $Q^{(mn)} = {\rm{Tr}}\{ q r^{(mn)} \}$. If $n$ is odd,
$Q^{(mn)}$ vanish since we assume that $Q$ is time-even. All
$Q^{(mn)}$ are real since $Q$ is Hermitian. The self-consistent
field becomes
\begin{eqnarray}
W\{R\}_{12} = \sum_{34} ( - \kappa q_{12} q_{34} + \kappa q_{14}
q_{32} ) R_{43} \nonumber \\
\approx - \kappa q_{12} \sum_{34} q_{34}
R_{43} = - \kappa q_{12} Q\{R\} ,     \label{W_QQ}
\end{eqnarray}
where we make the usual approximation keeping only the ``coherent''
summation. This is obvious in the harmonic order, where the
justification can be $r^{(10)}_{43} \sim q_{43}$; for higher orders
this approximation is discussed in Appendix \ref{app_qq_coherent}.
Substituting eq. (\ref{Q_exp}) into eq. (\ref{W_QQ}) we obtain the
expansion of $W\{R\}_{12}$.

Below we summarize the main results. Details including solutions for
$r^{(mn)}_{12}$ are given in Appendix \ref{app_qq}. In the
mean-field order we solve the HF equation (\ref{HF}):
\begin{eqnarray}
[f, \rho]_{12} = 0 ,~~~ f_{12} = Z_{12} - \kappa Q^{(00)} q_{12} .
\end{eqnarray}
Having in mind a spherical mean field, we assume that in the
solution $Q^{(00)} = Tr\{q \rho\} = 0$. Thus $f$ and $Z$ are the
same, $e_1 = \epsilon_1$.

In the harmonic order the RPA secular equation (\ref{RPA}) becomes:
\begin{eqnarray}
1 = - \kappa ~ \sum_{12} \frac{ |q_{12}|^2 ~ n_{12} e_{12}
}{(e_{12})^2 - \omega^2} .   \label{omega_QQ}
\end{eqnarray}
The normalization condition (\ref{norm}) becomes
\begin{eqnarray}
1 = - ~ (\kappa Q^{(10)})^2 \sum_{12} \frac{|q_{12}|^2 ~ n_{12}
e_{12}}{[(e_{12})^2 - \omega^2]^2} .   \label{norm_QQ}
\end{eqnarray}

For higher orders we give the leading order expressions in
$\omega^2$, following the procedure of Sec. \ref{sec_Soft}. In the
cubic order, eq. (\ref{L30_c}) becomes
\begin{eqnarray}
\Lambda^{(30)} \doteq  (\kappa Q^{(10)})^3 ~\Big(~ \sum_{e_1 \ne
e_2} \frac{[~ q , (\frac{n}{e} : q) ~]_{12} q_{21}}{e_{12}}
\nonumber \\
+ \sum_{e_1 = e_{1'}} \sum_2 \frac{q_{12} q_{21'}
q_{1'1} ~ n_{12}}{(e_{12})^2} ~\Big)  , ~~  \label{L30_c_QQ}
\end{eqnarray}
where we have introduced notations for the weight factors
$(\frac{n}{e} : q)_{12} \equiv \frac{n_{12}}{e_{12}} q_{12}$. In eq.
(\ref{L30_c_QQ}), the substitution of $\kappa Q^{(10)}$ by the
leading order of eq. (\ref{norm_QQ}) gives $\Lambda^{(30)}_c$.
Another equation (\ref{fix_f_w20}) becomes
\begin{eqnarray}
\Lambda^{(12)} \doteq - 2 ~ \frac{Q^{(20)}}{Q^{(10)}} + 4 (\kappa
Q^{(10)})^2 \Lambda^{(30)} \sum_{e_1 \ne e_2}
(\frac{n}{e^5}: q)_{12} q_{21} ~~~~~ \nonumber \\
+ 2 (\kappa Q^{(10)})^3 \cdot ~\Big(~ 2 \sum_{e_1 \ne e_2} \frac{[~
q , (\frac{n}{e} : q) ~]_{12} q_{21}}{(e_{12})^3} \nonumber \\
+ \sum_{e_1 \ne e_2} \frac{[q , (\frac{n}{e^2}: q)]_{12}
q_{21}}{(e_{12})^2} - \sum_{e_1 = e_{1'}} \sum_2 \frac{q_{12}
q_{21'} q_{1'1} ~ n_{12}}{(e_{12})^4} ~\Big) . ~~   \label{L12_c_QQ}
\end{eqnarray}
Eq. (\ref{L12_c_QQ}) determines $Q^{(20)}$ as a function of
$\Lambda^{(12)}$.

Summarizing the results in this order: there are two undetermined
parameters $\Lambda^{(12)}_c$ and $Q^{(02)}_c$; $\Lambda^{(30)}_c$
is fully determined; $Q^{(20)}_c$ and $r^{(20)}_{c 12}$,
$r^{(11)}_{c 12}$ are determined as a function of
$\Lambda^{(12)}_c$; $r^{(02)}_{c 12}$ is determined as a function of
$\Lambda^{(12)}_c$ and $Q^{(02)}_c$. In the present model
$Q^{(20)}_c$, $Q^{(02)}_c$ play the role of the ``undetermined
overall factor'' in $w^{(20)}_{12}$, $w^{(02)}_{12}$ of Sec.
\ref{sec_Soft}, respectively.

The quartic potential term (\ref{L40_c}) becomes
\begin{eqnarray}
\Lambda^{(40)} \doteq 2 (\kappa Q^{(10)})^2 (\Lambda^{(30)})^2
\sum_{e_1 \ne e_2} (\frac{n}{e^5}: q)_{12} q_{21}
\nonumber \\
- \Lambda^{(30)} \Lambda^{(12)}  \nonumber \\
+ (\kappa Q^{(10)})^3 \Lambda^{(30)} \cdot ~\{~ 2 \sum_{e_1 \ne e_2}
\frac{[~ q , (\frac{n}{e} : q) ~]_{12} q_{21}}{(e_{12})^3} \nonumber
\\
+ \sum_{e_1 \ne e_2} \frac{[q , (\frac{n}{e^2}: q)]_{12}
q_{21}}{(e_{12})^2} + \sum_{e_1 \ne e_2} \frac{[q , (\frac{n}{e^3}:
q)]_{12} q_{21}}{e_{12}} \nonumber \\
+ 2 \sum_{e_1 = e_{1'}}
\sum_{2} \frac{n_{12} q_{12} q_{21'} q_{1'1}}{(e_{12})^4} ~\}  \nonumber \\
+ 2 (\kappa Q^{(10)})^4 \cdot ~\{~  \sum_{e_1 \ne e_2} \sum_{e_3(\ne
e_2)} \frac{ q_{13} [~ q , (\frac{n}{e} : q) ~]_{32} q_{21}}{e_{12}
e_{32}} \nonumber \\
+ \sum_{e_1 \ne e_2} \sum_{e_{2'} (= e_2)}
\sum_3 \frac{ q_{12'}  q_{2'3} q_{32} ~
n_{23} q_{21}}{e_{12} (e_{23})^2}    \nonumber \\
+ \frac{2}{3} \sum_{e_1 = e_{1'}} \sum_{e_2(\ne e_1)} \frac{[~ q ,
(\frac{n}{e} : q) ~]_{12} q_{21'} q_{1'1}}{(e_{12})^2} \nonumber \\
+ \frac{1}{3} \sum_{e_1 = e_{1'}} \sum_{e_2(\ne e_1)} \frac{[q ,
(\frac{n}{e^2}: q)]_{12} q_{21'} q_{1'1}}{e_{12}} ~\} . ~~
\label{L40_c_QQ}
\end{eqnarray}
In eq. (\ref{L40_c_QQ}) there is an undetermined parameter
$\Lambda^{(12)}$.

\subsection{Two-Level Model  \label{sec_qq_example}}

Here the GDM method is compared with the exact diagonalization in a
simple two-level model (see 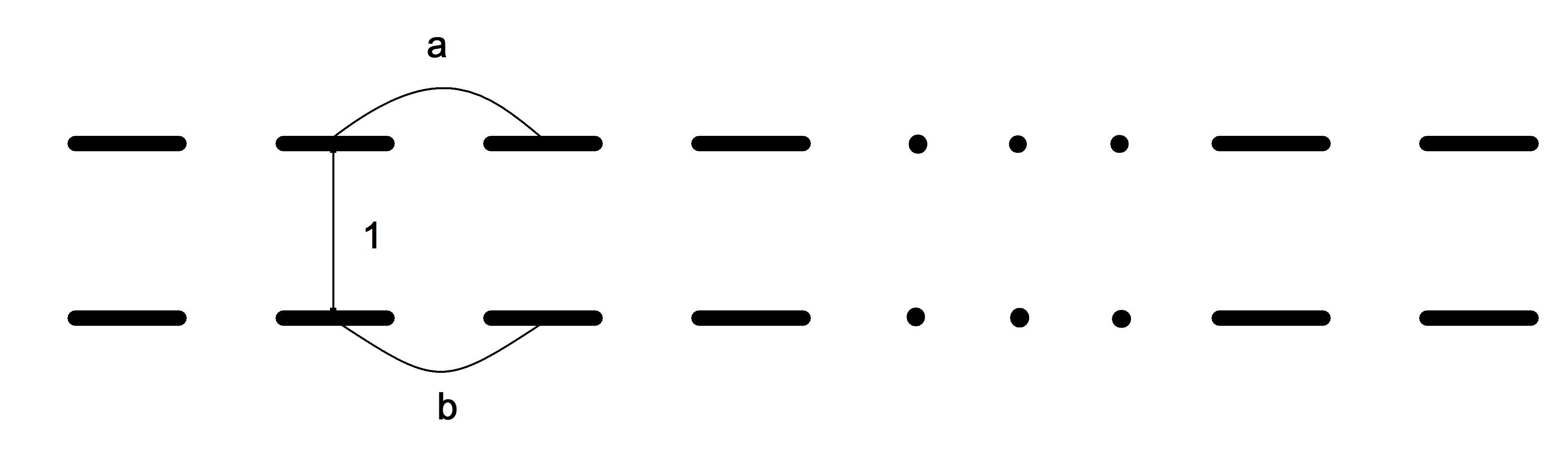). The model has two s.p.
levels with energies $\pm \frac{1}{2}$ (the spacing is the energy
unit), each with degeneracy $\Omega$. There are $N = \Omega$
particles. The nonzero matrix elements of $q$ are: vertical $q_{12}
= q_{21} = 1$, $q_{11'} = a/b$ for the \emph{nearest neighbors} of
the $+ \frac{1}{2} / - \frac{1}{2}$ s.p. levels (the leftmost and
rightmost s.p. levels are also connected by $a/b$). Each s.p. level
$1$ is connected to only a few (three) other s.p. levels by
$q_{12}$, thus the approximation in eq. (\ref{W_QQ}) is justified,
as explained in Appendix \ref{app_qq_coherent}. In summary the
interaction has three parameters: overall strength $\kappa$, and
ratios $a$, $b$.

In the mean-field order, $Q^{(00)} = \sum_1 n_1 q_{11} = 0$ since
$q_{11} = 0$. Hence s.p. energies $e$ are the same as $\epsilon$. In
the harmonic order, only the vertical $q_{12} = q_{21} = 1$ matrix
elements contribute. The RPA secular equation (\ref{omega_QQ})
becomes
\begin{eqnarray}
1 - \omega^2 = 2 \kappa \Omega .  \label{RPA_qq_tl}
\end{eqnarray}
Using eq. (\ref{RPA_qq_tl}) the normalization condition
(\ref{norm_QQ}) gives the collective amplitude
\begin{eqnarray}
(Q^{(10)})^2 = 2 \Omega .
\end{eqnarray}
In the cubic anharmonicity, $\Lambda^{(30)}_c = 0$ by eq.
(\ref{L30_c_QQ}), since there is no way to complete a three-body
loop. In the quartic anharmonicity, $\Lambda^{(40)}_c$ is calculated
from eq. (\ref{L40_c_QQ}):
\begin{eqnarray}
\Lambda^{(40)}_c = \frac{1}{\Omega} \cdot [ 1 - 2 (a-b)^2 ] .
\label{L40_c_qq_example}
\end{eqnarray}
The numerical diagonalization is done at $\Omega = 8$, thus
$\kappa_c = 1/16$ and $\Lambda^{(40)}_c = [1 - 2 (a-b)^2] / 8$.

\begin{figure*}
\includegraphics[width = 1.0\textwidth]{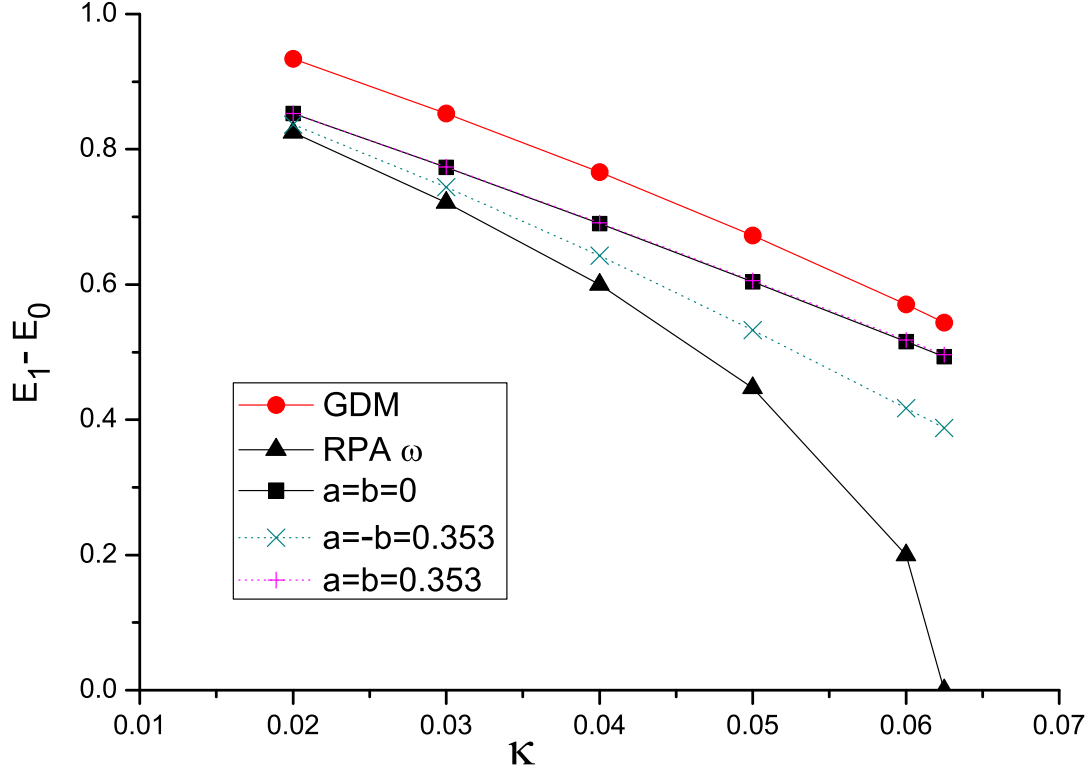}
\caption{\label{Fig_QQ1D} (Color online) The first excitation energy
$E_1 - E_0$ in the factorizable force model, as a function of
$\kappa$. The red circles result from diagonalizing $\frac{\omega^2
\alpha^2}{2} + \frac{\pi^2}{2}  + \frac{1}{4} \cdot
\frac{\alpha^4}{8}$ in the infinite phonon space. The black
triangles give the RPA frequency $\omega$, this corresponds to
diagonalizing $\frac{\omega^2 \alpha^2}{2} + \frac{\pi^2}{2}$. The
black squares, green crosses and purple pluses are the exact shell
model results with different values of $a$ and $b$ (two lines
``$a=b=0$'' and ``$a=b=0.353$'' closely overlap and are
indistinguishable on the figure). The second excitation energy $E_2
- E_0$ at $\kappa = \kappa_c = 1/16$ with $a=b=0$ is also in good
agreement; the exact one is $1.236$, while the GDM gives $1.269$.}
\end{figure*}

First we set the parameters $a = b = 0$. Fig. \ref{Fig_QQ1D} shows
the first excitation energy $E_1 - E_0$ as a function of $\kappa$
(for now ignore the two dot lines ``$a=b=0.353$'' and
``$a=-b=0.353$''). As $\kappa$ increases to the critical value, the
RPA frequency $\omega$ drops to zero, while $E_1 - E_0$ with the
quartic potential term $\Lambda^{(40)}$ remains finite and agrees
well with the exact results. This term restores the stability of the
system near the critical point. We emphasize that we have replaced
$\Lambda^{(40)}$ by $\Lambda^{(40)}_c = 1/8$ in eq. (\ref{Lmn_exp}),
thus we are making a big mistake when $\omega^2$ is large. However,
it does not matter too much since in this region $\omega^2 \alpha^2
/ 2$ dominates over $\Lambda^{(40)} \alpha^4 / 4$.

Next we consider the case of nonzero $a$ and $b$. Both the exact
$E_1 - E_0$ and our collective Hamiltonian are invariant under the
change $(a,b) \rightarrow (-a,-b)$, thus it is enough to consider
only positive $a$. From the three lines of Fig. \ref{Fig_QQ1D}
``$a=b=0$'', ``$a=b=0.353$'' and ``$a=-b=0.353$'', we see that the
exact $E_1 - E_0$ depends on $a-b$, but is almost independent of
$a+b$. This is in agreement with our collective Hamiltonian:
$\omega$ is independent of $a$, $b$ (in leading order of
$1/\Omega$); $\Lambda^{(40)}$ depends on $a - b$ but not on $a + b$.
In the region of small $\kappa$, the $\omega^2$ potential term
dominates thus $E_1 - E_0$ depends weakly on $a-b$; whereas in the
region of $\kappa \approx \kappa_c$, the $\Lambda^{(40)}$ potential
term is important thus $E_1 - E_0$ depends relatively strongly on
$a-b$.

\begin{figure*}
\includegraphics[width = 1.0\textwidth]{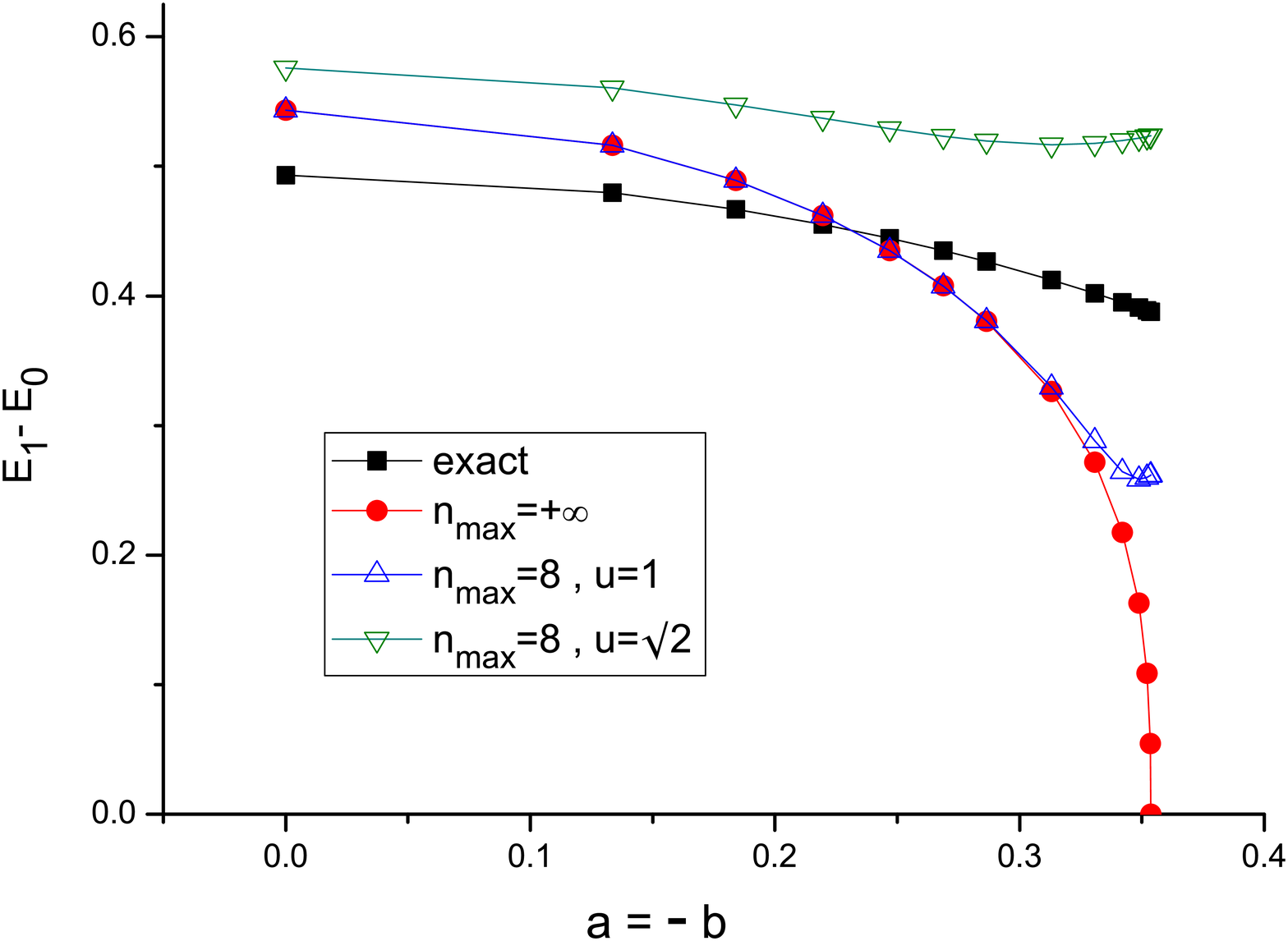}
\caption{\label{Fig_ab} (Color online) The first excitation energy
$E_1 - E_0$ in the factorizable force model, as a function of the
parameter $a = - b$, at the critical point $\omega = 0$ ($\kappa =
\kappa_c = 1/16$). The black squares show the exact shell model
results. The red circles are obtained by diagonalizing
$\frac{\pi^2}{2} + \frac{1}{4} \cdot \frac{[1 - 2 \cdot (a-b)^2]
\alpha^4}{8}$ in the infinite phonon space. The blue triangles and
the green inverted triangles are obtained by diagonalizing the same
Hamiltonian in two different finite phonon spaces, specified by
$n_{\rm{max}}$ and $u$; $u$ is the canonical transformation
parameter defined in eq. (\ref{can_tran}), $n_{\rm{max}}$ is the
maximal number of phonons. }
\end{figure*}

As $a = - b$ increases, $\Lambda^{(40)}_c$ decreases. At some point
$\Lambda^{(40)}_c$ becomes small numerically and $\frac{\pi^2}{2} +
\frac{\Lambda_c^{(40)} \alpha^4}{4}$ no longer describes the
behavior of the system near the critical point. First, other
anharmonic terms, suppressed by powers of $1/\Omega$, may become
important (see Appendix \ref{app_qq_40_dorm}). Second, even if there
are no other anharmonicities, the description breaks down because
the increasing zero-point-vibrations $\langle \alpha^2 \rangle$ will
exceed the range of the physical space, as discussed in Sec.
\ref{sec_Lipkin_numerical}. The current model has a larger
vibrational amplitude than the Lipkin model due to a smaller
$\Lambda^{(40)}_c$ ($\sim 1/\Omega$ verses $4/\Omega$). Fig.
\ref{Fig_ab} shows $E_1 - E_0$ as a function of the parameter $a = -
b$ at the critical point $\omega = 0$ ($\kappa = \kappa_c = 1/16$).
$E_1 - E_0$ depends on the space in which we diagonalize
$\frac{\pi^2}{2} + \frac{\Lambda_c^{(40)} \alpha^4}{4}$. Unlike in
the Lipkin model, we do not know a priori what the physical space is
in the current model. But it should be similar to that of the Lipkin
model with $8$ particles. Thus we choose $n_{\rm{max}} = 8$ for both
two finite spaces, each with a reasonable $u$ of eq.
(\ref{can_tran}). When $a = - b$ is small, say, less than $0.2$,
$E_1 - E_0$ of different spaces are close and all follow the trend
of the exact $E_1 - E_0$. When $a = - b$ is large, $E_1 - E_0$ of
different spaces differ substantially, implying that $\langle
\alpha^2 \rangle$ has reached the edge of the physical space, thus
the bosonic approach becomes invalid. If in the current model we
increase the collectivity $\Omega = N$, it is expected that $E_1 -
E_0$ from the GDM method will agree with the exact $E_1 - E_0$ up to
a larger value of $a = - b$.

In summary, near the critical point $\omega^2 \approx 0$, the next
even potential term $\Lambda^{(40)}$ dominates the dynamics of the
system, provided it is positive and large. $\Lambda^{(40)}$ should
be large enough such that other anharmonicities were negligible, and
zero-point vibrations $\langle \alpha^2 \rangle$ were within the
finite physical boson space. A larger collectivity factor $\Omega$
helps both, since other anharmonicities are suppressed by powers of
$\Omega^{-1}$ (see Appendix \ref{app_qq_40_dorm}), and the range of
the physical space grows as $\Omega$.

\section{Realistic Nuclear Application  \label{sec_real_nucl}}

There are three complications in realistic applications of the GDM
method. A realistic nucleus has two kinds of fermions; symmetries,
e.g. rotational invariance, need to be respected; pairing
correlations should be considered.

As in the BCS theory we substitute the original system by a
grand-canonical ensemble, in which the chemical potential is fixed
by the average particle number of the ground state in the mean-field
order. In this case we need to consider e.o.m. of not only
$a_2^\dagger a_1$ but also $a_2 a_1$. A good treatment of the
superfluid ground state, on top of which collective excitations are
formed, is essential.

The collective mode operators $\alpha_{\lambda \mu}$, $\pi_{\lambda
\mu}$ have quantum numbers corresponding to symmetries of the
Hamiltonian. In this section we keep only the quadrupole mode which
is the most important one at low energy. The case of interacting
modes (quadrupole and octupole) is discussed briefly in Appendix
\ref{app_mode_coupling}.

This section is a straightforward generalization of Sec.
\ref{sec_GDM}. The details of the derivation are given in Appendix
\ref{app_real_nucl}.

\subsection{Preparation}

The microscopic fermionic Hamiltonian for the canonical ensemble is
still given by eq. (\ref{H_f}): we include the $ - \mu \hat{N}$ term
in $\hat{Z}$, and the s.p. index $1$, $2$ ... can run over protons
and neutrons. Isospin may not be conserved for some effective
interactions. We do not write $V$ in the form $V^J_{(j_1 j_2),(j_3
j_4)}$; $Z_{12}$ and $V_{1234}$ carry all the symmetries of $H$
implicitly.

Now the reference state $|\Phi\rangle$ does not have definite
particle number,
\begin{eqnarray}
\langle \Phi | a_1^\dagger a_2 | \Phi \rangle \equiv \rho_{21} ,~~~
\langle \Phi | a_1 a_2 | \Phi \rangle \equiv \kappa_{21} .
\label{dm_def_mu}
\end{eqnarray}
$\kappa$ is the pair correlator \cite{pair_correlator}. Also we need
two generalized density matrix operators
\begin{eqnarray}
R_{12} \equiv a_2^\dagger a_1 ,~~~  K_{12} \equiv a_2 a_1 ,
\label{R_K_def_mu}
\end{eqnarray}
and two self-consistent field operators
\begin{eqnarray}
W\{R\}_{12} \equiv \sum_{34} V_{1432} R_{34} ,~~~   f\{R\} \equiv Z + W\{R\} ,  \label{W_def_mu} \\
\Delta\{K\}_{12} \equiv \frac{1}{2} \sum_{34} V_{1234} K_{43} .
\label{Delta_def_mu}
\end{eqnarray}
It will be convenient to introduce ($R^T$, $f^T$ are transpose)
\begin{eqnarray}
D\{R,K\} \equiv \left(\begin{array}{cc}
         R & K \\
         K^\dagger & I - R^T \\
       \end{array} \right) , \nonumber \\
S\{R,K\} \equiv \left( \begin{array}{cc}
f\{R\} & \Delta\{K\} \\
\Delta^\dagger\{K\} & - f^T\{R\} \\
\end{array} \right) .   \label{D_S_def}
\end{eqnarray}

The collective mode operators $\alpha^\dagger_{\lambda\mu}$,
$\pi^\dagger_{\lambda\mu}$ carry quantum numbers of angular momentum
$\lambda$, its projection $\mu$, and parity $(-)^\lambda$. The
coordinate $\alpha^\dagger_{\lambda\mu}$ is time-even, and the
momentum $\pi^\dagger_{\lambda\mu}$ is time-odd. Their Hermitian
properties are
\begin{eqnarray}
\alpha_{\lambda \mu}^\dagger = (-)^{\lambda - \mu} \alpha_{\lambda -
\mu} ,~~~ \pi_{\lambda \mu}^\dagger = (-)^{\lambda - \mu}
\pi_{\lambda - \mu} .
\end{eqnarray}
The commutation relation is given by
\begin{eqnarray}
[\alpha_{\lambda \mu}^\dagger , \pi_{\lambda' \mu'}] = i ~
\delta_{\lambda \lambda'} \delta_{\mu \mu'} .
\end{eqnarray}
Here we consider only the quadrupole mode $\lambda = 2$, and drop
the label $\lambda$.

The collective Hamiltonian replacing eq. (\ref{H_b}) should be
written with correct vector coupling of the operators:
\begin{eqnarray}
H = E_0 + \frac{\omega^2}{2} \sqrt{5} (\alpha \times \alpha)^0_0 +
\frac{1}{2} \sqrt{5} (\pi \times \pi)^0_0 \nonumber \\
+ \frac{\Lambda^{(30)}}{6} \sqrt{5} \{ (\alpha \times \alpha)^2 ,
\alpha \}^0_0 + \frac{\Lambda^{(12)}}{4} \sqrt{5} \{ \alpha , (\pi
\times \pi)^2
\}^0_0 \nonumber \\
+ \frac{\Lambda^{(40)}}{4} \sqrt{5} ( (\alpha
\times \alpha)^0 \times (\alpha \times
\alpha)^0 )^0_0 \nonumber \\
+ \frac{\Lambda^{(04)}}{4} \sqrt{5} ( (\pi \times \pi)^0 \times (\pi
\times \pi)^0 )^0_0 \nonumber \\
+ \sum_{L = 0,2,4} \frac{\Lambda^{(22)}_L}{8} \sqrt{5} \{ (\alpha
\times \alpha)^L , (\pi \times \pi)^L \}^0_0 . \label{H_b_mu}
\end{eqnarray}
$H$ is Hermitian, time-even, invariant under rotation and inversion.

\subsection{Equations of Motion in the Collective Band  \label{sec_real_nucl_eom_band}}

Following the same procedure as in Sec. \ref{sec_GDM}, we find
e.o.m. replacing those in Sec.
\ref{sec_GDM_mf}-\ref{sec_GDM_quartic_anharm}. Matrices $D^{(mn)}$,
$S^{(mn)}$ are coefficients of expanding $D\{R,K\}$, $S\{R,K\}$ over
collective operators $\alpha^\dagger_\mu$, $\pi^\dagger_\mu$. In the
mean-field order we obtain the HFB equation
\begin{eqnarray}
[ S\{\rho,\kappa\} , D\{\rho,\kappa\} ] = [ S^{(00)} , D^{(00)} ] =
0 .   \label{HFB}
\end{eqnarray}
In the harmonic order we obtain the QRPA equations
\begin{eqnarray}
\underline{\pi^\dagger_\mu} : ~~~~~~~~~ i D^{(10)}_\mu = [S^{(00)} ,
D^{(01)}_\mu] +
[S^{(01)}_\mu , D^{(00)}] ,   \label{QRPA1} \\
\underline{\alpha^\dagger_\mu} : ~~ - i \omega^2 D^{(01)}_\mu =
[S^{(00)} , D^{(10)}_\mu] + [S^{(10)}_\mu , D^{(00)}] .
\label{QRPA2}
\end{eqnarray}
In the cubic order:
\begin{eqnarray}
\underline{ (\alpha^\dagger \times \alpha^\dagger)^L_\mu / 2
 , L
= 0,2,4: } ~~~~~~~~~~~~~~~~~~~~~~~~~~~~~~ \nonumber \\
- 2 i \omega^2 D^{(11)}_{L\mu} - 2 i \delta_{L,2} \Lambda^{(30)}
D^{(01)}_\mu = \nonumber \\
{[}S^{(00)} , D^{(20)}_{L\mu}] + [S^{(20)}_{L\mu} , D^{(00)}] + 2 ~
[S^{(10)} , D^{(10)}]^{L}_\mu ,~   \label{Cubic1_mu} \\
\underline{ (\pi^\dagger \times \pi^\dagger)^L_\mu / 2
, L = 0,2,4: } ~~~~~~~~~~~~~~~~~~~~~~~~~~~~~~ \nonumber \\
2 i D^{(11)}_{L\mu} - i \delta_{L,2} \Lambda^{(12)} D^{(01)}_\mu =
\nonumber \\
{[}S^{(00)} , D^{(02)}_{L\mu}] + [S^{(02)}_{L\mu} , D^{(00)}] + 2 ~
[S^{(01)} ,
D^{(01)}]^{L}_\mu  ,~    \label{Cubic2_mu} \\
\underline{ \{ \alpha^\dagger , \pi^\dagger \}^L_\mu / 2
, L = 0,1,2,3,4: } ~~~~~~~~~~~~~~~~~~~~~~~~~~ \nonumber \\
- i \delta_{L,{\rm{even}}} \omega^2 D^{(02)}_{L\mu} + i
\delta_{L,{\rm{even}}} D^{(20)}_{L\mu} + i
\delta_{L,2} \Lambda^{(12)} D^{(10)}_\mu  \nonumber \\
= {[}S^{(00)} , D^{(11)}_{L\mu}] + [S^{(11)}_{L\mu} , D^{(00)}]
\nonumber \\
+ [S^{(10)} , D^{(01)}]^{L}_\mu - [D^{(10)} , S^{(01)}]^{L}_\mu .~
\label{Cubic3_mu}
\end{eqnarray}
In the quartic order:
\begin{small}
\begin{eqnarray}
\underline{ \{ (\alpha^\dagger \times \alpha^\dagger)^{l_L} ,
\alpha^\dagger\}^L_\mu / 6, L = 0,2,3,4,6: } ~~~~~~~~  \nonumber \\
- \frac{3 i}{2} \omega^2 \sum_{l=0,2,4} D^{(21)}_{Ll\mu} \cdot
\gamma^L_{l,l_L} - 3 i (-)^{L} \Lambda^{(30)}
D^{(11)}_{L\mu} \cdot \gamma^L_{2,l_L} \nonumber \\
- 3 i \delta_{L 2} \Lambda^{(40)} D^{(01)}_\mu \cdot
\gamma^{L=2}_{0,l_L} = [ S^{(00)} ,
D^{(30)}_L ] + [ S^{(30)}_L , D^{(00)} ] \nonumber \\
+ \frac{3}{2} \sum_{l=0,2,4} ( [ S^{(20)}_l , D^{(10)} ]^L_\mu - [
D^{(20)}_l , S^{(10)} ]^L_\mu ) \cdot \gamma^L_{l,l_L} ,~
\label{Quartic1_mu}
\end{eqnarray}
\end{small}
\begin{small}
\begin{eqnarray}
\underline{ \{ (\alpha^\dagger \times \alpha^\dagger)^l
, \pi^\dagger\}^L_\mu /4 ; l = 0,2,4; L = 0,1,2,3,4,5,6: } \nonumber \\
- 2 i (-)^{L} \omega^2 \sum_{l'=0,2,4} D^{(12)}_{Ll'\mu} \cdot
g^L_{l,l'} + \frac{2 i}{3} ~ \delta_{l,l_L} D^{(30)}_{L\mu}
\nonumber \\
+ \frac{4~i}{3} D^{(30)}_{L\mu} \cdot g^L_{l,l_L} - 2 i ~
\delta_{l2} ~ \Lambda^{(30)} D^{(02)}_{L\mu} \nonumber \\
+ 2 ~ i \Lambda^{(12)}
D^{(20)}_{L\mu} \cdot g^L_{l,2} + i ~ \delta_{L2} ~ \Lambda^{(22)}_l D^{(10)}_\mu  \nonumber \\
= [ S^{(00)} , D^{(21)}_{Ll\mu} ] + [ S^{(21)}_{Ll\mu} , D^{(00)} ]
\nonumber \\
+ [ S^{(20)}_l , D^{(01)} ]^L_\mu - [ D^{(20)}_l ,
S^{(01)}
]^L_\mu   \nonumber \\
+ 2 \sum_{l'=0,1,2,3,4} ( [ S^{(11)}_{l'} , D^{(10)} ]^L_\mu - [
D^{(11)}_{l'} , S^{(10)} ]^L_\mu ) \cdot g^L_{l,l'} ,
\label{Quartic2_mu}
\end{eqnarray}
\end{small}
\begin{small}
\begin{eqnarray}
\underline{ \{ \alpha^\dagger , (\pi^\dagger \times
\pi^\dagger)^l \}^L_\mu /4 ; l = 0,2,4; L = 0,1,2,3,4,5,6: } \nonumber \\
- \frac{2i}{3} \delta_{l,l_L} (-)^{L} \omega^2 D^{(03)}_{L\mu} -
\frac{4i}{3} \omega^2 D^{(03)}_{L\mu}   (-)^{L - l_L} \cdot
g^L_{l,l_L} \nonumber \\
+ 2 i  (-)^{L} \sum_{l'=0,2,4} D^{(21)}_{Ll'\mu} \cdot g^L_{l,l'} -
i \delta_{l 2} \Lambda^{(12)} D^{(11)}_{L\mu} \nonumber \\
+ 2 i (-)^{L} \Lambda^{(12)} D^{(11)}_{L\mu} \cdot g^L_{l,2}
- i ~ \delta_{L2} ~ \Lambda^{(22)}_l D^{(01)}_\mu  \nonumber \\
= [ S^{(00)} , D^{(12)}_{Ll\mu} ] + [ S^{(12)}_{Ll\mu} , D^{(00)} ]
\nonumber \\
+ [ S^{(10)} , D^{(02)}_l ]^L_\mu - [ D^{(10)} ,
S^{(02)}_l
]^L_\mu   \nonumber \\
+ 2 \sum_{l'=0,1,2,3,4} ( [ S^{(11)}_{l'} , D^{(01)} ]^L_\mu - [
D^{(11)}_{l'} , S^{(01)} ]^L_\mu ) (-)^{L - l'} g^L_{l,l'} ,~~~
\label{Quartic3_mu}
\end{eqnarray}
\end{small}
\begin{small}
\begin{eqnarray}
\underline{ \{ (\pi^\dagger \times \pi^\dagger)^{l_L} ,
\pi^\dagger \}^L_\mu /6 , L = 0,2,3,4,6 : } ~~~~~~~~~~~~~~~~~~~~~  \nonumber \\
\frac{3 i}{2} (-)^{L} \sum_{l=0,2,4} D^{(12)}_{Ll\mu} \cdot
\gamma^L_{l,l_L} - \frac{3 i}{2} \Lambda^{(12)} D^{(02)}_{L\mu}
\cdot \gamma^L_{2,l_L} \nonumber \\
+ 3 i \delta_{L2} \Lambda^{(04)} D^{(10)}_\mu \cdot
\gamma^{L=2}_{0,l_L} = [ S^{(00)} , D^{(03)}_L ] + [ S^{(03)}_L ,
D^{(00)} ] \nonumber
\\
+ \frac{3}{2} \sum_{l=0,2,4} ( [ S^{(02)}_l , D^{(01)} ]^L_\mu  - [
D^{(02)}_l , S^{(01)} ]^L_\mu ) \cdot \gamma^L_{l,l_L} .~~
\label{Quartic4_mu}
\end{eqnarray}
\end{small}
The numerical coefficients $\gamma^L_{l,l'}$ and $g^L_{l,l'}$ are
defined by
\begin{eqnarray}
\{ (\alpha \times \alpha)^l , \alpha \}^L_\mu = \gamma^L_{l,l'}
\cdot \{ (\alpha \times \alpha)^{l'} , \alpha \}^L_\mu ,~ (
\gamma^L_{l,l} = 1 ) ,    \label{gamma_def}  \\
\frac{1}{8} \{ \{\alpha , \pi\}^{l'} , \alpha \}^L_\mu =
\sum_{l=0,2,4} g^L_{l,l'} \cdot \frac{1}{4} \{ (\alpha \times
\alpha)^l , \pi\}^L_\mu . ~ \label{g_def1}
\end{eqnarray}
Values of $\gamma^{L}_{l,l'}$ and $g^L_{l,l'}$ are given in Appendix
\ref{app_g_gamma}. $l_L$ in $\{ (\alpha^\dagger \times
\alpha^\dagger)^{l_L} , \alpha^\dagger\}^L_\mu /6$ is the choice of
basis, different choices of $l_L$ do not influence results.

There exists a relation involving cubic and quartic anharmonicities,
replacing eq. (\ref{quartic_sol_con}). Setting $L=2$, keeping only
$(D/S)^{(30/21/12/03)}$ and $\Lambda^{(40)}$, $\Lambda^{(22)}_l$,
$\Lambda^{(04)}$ terms, $f_{l_L} \times$ eq. (\ref{Quartic1_mu}) $+
\frac{1}{2} \omega^2 \sum_{l=0,2,4} f_l \times$ eq.
(\ref{Quartic3_mu}) gives
\begin{eqnarray}
\underline{ L = 2: } ~~~ - i \omega^2 \cdot \Big(~ \frac{1}{2}
\sum_{l=0,2,4} f_l \cdot D^{(21)}_{Ll\mu} + \omega^2 f_{l_L} \cdot
D^{(03)}_{L\mu} ~\Big) \nonumber
\\
= [~ S^{(00)} ~,~ \Big(~ f_{l_L} \cdot D^{(30)}_L +  \frac{1}{2}
\omega^2
\sum_{l=0,2,4} f_l \cdot D^{(12)}_{Ll\mu} ~\Big) ~] \nonumber \\
+ [~ \Big(~ f_{l_L} \cdot S^{(30)}_L + \frac{1}{2} \omega^2
\sum_{l=0,2,4} f_l \cdot S^{(12)}_{Ll\mu} ~\Big) ~,~ D^{(00)} ~]
\nonumber
\\
+ 3 i \Lambda^{(40)} D^{(01)}_\mu \cdot f_0 + \frac{i}{2} \omega^2
\sum_{l=0,2,4} f_l \cdot \Lambda^{(22)}_l D^{(01)}_\mu + \ldots ~~~
\label{quar_sol1_mu}
\end{eqnarray}
$\frac{1}{2}  \sum_{l=0,2,4} f_l \times$ eq. (\ref{Quartic2_mu}) $+
\omega^2 f_{l_L}
 \times$ eq. (\ref{Quartic4_mu}) gives
\begin{eqnarray}
\underline{ L = 2 : } ~~~~~~~~~ i \cdot \Big(~ f_{l_L} \cdot
D^{(30)}_{L\mu} + \frac{1}{2} \omega^2 \sum_{l=0,2,4} f_l \cdot
D^{(12)}_{Ll\mu}
~\Big) \nonumber \\
= [~ S^{(00)} ~,~ \Big(~ \frac{1}{2} \sum_{l=0,2,4} f_l \cdot
D^{(21)}_{Ll\mu} +  \omega^2 f_{l_L} \cdot D^{(03)}_L ~\Big) ~]
\nonumber
\\
+ [~ \Big(~ \frac{1}{2} \sum_{l=0,2,4} f_l \cdot S^{(21)}_{Ll\mu} +
\omega^2 f_{l_L} \cdot S^{(03)}_L ~\Big) ~,~ D^{(00)} ~] \nonumber \\
- \frac{i}{2} \sum_{l=0,2,4} f_l \cdot \Lambda^{(22)}_l D^{(10)}_\mu
- 3 i \omega^2 \Lambda^{(04)} D^{(10)}_\mu \cdot f_0 + \ldots ~~~
\label{quar_sol2_mu}
\end{eqnarray}
where $f_l$ is defined in eq. (\ref{f_val}). A solvability condition
exists because the variable parts of eqs. (\ref{quar_sol1_mu}) and
(\ref{quar_sol2_mu}) have the same structure as the QRPA equations
(\ref{QRPA1}) and (\ref{QRPA2}).

Following the procedure in Sec. \ref{sec_Soft}, we can obtain
expressions of $\Lambda^{(30)}_c$ and $\Lambda^{(40)}_c$. In the
next section we do this explicitly for the quadrupole plus pairing
model.

\section{Quadrupole plus Pairing Model  \label{sec_QQPP}}

In this section the GDM method is applied to the quadrupole plus
pairing Hamiltonian. As was understood long ago \cite{P_Q1, P_Q2},
this model combines the most important nuclear collective phenomena
in particle-particle (pairing) and particle-hole (quadrupole mode)
channels. The approximate analytical results of the GDM method are
compared below with the exact results of the shell model
diagonalization. The operator of multiple moment is defined as
\begin{eqnarray}
Q^\dagger_{\lambda \mu} \{R\} = Tr\{q^\dagger_{\lambda \mu} R\} =
\sum_{12} q^\dagger_{\lambda \mu12} a_1^\dagger a_2 ,   \\
q_{\lambda\mu}^\dagger = f_\lambda(r) \cdot i^\lambda ~
Y_{\lambda\mu}(\theta,\phi) ,  \label{q_def_mu}
\end{eqnarray}
where $f_\lambda(r)$ is real. The definition of eq. (\ref{q_def_mu})
differs from the ``usual'' one in two aspects: a factor $i^\lambda$
is included, and $q_{\lambda\mu}^\dagger \sim Y_{\lambda\mu}$
instead of $q_{\lambda\mu}$, thus $q_{\lambda\mu}^\dagger$ creates
projection $\mu$. The Hermitian properties are
\begin{eqnarray}
q^\dagger_{\lambda \mu} = (-)^{\lambda - \mu} q_{\lambda \mu} ,~~~
Q^\dagger_{\lambda \mu} = (-)^{\lambda - \mu} Q_{\lambda \mu} .
\label{Q_herm_mu}
\end{eqnarray}
The pairing operators $P$ and $P^\dagger$ are defined by
\begin{eqnarray}
P^\dagger = \frac{1}{2} \sum_{1} a_1^\dagger a_{\tilde{1}}^\dagger
,~~~ P = \frac{1}{2} \sum_{1} a_{\tilde{1}} a_1 ,
\end{eqnarray}
where $\tilde{1}$ is the time-reversed s.p. level of $1$. $P$ has
angular momentum $0$ and positive parity. $P + P^\dagger$ is
time-even, $P - P^\dagger$ is time-odd.

The quadrupole plus pairing Hamiltonian is (dropping $\lambda = 2$)
\begin{eqnarray}
H = \sum_1 (\epsilon_1 - \mu) a_1^\dagger a_1 - \frac{G}{4}
\sum_{12}
a_1^\dagger a_{\tilde{1}}^\dagger a_{\tilde{2}} a_2  \nonumber \\
+ \frac{1}{4} \sum_{1234} \sum_{\mu} ( - \kappa ~ q^\dagger_{\mu 14}
q_{\mu 23} + \kappa ~ q^\dagger_{\mu 13} q_{\mu 24} ) a_1^\dagger
a_2^\dagger a_3 a_4 .  \label{H_QP_mu}
\end{eqnarray}
Approximately, this Hamiltonian can be written as $H \approx \sum_1
(\epsilon_1 - \mu) a_1^\dagger a_1 - G P^\dagger P - \frac{1}{2}
\kappa \sum_{\mu} Q_\mu^\dagger Q_\mu$. The difference is in a
one-body term originating from the $Q \cdot Q$ part. $H$ is
Hermitian and time-even, which implies real $G$, $\kappa$,
$\epsilon_1 = \epsilon_{\tilde{1}}$. In a realistic nucleus there
are protons and neutrons; formally we can still use eq.
(\ref{H_QP_mu}) if the quadrupole force strengths are the same for
proton-proton, neutron-neutron, and proton-neutron ($\kappa_p =
\kappa_n = \kappa_{pn} = \kappa$), while remembering the pairing is
treated for protons and neutrons separately ($G_p \ne G_n$). We will
assume this is the case.

\subsection{The GDM Method}

\subsubsection{BCS}

In the pairing plus quadrupole model the HFB equation (\ref{HFB})
becomes the BCS equation:
\begin{eqnarray}
\Delta \cdot \Big(~ 1 - \frac{G}{4} \sum_{1} \frac{1}{E_1} ~\Big) = 0 ,    \label{gap_eq}  \\
e_1 = \epsilon_1 - \mu - G (v_1)^2 , \\
E_1 = \sqrt{ (e_1)^2 + (\Delta)^2 } ,  \\
(u_1)^2 = \frac{1}{2} ~ ( 1 + \frac{e_1}{E_1} ) ,~~~ (v_1)^2 =
\frac{1}{2} ~ ( 1 - \frac{e_1}{E_1} ) ,  \\
N = \sum_1 (v_1)^2 .    \label{mu_eq}
\end{eqnarray}
BCS amplitudes $u_{1} = u_{\tilde{1}}$, $v_{1} = v_{\tilde{1}}$ are
real. Pairing energy $\Delta$ is a real number, not to be confused
with the field $\Delta\{K\}$ in eq. (\ref{Delta_def_mu}) that is an
operator matrix. $E_1$ is the quasiparticle energy. The chemical
potential $\mu$ is fixed by eq. (\ref{mu_eq}). The gap equation
(\ref{gap_eq}) has a non-trivial solution $\Delta > 0$ only if $G$
is greater than its critical value $G_c$ \cite{P_Q2}. For
convenience we introduce:
\begin{eqnarray}
\xi^{(2)}_{\mu 12} \equiv \frac{(u_1 v_2 + u_2 v_1)}{(E_1 + E_2)^2}
q_{\mu 12} ,~ \eta_{\mu 12} \equiv (u_1 u_2 - v_1 v_2) q_{\mu 12}
.~~~
\end{eqnarray}

\subsubsection{QRPA}

The QRPA secular equation corresponding to eq. (\ref{omega_QQ}) is
given by
\begin{eqnarray}
1 = \kappa \sum_{12}  \frac{ (E_1 + E_2) ~ |\xi_{\mu 21}|^2 }{(E_1 +
E_2)^2 - \omega^2} .   \label{omega_QP_mu}
\end{eqnarray}
The solution $\omega^2$ is independent of $\mu$. Results in the form
of reduced matrix elements are given in Appendix \ref{app_QQPP_mu}.
The normalization condition corresponding to eq. (\ref{norm_QQ}) is
\begin{eqnarray}
1 = (\kappa Q^{(10)})^2 ~ \sum_{12} \frac{(E_1 + E_2) ~ |\xi_{\mu
21}|^2 }{[(E_1 + E_2)^2 - \omega^2]^2} . \label{norm_QP_mu}
\end{eqnarray}

\subsubsection{Cubic Anharmonicity}

The cubic anharmonicity corresponding to eq. (\ref{L30_c_QQ}) is
\begin{eqnarray}
\Lambda^{(30)} \doteq  3 (\kappa Q^{(10)})^3 \sum_{n_1 j_1 n_2 j_2
n_3 j_3} \nonumber \\
\sqrt{\frac{(2 j_1 + 1)(2 j_2 + 1)(2 j_3 +
1)}{5}} \cdot \left\{
                                          \begin{array}{ccc}
                                            2   & 2   & 2 \\
                                            j_1 & j_2 & j_3 \\
                                          \end{array}
                                        \right\}
\nonumber \\
\cdot ~ \xi^{(1) \dagger}_{\| 12} ~ \xi^{{(1)} \dagger}_{\| 23} ~
\eta^{\dagger}_{\| 31} ,   \label{L30_c_red_QP_mu}
\end{eqnarray}
where $\xi^{(1) \dagger}_{\| 12} \equiv \langle n_1 j_1 \| \xi^{(1)
\dagger} \| n_2 j_2 \rangle$ is the reduced matrix element, the
convention for which is given in Appendix \ref{app_convention}.
$n_1$ combines all other quantum numbers specifying a s.p. level,
except $j_1$.

We give the expression of $P^{(20)}$ which will appear in
$\Lambda^{(40)}_c$:
\begin{eqnarray}
P^{(20)} \cdot [~ 1 - G \sum_{n_1 j_1} (2 j_1 + 1) \frac{
[(u_1)^2 - (v_1)^2]^2}{4 E_1} ~] \doteq \nonumber \\
- (\kappa Q^{(10)})^2 \sum_{n_1 j_1 n_2 j_2} (2 j_1 + 1) \cdot
\sqrt{2 j_2 + 1} \left\{
                                          \begin{array}{ccc}
                                            2   & 2   & 0   \\
                                            j_1 & j_1 & j_2 \\
                                          \end{array}
                                        \right\} \nonumber \\
\cdot ~\Big[~ 2 u_1 v_1  \cdot \xi^{{(1)} \dagger}_{\| 12}
\xi^{{(1)} \dagger}_{\| 21} - ~ \frac{(u_1)^2 - (v_1)^2}{E_1} \cdot
\eta^{\dagger}_{\| 12} \xi^{{(1)} \dagger}_{\| 21} ~\Big] .~~~
\label{P20_c_red_mu}
\end{eqnarray}
$P^{(20)}$ is divergent when $G$ is greater than but close to $G_c$.
In this region of the pairing phase transition, $\Delta$ is small,
and $P^{(20)} \sim 1/\Delta$. The GDM $+$ BCS method is not valid in
this region: in the mean-field order the BCS solution already fails,
as is well known.

\subsubsection{Quartic Anharmonicity}

The quartic anharmonicity corresponding to eq. (\ref{L40_c_QQ}) is
\begin{small}
\begin{eqnarray}
f_0 \cdot \Lambda^{(40)} \doteq - 2 ~ f_2 \cdot (\kappa Q^{(10)})^2
(\Lambda^{(30)})^2 ~ {\rm{Tr}} \Big[ \xi_{\mu} \xi^{{(5)}
\dagger}_{\mu} \Big] \nonumber \\
- f_2 \cdot \Lambda^{(30)} \Lambda^{(12)}  \nonumber \\
- f_2 \cdot \Lambda^{(30)} (\kappa Q^{(10)})^3 \Big( ~ {\rm{Tr}}
\Big[ \{ \xi^{(1)} , \xi^{(3)} \}^{L=2}_{\mu} \eta^{\dagger}_{\mu} \Big] \nonumber \\
+ {\rm{Tr}} \Big[ \{ \eta , \xi^{(3)} \}^{L=2}_{\mu} \xi^{{(1)}
\dagger}_{\mu} \Big] + 2 ~ {\rm{Tr}} \Big[ \{ \eta , \xi^{(1)}
\}^{L=2}_{\mu} \xi^{{(3)}
\dagger}_{\mu} \Big] \nonumber \\
+ {\rm{Tr}} \Big[ \{ \eta , \xi^{{(2)}} \}^{L=2}_{\mu} \xi^{{(2)}
\dagger}_{\mu} \Big] ~ \Big) \nonumber
\\
- f_0 \cdot G ~ P^{(20)} (\kappa Q^{(10)})^2 ~ \sum_{12} (u_1
v_1 + u_2 v_2) ~ \xi_{\mu 12} \xi^{{(2)} \dagger}_{\mu 21}  \nonumber \\
+ f_0 \cdot G ~ P^{(20)} (\kappa Q^{(10)})^2 \nonumber \\
\cdot \sum_{12} \{ \frac{(u_1)^2 - (v_1)^2}{2 E_1} +
\frac{(u_2)^2 - (v_2)^2}{2 E_2} \} \xi^{(1)}_{\mu 12} \eta^{ \dagger}_{\mu 21}  \nonumber \\
+ (\kappa Q^{(10)})^4 \sum_{l=0,2,4} f_l \cdot \Big(  {\rm{Tr}}
\Big[ \{ \xi , ( \xi^{(1)} \times \xi^{(1)} )^{l} \}^{L=2}_{\mu}
\xi^{{(1)}
\dagger}_{\mu} \Big] \nonumber \\
- {\rm{Tr}} \Big[ \{ \eta , \{ \eta , \xi^{{(1)}} \}^{l,(1)}
\}^{L=2}_{\mu} \xi^{{(1)} \dagger}_{\mu} \Big] \nonumber \\
- {\rm{Tr}} \Big[ \{ \xi^{(1)} , \{ \eta , \xi^{{(1)}} \}^{l,(1)}
\}^{L=2}_{\mu} \eta^{ \dagger}_{\mu} \Big] ~\Big) ,
\label{L40_c_QP_mu}
\end{eqnarray}
\end{small}
where $\{\eta , \xi^{(1)}\}^{l,(1)}_{12} = \{\eta ,
\xi^{(1)}\}^{l}_{12} / (E_1 + E_2)$. There is an undetermined
parameter $\Lambda^{(12)}$ in eq. (\ref{L40_c_QP_mu}). Values of
numerical factors $f_l$ are given in Appendix \ref{app_g_gamma}.

\subsection{Comparison with Exact Results  \label{sec_QQPP_example}}

We compare the results of our method in a semi-realistic model with
those of NuShellX \cite{MSU_Nu}. There are $10$ fermions of one kind
and four s.p. levels with energies:
\begin{eqnarray}
\left.
  \begin{array}{c|c|c|c|c}
  {\rm{s.p. ~ levels}}    & 1p\frac{1}{2}  & 0f\frac{7}{2} & 1p\frac{3}{2} & 0f\frac{5}{2} \\
  \epsilon ~ {\rm{(MeV)}} & -0.1 & 0.0 & 1.0 & 1.1 \\
  \end{array}  \nonumber
\right.
\end{eqnarray}
We take the radial wavefunctions to be harmonic oscillator ones. In
eq. (\ref{q_def_mu}) we take $f(r)$ to be $r^2$ so $q^\dagger_\mu =
- \hat{r}^2 Y_{2\mu}(\hat{\theta} , \hat{\phi})$. For convenience we
make $q^\dagger_\mu$ dimensionless by combining its original
dimension with $\kappa$ (see the end of Appendix
\ref{app_convention}). The model space is similar to the realistic
$pf$-shell, but the $1p\frac{1}{2}$ and $1p\frac{3}{2}$ levels are
inverted to increase collectivity: in the current case the $q$
matrix elements ($q_{1p\frac{1}{2},0f\frac{5}{2}}$ and
$q_{1p\frac{3}{2},0f\frac{7}{2}}$) are large between the s.p. levels
above and below the Fermi surface.

\begin{figure*}
\includegraphics[width = 1.0\textwidth]{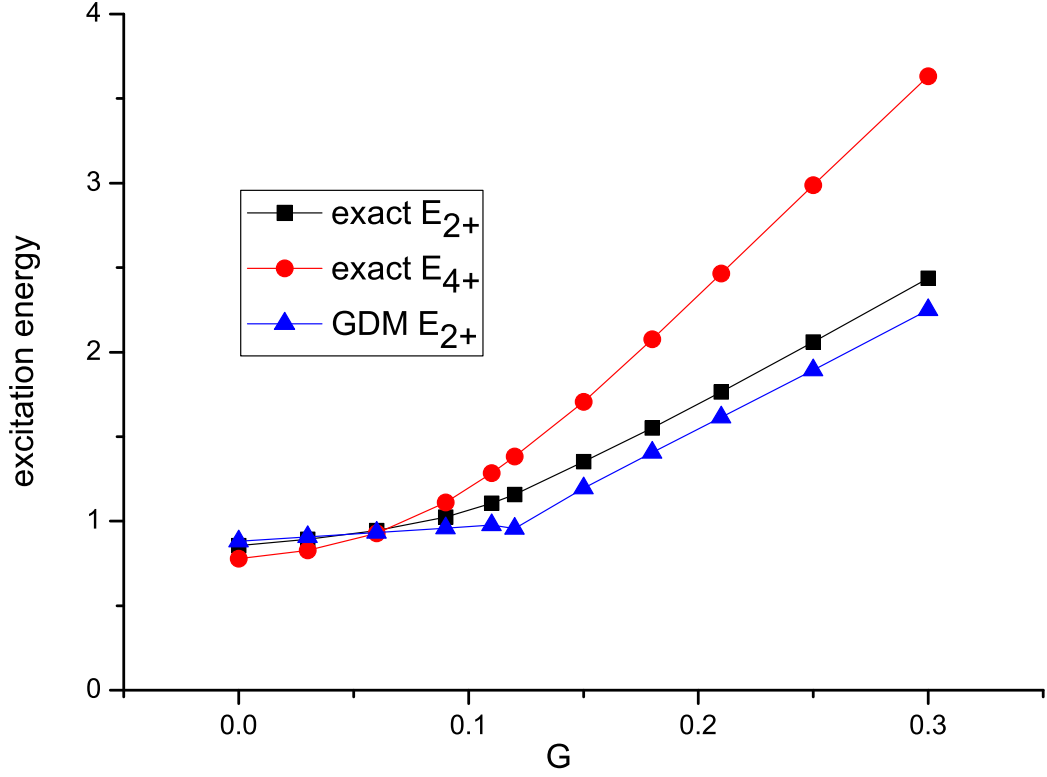}
\caption{\label{Fig_QQ_PP3D} (Color online) Excitation energies
(from Table \ref{Table_2}) in the quadrupole plus pairing model as a
function of the pairing strength $G$, at the critical point $\omega
= 0$ ($\kappa = \kappa_c$). The black squares and red circles show
the exact excitation energy of the first $2^+$ and $4^+$ state,
respectively, ``NuShellX $E_{2_1^+}$'' and ``NuShellX $E_{4_1^+}$''
in Table \ref{Table_2}. The blue triangles give ``GDM $E_{2^+}$''
from Table \ref{Table_2}. }
\end{figure*}

We did a set of calculations with increasing pairing strength $G$.
At each value of $G$, the strength $\kappa$ of the $Q \cdot Q$ force
is taken to be at the critical value $\kappa_c$ such that the RPA
frequency $\omega^2 = 0$. The results are summarized in Table
\ref{Table_2}. For clarity, we draw the last three lines of Table
\ref{Table_2} as Fig. \ref{Fig_QQ_PP3D}. The coefficient
$\Lambda^{(40)}$ in Table \ref{Table_2} is calculated by eq.
(\ref{L40_c_QP_mu}) setting $\Lambda^{(12)} = 0$ (dropping the $-
f_2 \cdot \Lambda^{(30)} \Lambda^{(12)}$ term). A non-zero term
$\Lambda^{(12)}$ in its reasonable range does not influence
$\Lambda^{(40)}$ much, since in the current model $\Lambda^{(30)}$
is small due to the approximate symmetry with respect to the Fermi
surface (see Table \ref{Table_2}). Then ``GDM $E_{2^+}$'' is
calculated by diagonalizing eq. (\ref{H_b_mu}), setting
$\Lambda^{(12)} = \Lambda^{(04)} = \Lambda^{(22)}_L = 0$ ($\omega^2
= 0$ since $\kappa$ takes its critical value).

The critical value of the pairing strength $G_c$ is around $0.11
\sim 0.12$ MeV. When $G < G_c$, the BCS solution $\Delta = 0$, and
$\mu$ can be anywhere between $\epsilon_{0f\frac{7}{2}} = 0$ and
$\epsilon_{1p\frac{3}{2}} = 1.0$ MeV. We checked that in this case
our results (\ref{omega_QP_mu}-\ref{L40_c_QP_mu}) do not depend on
the choice of $\mu$. In Table \ref{Table_2} we fix $\mu$ at $0.5$
MeV. In the region where $G$ is greater than but close to $G_c$, our
method is invalid as discussed under eq. (\ref{P20_c_red_mu}). This
is illustrated in Fig. \ref{Fig_QQ_PP3D} by the `kink' on the ``GDM
$E_{2^+}$'' curve near $G \sim 0.12$.

In Fig. \ref{Fig_QQ_PP3D} ``exact $E_{2^+}$'' and ``exact
$E_{4^+}$'' are the exact results by NuShellX. At $G = 0$ the first
excited state is $4^+$ instead of $2^+$. In this case the $4^+$
state is a single-particle excitation from $0f\frac{7}{2}$ to
$0f\frac{5}{2}$; the $2^+$ state is a collective state with
approximately half holes in $1p\frac{1}{2}$ and $0f\frac{7}{2}$
levels, half particles in $1p\frac{3}{2}$ and $0f\frac{5}{2}$. As
$G$ increases, the collective $2^+$ state becomes the first excited
state. When $G$ is large enough, $\Delta$ dominates over the
original s.p. spacing $\epsilon$, and the results become stable. As
an example, at $G = 0.30$ MeV, the quasiparticle continuum starts at
$\sim 3.5$ MeV; from the second excited state $0^+$ at $3.506$ MeV
to $4.153$ MeV there are $15$ states with $J^P = 0^+, 2^+, 4^+,
6^+$. The first excited state $2^+$ at $2.438$ MeV should be
identified as a collective state, stabilized at around sixty
percents within the gap.

It is seen in Fig. \ref{Fig_QQ_PP3D} that ``GDM $E_{2^+}$'' agrees
well with the exact result ``exact $E_{2^+}$'' in general. On the $G
< G_c$ side, our $E_{2^+}$ does increase with $G$ although not
rapidly enough. On the $G > G_c$ side, when $\Delta$ is not too
small, the agreement is very good.\\

In summary, this section shows the potential of the GDM method in
doing realistic calculations. In medium and heavy nuclei the pairing
gap $\sim 2 \Delta \sim 2$ MeV, the critical region is approximately
bounded by $| \omega | < 1~{\rm{MeV}}$. Nuclei on the $\omega^2 < 0$
side are gamma-unstable. On the $\omega^2 > 0$ side, the whole
region can be calculated as in Fig. \ref{Fig_QQ1D} [explained in the
paragraph under eq. (\ref{L40_c_qq_example})].

\section{Conclusions  \label{sec_conclusion}}

The GDM method is promising in solving the longstanding problem:
constructing the collective bosonic Hamiltonian microscopically. The
procedure is straightforward and consistent. Results of the lowest
orders, the well-known HFB and QRPA equations, give us confidence to
proceed to higher order anharmonicities. The anharmonicities are
important as the harmonic potential $\omega^2 \alpha^2 / 2$ becomes
small or negative when going away from closed shells. The GDM method
provides a unified description of different collective phenomena,
including soft vibrational modes of large amplitudes, gamma-unstable
potential and transition to static deformation. It maps the exact
fermionic e.o.m. onto the dynamics generated by approximate
collective operators. Here we used the phonon-like operators; other
possibilities include rotational dynamics and the dynamics
corresponding to the symplectic symmetry or other group-theoretical
models. In such cases the GDM expansion should be based on the group
generators.

Sec. \ref{sec_GDM} discusses the general procedure of the GDM
method. In each order, a set of coupled linear equations is solved
in terms of lower order results. At the current stage the
anharmonicities are not completely fixed; we find only one relation
(\ref{quartic_sol_con}) involving the cubic and quartic
anharmonicities, appearing in the third order as a solvability
condition. In Sec. \ref{sec_Soft} it is shown that around the
critical point $\omega^2 \approx 0$, we are able to determine the
cubic potential $\Lambda^{(30)}$ (\ref{L30_c}) and the quartic
potential $\Lambda^{(40)}$ (\ref{L40_c}). $\Lambda^{(40)}$ should be
responsible for restoring the stability of the system near the
critical point, if it is positive and large. This idea is then
tested in three models of increasing complexity: the Lipkin model
(Sec. \ref{sec_Lipkin}), model with factorizable forces (Sec.
\ref{sec_qq}), and the quadrupole plus pairing model (Sec.
\ref{sec_QQPP}). The GDM method is only responsible for calculating
$\Lambda^{(mn)}$; other conditions are needed if the resultant
bosonic Hamiltonian is used to reproduce the spectrum of the
original fermionic Hamiltonian, as discussed in the last two
paragraphs of Sec. \ref{sec_Lipkin_numerical}. If these conditions
are not fulfilled, the approach of effective bosonic Hamiltonian
encounters serious difficulties. The conditions for quartic
potential ($\Lambda^{(40)} \alpha^4 / 4$) dominance near the
critical point are discussed in the last paragraph of Sec.
\ref{sec_qq_example}.

Calculations for realistic nuclei are in progress. However, the
pairing correlations need to be treated better than in the BCS
framework, because anharmonicities are sensitive to the occupation
numbers ($u_1$, $v_1$) of the superfluid ground state. Unlike the
QRPA secular equation (\ref{omega_QP_mu}), where terms in the
summation contribute coherently, in the expressions of
anharmonicities (\ref{L30_c_red_QP_mu}) and (\ref{L40_c_QP_mu})
different terms may cancel. $\Lambda^{(30)}$ and $\Lambda^{(40)}$
depend on the balancing above and below the Fermi surface, thus they
are sensitive to the occupation numbers ($u_1$, $v_1$). Work is also
in progress about the role of $\Lambda^{(30)}$ on deformation, as
well as the quadrupole-octupole coupling in the presence of a
low-lying octupole mode. The realistic effective interactions
(better than the quadrupole plus pairing Hamiltonian) are to be used
in the calculation. The present paper sets the scene for the GDM
method in the sense that it is seen explicitly there are no
contradictions in the solutions (Sec. \ref{sec_GDM} and
\ref{sec_real_nucl}), although at the current stage we find only one
constraint (\ref{quartic_sol_con}) on the anharmonicities. New
constraints, if found, would fix the anharmonicities completely.\\

{\bf Acknowledgements:}  The author gratefully expresses many thanks
to his advisor Vladimir Zelevinsky, who provided superior guidance
and help during the whole work. Support from the NSF grant
PHY-0758099 is acknowledged. The author is also thankful to the
National Superconducting Cyclotron Laboratory and Department of
Physics and Astronomy at Michigan State University.

\appendix

\section{Three-Body Force  \label{app_3b}}

It is straightforward to include three-body forces in the
formulation. The microscopic Hamiltonian (\ref{H_f}) includes a new
(anti-symmetrized) term
\begin{eqnarray}
H^{(3)} = \frac{1}{36} \sum_{123456} G_{123456} a_1^\dagger
a_2^\dagger a_3^\dagger a_4 a_5 a_6 .    \label{H_f_3b}
\end{eqnarray}
Under the definition
\begin{eqnarray}
G\{R\}_{1256} \equiv \sum_{34} G_{123456} R_{43} , \nonumber \\
G\{R,R\}_{14} \equiv \sum_{23} G\{R\}_{1234} R_{32}  ,
\end{eqnarray}
the normal ordering Hamiltonian (\ref{H_norm_order}) acquires new
terms,
\begin{eqnarray}
\langle \Phi | H^{(3)} | \Phi \rangle = \frac{1}{6} {\rm{Tr}} [
G\{\rho,\rho\} \rho ]  ,\nonumber \\
f^{(3)} = \frac{1}{2}
G\{\rho,\rho\} ,~~~ V^{(3)} = G\{\rho\}  ,
\end{eqnarray}
and a term $\frac{1}{36} \sum_{123456} G_{123456} N[a_1^\dagger
a_2^\dagger a_3^\dagger a_4 a_5 a_6]$. In the e.o.m. (\ref{e.o.m.})
$f$ and $V$ are replaced by the new ones including $f^{(3)}$ and
$V^{(3)}$ ($W\{R\}$ is calculated from the new $V$), and there are
two additional terms:
\begin{eqnarray}
+ \frac{1}{4} [ G\{R^{2N}\}, \rho ]_{12} + \frac{1}{12} \sum_{34567}
( G_{134567} N[a_2^\dagger a_3^\dagger a_4^\dagger a_5 a_6 a_7]
\nonumber \\
- G_{765432} N[a_7^\dagger a_6^\dagger a_5^\dagger a_4 a_3 a_1] )
,~~ \label{eom_3b}
\end{eqnarray}
where
\begin{eqnarray}
G\{R^{2N}\}_{16} \equiv \sum_{2345} G_{123456} N[a_2^\dagger
a_3^\dagger a_4 a_5] .
\end{eqnarray}

Formally the HF and RPA equations are the same as before, replacing
$f$ and $W\{R\}$ by the new ones.

\section{Saturation Principle for Section \ref{sec_GDM}   \label{app_sat}}

Keeping only one-body terms in $\alpha$ and $\pi$,
\begin{eqnarray}
\alpha = \sum_{12} x_{12} R_{21} = \sum_{12} x_{12} a_1^\dagger a_2  ,  \label{a_1b} \\
\pi = \sum_{12} p_{12} R_{21} = \sum_{12} p_{12} a_1^\dagger a_2  ,
\label{p_1b}
\end{eqnarray}
we have the following identities in the full space:
\begin{eqnarray}
[ R_{12} , \alpha ] = [x, R]_{12} ,   \label{sat_x} \\
{[} R_{12} , \pi ] = [p, R]_{12} ,    \label{sat_p}
\end{eqnarray}
where $[x, R]_{12} = \sum_3 ( x_{13} R_{32} - R_{13} x_{32} )$.
Similarly to the manipulation of eq. (\ref{e.o.m.}), we project eqs.
(\ref{sat_x},\ref{sat_p}) onto the collective subspace. Since
$\alpha$ and $\pi$ are collective operators, we can substitute $R$
by its boson expansion (\ref{R_exp}). After calculating commutators
on the l.h.s. , we equate coefficients of the same phonon structure:
$1$, $\alpha$, $\pi$, $\frac{\alpha^2}{2}$ ... Eq. (\ref{sat_x})
gives
\begin{eqnarray}
- i r^{(01)}_{12} = [x, \rho]_{12} = x_{12} (n_2 - n_1) ,  \label{x_ele} \\
- i r^{(02)} = {[ x, r^{(01)} ]} ,~~~ - i r^{(11)} = {[x, r^{(10)}]
} .  \label{x_higher}
\end{eqnarray}
Eq. (\ref{sat_p}) gives
\begin{eqnarray}
i r^{(10)}_{12} = [p, \rho]_{12} = p_{12} (n_2 - n_1) ,  \label{p_ele} \\
i r^{(20)} = {[ p, r^{(10)} ]} ,~~~ i r^{(11)} = {[p, r^{(01)}] } .
\label{p_higher}
\end{eqnarray}
Only the $n_1 \ne n_2$ matrix elements of $x$ and $p$ are determined
from eqs. (\ref{x_ele}) and (\ref{p_ele}). Higher order expressions
(\ref{x_higher}) and (\ref{p_higher}) are approximate, saying that
$r^{(20/11/02)}$ are completely fixed by the the harmonic order
solutions. In fact the two expressions of $r^{(11)}$ are not
consistent with each other. These defects are due to the neglected
many-body components in eqs. (\ref{a_1b}) and (\ref{p_1b}), as
explained in Appendix \ref{app_ap_mb}. The approximate expressions
(\ref{x_higher}) and (\ref{p_higher}) are used below to derive
expressions of $r^{(mn)}_{1234}$ in terms of $r^{(mn)}_{12}$.

In the full space we also have
\begin{eqnarray}
[ N[a_4^\dagger a_3^\dagger a_2 a_1], \alpha ] = \nonumber \\
- \sum_5 x_{25} N[a_4^\dagger a_3^\dagger a_1 a_5] + \sum_5 x_{15}
N[a_4^\dagger a_3^\dagger a_2 a_5] \nonumber \\
- \sum_6 x_{64} N[a_6^\dagger a_3^\dagger a_2 a_1] + \sum_{6} x_{63}
N[a_6^\dagger a_4^\dagger a_2 a_1] \nonumber
\\
+ i r^{(01)}_{24} N[a_3^\dagger a_1] - i r^{(01)}_{23} N[a_4^\dagger
a_1] \nonumber \\
- i r^{(01)}_{14} N[a_3^\dagger a_2] + i
r^{(01)}_{13} N[a_4^\dagger a_2] ,    \label{sat_4x}
\end{eqnarray}
where we have used eq. (\ref{x_ele}). Similarly
\begin{eqnarray}
[ N[a_4^\dagger a_3^\dagger a_2 a_1], \pi ] = \nonumber \\
- \sum_5
p_{25} N[a_4^\dagger a_3^\dagger a_1 a_5] + \sum_5 p_{15}
N[a_4^\dagger a_3^\dagger a_2 a_5] \nonumber \\
- \sum_6 p_{64} N[a_6^\dagger a_3^\dagger a_2 a_1] + \sum_{6} p_{63}
N[a_6^\dagger a_4^\dagger a_2 a_1] \nonumber
\\- i r^{(10)}_{24}
N[a_3^\dagger a_1] + i r^{(10)}_{23} N[a_4^\dagger a_1] \nonumber
\\
+ i r^{(10)}_{14} N[a_3^\dagger a_2] - i r^{(10)}_{13}
N[a_4^\dagger a_2] ,   \label{sat_4p}
\end{eqnarray}
where we have used eq. (\ref{p_ele}). Again we project eqs.
(\ref{sat_4x}) and (\ref{sat_4p}) onto the collective subspace, then
substitute the expansions (\ref{R_exp}) and (\ref{RR_exp}). Both the
l.h.s. and the r.h.s. have no constant terms. This justifies the
assumption under eq. (\ref{RR_exp}): terms linear in $\alpha$ and
$\pi$ are absent in the expansion (\ref{RR_exp}) of $N[a_4^\dagger
a_3^\dagger a_2 a_1]$, since they generate constant terms in the
l.h.s. of eqs. (\ref{sat_4x}) and (\ref{sat_4p}). The $\alpha$ and
$\pi$ terms of eqs. (\ref{sat_4x}) and (\ref{sat_4p}) give
\begin{eqnarray}
r^{(20)}_{1234} = r^{(10)}_{14} r^{(10)}_{23} - r^{(10)}_{13}
r^{(10)}_{24} - r^{(10)}_{24} r^{(10)}_{13} + r^{(10)}_{23}
r^{(10)}_{14}  \label{r20_1234} ,~~~ \\
r^{(02)}_{1234} = r^{(01)}_{14} r^{(01)}_{23} - r^{(01)}_{13}
r^{(01)}_{24} - r^{(01)}_{24} r^{(01)}_{13} + r^{(01)}_{23}
r^{(01)}_{14}  \label{r02_1234} ,~~~ \\
r^{(11)}_{1234} = r^{(10)}_{14} r^{(01)}_{23} - r^{(10)}_{13}
r^{(01)}_{24} - r^{(10)}_{24} r^{(01)}_{13} + r^{(10)}_{23}
r^{(01)}_{14}  \label{r11_1234} .~~~~
\end{eqnarray}
We mention that eq. (\ref{sat_4x}) and (\ref{sat_4p}) give the same
expression of $r^{(11)}_{1234}$ (\ref{r11_1234}). Using eqs.
(\ref{x_higher}), (\ref{p_higher}) and
(\ref{r20_1234}-\ref{r11_1234}), the $\alpha^2/2$, $\{\alpha ,
\pi\}/2$, $\pi^2/2$ terms of eqs. (\ref{sat_4x}) and (\ref{sat_4p})
give
\begin{small}
\begin{eqnarray}
r^{(30)}_{1234} = \frac{3}{2} ( r^{(20)}_{14} r^{(10)}_{23} -
r^{(20)}_{13} r^{(10)}_{24} -
r^{(20)}_{24} r^{(10)}_{13} + r^{(20)}_{23} r^{(10)}_{14} ) ,~~~  \label{r30_1234} \\
r^{(03)}_{1234} = \frac{3}{2} ( r^{(02)}_{14} r^{(01)}_{23} -
r^{(02)}_{13} r^{(01)}_{24} -
r^{(02)}_{24} r^{(01)}_{13} + r^{(02)}_{23} r^{(01)}_{14} ) ,~~~  \label{r03_1234} \\
r^{(21)}_{1234} = r^{(20)}_{14} r^{(01)}_{23} - r^{(20)}_{13}
r^{(01)}_{24} - r^{(20)}_{24} r^{(01)}_{13} + r^{(20)}_{23}
r^{(01)}_{14}
\nonumber \\
+ 2 r^{(11)}_{14} r^{(10)}_{23} - 2 r^{(11)}_{13} r^{(10)}_{24} - 2
r^{(11)}_{24} r^{(10)}_{13} + 2 r^{(11)}_{23} r^{(10)}_{14} ,~~~  \label{r21_1234} \\
r^{(12)}_{1234} = r^{(10)}_{14} r^{(02)}_{23} - r^{(10)}_{13}
r^{(02)}_{24} - r^{(10)}_{24} r^{(02)}_{13} + r^{(10)}_{23}
r^{(02)}_{14} \nonumber \\
+ 2 r^{(11)}_{14} r^{(01)}_{23} - 2 r^{(11)}_{13} r^{(01)}_{24} - 2
r^{(11)}_{24} r^{(01)}_{13} + 2 r^{(11)}_{23} r^{(01)}_{14} .~~~~
\label{r12_1234}
\end{eqnarray}
\end{small}
Eq. (\ref{sat_4x}) and eq. (\ref{sat_4p}) give the same expression
of $r^{(21)}_{1234}$ (\ref{r21_1234}) and $r^{(12)}_{1234}$
(\ref{r12_1234}). The results (\ref{r20_1234}-\ref{r12_1234})
generalize the so-called linearization of e.o.m. method,
\begin{eqnarray}
a_4^\dagger a_3^\dagger a_2 a_1 \rightarrow \rho_{14} a_3^\dagger
a_2 - \rho_{13} a_4^\dagger a_2 - \rho_{24} a_3^\dagger a_1 +
\rho_{23} a_4^\dagger a_1 .~~~~
\end{eqnarray}

The normalization of the RPA solution $r^{(10)}$, $r^{(01)}$ is
determined by the commutator $[ \alpha , \pi ] = i$. Under the
one-body assumption (\ref{a_1b}) and (\ref{p_1b}),
\begin{eqnarray}
i = [\alpha,\pi] = {\rm{Tr}}\{[x,p]R\} \nonumber \\
= {\rm{Tr}}\{[x,p]\rho\} + {\rm{Tr}}\{[x,p] r^{(10)} \} \alpha +
\ldots  \label{xp_com}
\end{eqnarray}
The constant term of eq. (\ref{xp_com}) gives eq. (\ref{norm}). The
higher order terms of eq. (\ref{xp_com}) should vanish, as discussed
in Appendix \ref{app_ap_mb}.

\section{Many-Body Mode Operators  \label{app_ap_mb}}

Outside the harmonic regime the mode operators $\alpha$ and $\pi$
have many-body components. Here we write down the results for
$\alpha$ only, $\pi$ is treated similarly. The structure of $\alpha$
replacing eq. (\ref{a_1b}) is
\begin{eqnarray}
\alpha = \sum_{12} x_{12} a_1^\dagger a_2 + \frac{1}{4} \sum_{1234}
\tilde{x}_{1234} N[a_1^\dagger a_2^\dagger a_3 a_4] \nonumber \\
+ \frac{1}{9} \sum_{123456} \tilde{\tilde{x}}_{123456} N[a_1^\dagger
a_2^\dagger a_3^\dagger a_4 a_5 a_6] + \ldots ,    \label{a_mb}
\end{eqnarray}
where $\tilde{x}_{1234}$ and $\tilde{\tilde{x}}_{123456}$ are
anti-symmetrized structure coefficients. The saturation principle
replacing eq. (\ref{sat_x}) is
\begin{eqnarray}
[ N[a_2^\dagger a_1], \alpha ] = [x , \rho]_{12} + [x , R^N]_{12} +
[
\tilde{x}\{R^N\}, \rho ]_{12} \nonumber \\
+ \frac{1}{2} \sum_{345} (\tilde{x}_{1345}N[a_2^\dagger a_3^\dagger
a_4 a_5] - \tilde{x}_{5432} N[a_5^\dagger a_4^\dagger a_3 a_1])
\nonumber \\
+ [ \tilde{\tilde{x}}\{R^{2,N}\}, \rho ]_{12} + \frac{1}{3}
\sum_{34567} (\tilde{\tilde{x}}_{134567} N[a_2^\dagger
a_3^\dagger a_4^\dagger a_5 a_6 a_7] \nonumber \\
- \tilde{\tilde{x}}_{765432} N[a_7^\dagger a_6^\dagger a_5^\dagger
a_4 a_3 a_1]) + \ldots ,
\end{eqnarray}
where
\begin{eqnarray}
\tilde{x}\{R^N\}_{14} \equiv \sum_{23} \tilde{x}_{1234} N[a_2^\dagger a_3] ,     \\
\tilde{\tilde{x}}\{R^{2,N}\}_{16} \equiv \sum_{2345}
\tilde{\tilde{x}}_{123456} N[a_2^\dagger a_3^\dagger a_4 a_5] .
\end{eqnarray}
Comparing coefficients of the same phonon structure we obtain
\begin{eqnarray}
- i r^{(01)} = [x , \rho] ,  \label{mb_1}
\end{eqnarray}
and
\begin{eqnarray}
- i r^{(02)} = [x, r^{(01)}] + [\tilde{x}^{(01)} , \rho]  ,  \label{mb_2} \\
- i r^{(11)} = [x, r^{(10)}] + [\tilde{x}^{(10)} , \rho]  ,
\label{mb_3}
\end{eqnarray}
and
\begin{eqnarray}
- i r^{(21)}_{12} = [x, r^{(20)}]_{12} + 2 [\tilde{x}^{(10)},
r^{(10)}]_{12} \nonumber \\
+ [\tilde{x}^{(20)} + \tilde{\tilde{x}}^{(20)}, \rho]_{12} ,~~   \label{mb_4} \\
- i r^{(12)}_{12} = [x, r^{(11)}]_{12} + [\tilde{x}^{(10)},
r^{(01)}]_{12}  \nonumber \\
+ [\tilde{x}^{(01)}, r^{(10)}]_{12} +
[\tilde{x}^{(11)} + \tilde{\tilde{x}}^{(11)},
\rho]_{12} ,     \label{mb_5} \\
- 2 i r^{(03)}_{12} = [x, r^{(02)}]_{12} + 2 [\tilde{x}^{(01)},
r^{(01)}]_{12} \nonumber \\
+ [\tilde{x}^{(02)} + \tilde{\tilde{x}}^{(02)}, \rho]_{12} .
\label{mb_6}
\end{eqnarray}
From eqs. (\ref{mb_1}-\ref{mb_6}) the structure coefficients $x$,
$\tilde{x}$ and $\tilde{\tilde{x}}$ of $\alpha$ are determined by
the e.o.m. solutions $r^{(mn)}$, order by order. For
self-consistency, substituting them into eq. (\ref{a_mb}) should
give $\alpha$,
\begin{eqnarray}
\alpha = {\rm{Tr}}\{x\rho\} + {\rm{Tr}}\{x r^{(10)}\} \alpha +
{\rm{Tr}}\{x r^{(01)}\}
\pi \nonumber \\
+ ({\rm{Tr}}\{x r^{(20)}\} + {\rm{Tr}}\{\tilde{x}^{(10)} r^{(10)}\})
\frac{\alpha^2}{2} \nonumber \\
+ ({\rm{Tr}}\{x r^{(02)}\} +
{\rm{Tr}}\{\tilde{x}^{(01)} r^{(01)}\}) \frac{\pi^2}{2} + \ldots ,
\label{a_self}
\end{eqnarray}
which means that all other coefficients vanish, except ${\rm{Tr}}\{x
r^{(10)}\} = 1$. ${\rm{Tr}}\{x\rho\} = 0$ implies that diagonal
matrix elements $x_{11} = 0$. ${\rm{Tr}}\{x r^{(01)}\} = 0$ is
satisfied identically by eq. (\ref{mb_1}). ${\rm{Tr}}\{x r^{(10)}\}
= 1$ is identical to the normalization condition (\ref{norm}). For
higher order coefficients in eq. (\ref{a_self}), some are
identically zero, e.g. the $\pi^2 / 2$ coefficient by eqs.
(\ref{mb_1}) and (\ref{mb_2}); some impose new constraints, e.g. the
vanishing of the $\alpha^2 / 2$ coefficient implies
\begin{eqnarray}
{\rm{Tr}}\{x r^{(20)}\} + {\rm{Tr}}\{r^{(11)}p\} - i
{\rm{Tr}}\{[x,r^{(10)}]p\} = 0
\end{eqnarray}
In the Lipkin model we have checked that these constraints are
satisfied identically, up to the $\alpha^3$, $\{\alpha^2,\pi\}$,
$\{\alpha,\pi^2\}$ and $\pi^3$ terms.

These many-body components should be kept in mind if we want to
compare the bosonic wavefunction with the shell-model wavefunction.

\section{Coherent Summation  \label{app_qq_coherent}}

The factorizable force model has an \emph{analytical} solution only
if we neglect the ``incoherent'' terms in eq. (\ref{W_QQ}), as is
usually assumed in such models. Here we consider its justification
beyond the harmonic order. The exact expression of $W\{R\}_{12}$ is
\begin{eqnarray}
w^{(mn)}_{12} = \sum_{34} ( - \kappa q_{12} q_{34} + \kappa q_{14}
q_{32} ) r^{(mn)}_{43} \nonumber \\
= - \kappa q_{12} \sum_{34}
q_{34} r^{(mn)}_{43} + \kappa \sum_{34} q_{14} r^{(mn)}_{43} q_{32}
. \nonumber
\end{eqnarray}
An observable is given by a trace of $w^{(mn)}_{12}$ with some
operator(s) $t$:
\begin{eqnarray}
O \sim {\rm{Tr}}[ t w^{(mn)} ] = \nonumber \\
- \kappa \cdot {\rm{Tr}}[ t q ] \cdot {\rm{Tr}} [ q r^{(mn)} ] +
\kappa \cdot {\rm{Tr}}[ q t q r^{(mn)} ] .~~  \label{o_trace}
\end{eqnarray}
Quite generally, operator $q_{12}$ has the following property: for a
given s.p. level $1$, $q_{12}$ essentially vanishes except for a few
s.p. level $2$. For the realistic quadrupole moment operator $q_\mu
= r^2 Y_{2\mu}$, it is ensured by the selection rules with respect
to $r^2$, $L=2$ and $\mu$. If $q_{12}$ has the above property, a
trace grows linearly with the collectivity factor $\Omega$,
independently of the number of operators $q$ inside. Hence in eq.
(\ref{o_trace}) the incoherent sum is smaller by a factor of
$1/\Omega$ than the coherent one. The approximation of keeping only
coherent terms is valid when the collectivity $\Omega$ is large.

\section{Details of Factorizable Force Model  \label{app_qq}}

Here we supply the details for Sec. \ref{sec_qq_GDM}. In the
harmonic order we solve the RPA equation. The formal solutions
(\ref{RPA_r10}) and (\ref{RPA_r01}) become
\begin{eqnarray}
r^{(10)}_{12} = \frac{- \kappa Q^{(10)} q_{12} }{(e_{12})^2 -
\omega^2} n_{12} e_{12} , ~ r^{(01)}_{12} = \frac{- \kappa Q^{(10)}
q_{12} }{(e_{12})^2 - \omega^2} i n_{12} .~ \label{r10_01_QQ}
\end{eqnarray}
Then $Q^{(01)} = Tr\{q r^{(01)}\} = 0$, as it should be. From
$Q^{(10)} = Tr\{q r^{(10)}\} \ne 0$ we obtain the RPA secular
equation (\ref{omega_QQ}). The $n_1 \ne n_2$ matrix elements of $x$
and $p$ are given by eqs. (\ref{x_ele}) and (\ref{p_ele}):
\begin{eqnarray}
x_{12} = \frac{\kappa Q^{(10)}q_{12}}{(e_{12})^2-\omega^2} ,~ p_{12}
= \frac{\kappa Q^{(10)}q_{12}}{(e_{12})^2-\omega^2} i e_{12} ,~ (n_1
\ne n_2) .~~
\end{eqnarray}

The leading order of eq. (\ref{r10_01_QQ}) is
\begin{eqnarray}
r^{(10)} \doteq - \kappa Q^{(10)} (\frac{n}{e} : q) ,~~ r^{(01)}
\doteq - i \kappa Q^{(10)} (\frac{n}{e^2}: q) .
\end{eqnarray}
The leading order of the RPA secular equation (\ref{omega_QQ}) is
\begin{eqnarray}
1 \doteq - \kappa \sum_{12} \frac{ |q_{12}|^2 n_{12} }{e_{12}} .
\label{omega_c_QQ}
\end{eqnarray}
The leading order of the normalization condition (\ref{norm_QQ}) is
\begin{eqnarray}
1 \doteq - (\kappa Q^{(10)})^2 \sum_{12} \frac{|q_{12}|^2
n_{12}}{(e_{12})^3} .    \label{norm_c_QQ}
\end{eqnarray}

In the cubic order, the $e_1 = e_{1'}$ matrix elements are given by
eq. (\ref{sat_11p}):
\begin{eqnarray}
r^{(20)}_{11'} \doteq - 2 (\kappa Q^{(10)})^2 \sum_2  \frac{q_{12}
q_{21'} ~ n_{12}}{(e_{12})^2} ,  \label{r20_11p_QQ} \\
r^{(02)}_{11'} \doteq - 2 (\kappa Q^{(10)})^2 \sum_2 ~
\frac{q_{12} q_{21'} ~ n_{12}}{(e_{12})^4} ,  \label{r02_11p_QQ} \\
r^{(11)}_{11'} = 0 .  ~~~~~~~~~~~~~~~~     \label{r11_11p_QQ}
\end{eqnarray}
The $e_1 \ne e_2$ matrix elements are determined from eqs.
(\ref{Cubic_1}-\ref{Cubic_3}),
\begin{eqnarray}
- 2 i \Lambda^{(30)} r^{(01)}_{12} \doteq e_{12}
r^{(20)}_{12} + \kappa Q^{(20)} n_{12} q_{12} \nonumber \\
- 2 \kappa Q^{(10)} [q , r^{(10)}]_{12} ,  \\
i r^{(20)}_{12} + i \Lambda^{(12)} r^{(10)}_{12} \doteq e_{12}
r^{(11)}_{12} - \kappa Q^{(10)} [q , r^{(01)}]_{12} ,
\\
2 i r^{(11)}_{12} - i \Lambda^{(12)} r^{(01)}_{12} = e_{12}
r^{(02)}_{12} + \kappa Q^{(02)} n_{12} q_{12} ,~
\end{eqnarray}
with the solution ($e_1 \ne e_2$)
\begin{eqnarray}
r^{(20)}_{12} \doteq  - \kappa Q^{(20)} (\frac{n}{e}:q)_{12} - 2
\kappa Q^{(10)} \Lambda^{(30)} (\frac{n}{e^3}: q)_{12} \nonumber \\
- 2 (\kappa Q^{(10)})^2 \frac{[ q , (\frac{n}{e} : q)
]_{12}}{e_{12}} ,~ \label{r20_12_QQ}
\\
r^{(11)}_{12} \doteq - i \kappa Q^{(20)} (\frac{n}{e^2}:q)_{12}
\nonumber \\
- i \kappa Q^{(10)} [ 2 \Lambda^{(30)} (\frac{n}{e^4}: q)_{12}  +
\Lambda^{(12)}
(\frac{n}{e^2} : q)_{12} ] \nonumber \\
- i (\kappa Q^{(10)})^2 \{ 2 \frac{[ q , (\frac{n}{e} : q)
]_{12}}{(e_{12})^2} + \frac{[q ,
(\frac{n}{e^2}: q)]_{12}}{e_{12}} \} ,~    \label{r11_12_QQ} \\
r^{(02)}_{12} \doteq - \kappa Q^{(02)} (\frac{n}{e} : q)_{12}  + 2
\kappa Q^{(20)} (\frac{n}{e^3} : q)_{12} \nonumber \\
+ \kappa Q^{(10)} [ 4 \Lambda^{(30)} (\frac{n}{e^5}: q)_{12} +
\Lambda^{(12)} (\frac{n}{e^3} : q)_{12} ] \nonumber
\\
+ 2 (\kappa Q^{(10)})^2 \{ 2 \frac{[ q , (\frac{n}{e} : q)
]_{12}}{(e_{12})^3} + \frac{[q , (\frac{n}{e^2}:
q)]_{12}}{(e_{12})^2} \} .~  \label{r02_12_QQ}
\end{eqnarray}
If we set $n_1 = n_2$ in eqs. (\ref{r20_12_QQ}-\ref{r02_12_QQ}), the
powers of $e_{12}$ in the denominators will be canceled, thus
$r^{(20/11/02)}_{12}$ are finite in the limit $e_1 \approx e_2$, as
they should be. Moreover, if we set $e_2 = e_{1'} = e_1$ in the
resultant expressions, we obtain eqs.
(\ref{r20_11p_QQ}-\ref{r11_11p_QQ}), derived from the saturation
principle. This is also true in the case of a general $V_{1234}$.
With the solutions (\ref{r20_11p_QQ}) and (\ref{r20_12_QQ}) we can
calculate $Q^{(20)}$,
\begin{eqnarray}
Q^{(20)} = \sum_{e_1 \ne e_2} r^{(20)}_{12} q_{21} + \sum_{e_1 =
e_{1'}} r^{(20)}_{11'} q_{1'1} \doteq \nonumber \\
Q^{(20)} + 2 \frac{\Lambda^{(30)}}{\kappa Q^{(10)}} - 2 (\kappa
Q^{(10)})^2 \sum_{e_1 \ne e_2} \frac{[ q , (\frac{n}{e}
: q) ]_{12} q_{21}}{e_{12}} \nonumber \\
- 2 (\kappa Q^{(10)})^2 \sum_{e_1 = e_{1'}} \sum_2 \frac{q_{12}
q_{21'} q_{1'1} n_{12}}{(e_{12})^2} ,~~
\end{eqnarray}
where we have used eqs. (\ref{omega_c_QQ}) and (\ref{norm_c_QQ}).
Canceling $Q^{(20)}$ from both sides we obtain eq. (\ref{L30_c_QQ}).
Similarly from $Q^{(02)} = \sum_{e_1 \ne e_2} r^{(02)}_{12} q_{21} +
\sum_{e_1 = e_{1'}} r^{(02)}_{11'} q_{1'1}$ we obtain eq.
(\ref{L12_c_QQ}).

In the quartic order, the leading $e_1 = e_{1'}$ matrix element
$r^{(30)}_{11'}$ is determined from eq. (\ref{quartic_2}),
\begin{eqnarray}
2 i r^{(30)}_{11'} - 2 i \Lambda^{(30)} r^{(02)}_{11'} + 2 i
\Lambda^{(12)} r^{(20)}_{11'} \doteq  \nonumber \\
- \kappa Q^{(20)}
[q , r^{(01)}]_{11'} - 2 \kappa Q^{(10)} [q , r^{(11)}]_{11'}   ,
\end{eqnarray}
with the solution ($e_1 = e_{1'}$)
\begin{eqnarray}
r^{(30)}_{11'} \doteq - 2 (\kappa Q^{(10)})^2 \Lambda^{(30)} \sum_2
\frac{q_{12} q_{21'} n_{12}}{(e_{12})^4} \nonumber \\
+ 2 (\kappa
Q^{(10)})^2 \Lambda^{(12)} \sum_2 \frac{q_{12}
q_{21'} n_{12}}{(e_{12})^2}  \nonumber \\
- \kappa Q^{(10)} \kappa Q^{(20)} \sum_2 \frac{q_{12} q_{21'}
n_{12}}{(e_{12})^2} + i \kappa Q^{(10)} [q , r^{(11)}]_{11'} .~~~
\end{eqnarray}
The leading $e_1 \ne e_2$ matrix element $r^{(30)}_{12}$ is
determined from eq. (\ref{quartic_1}),
\begin{eqnarray}
- 3 i \Lambda^{(30)} r^{(11)}_{12} - 3 i \Lambda^{(40)}
r^{(01)}_{12} \doteq e_{12} r^{(30)}_{12} + \kappa Q^{(30)} n_{12}
q_{12} \nonumber
\\
- \frac{3}{2} \kappa Q^{(20)} [q , r^{(10)}]_{12} - \frac{3}{2}
\kappa Q^{(10)} [q , r^{(20)}]_{12} ,~~~
\end{eqnarray}
with the solution ($e_1 \ne e_2$)
\begin{eqnarray}
r^{(30)}_{12} \doteq - \kappa Q^{(30)} (\frac{n}{e}:q)_{12} - 3
\kappa Q^{(10)}
\Lambda^{(40)} (\frac{n}{e^3}: q)_{12}  \nonumber \\
- 3 i \Lambda^{(30)} \frac{r^{(11)}_{12}}{e_{12}} - \frac{3}{2}
\kappa Q^{(10)} \kappa Q^{(20)} \frac{[ q , (\frac{n}{e} : q)
]_{12}}{e_{12}} \nonumber \\
+ \frac{3}{2} \kappa Q^{(10)} \frac{[q , r^{(20)}]_{12}}{e_{12}} .~
\end{eqnarray}
Then from $Q^{(30)} = \sum_{e_1 \ne e_2} r^{(30)}_{12} q_{21} +
\sum_{e_1 = e_{1'}} r^{(30)}_{11'} q_{1'1}$ we obtain eq.
(\ref{L40_c_QQ}).

The solutions $r^{(mn)}_{12}$ are needed if we want to calculate the
transitions of the operator $a_2^\dagger a_1$ from eq.
(\ref{R_exp}).

\section{Quartic Potential Dominance  \label{app_qq_40_dorm}}

Around the critical point $\omega^2 \approx 0$ the stability of the
system is restored by higher order anharmonicities. We assume that
the quartic potential term $\Lambda^{(40)} \alpha^4 / 4$ is
dominate, and study the conditions for this to be true. Under the
rescaling of $\alpha$ and $\pi$
\begin{eqnarray}
\bar{\alpha} = (\Lambda^{(40)})^{\frac{1}{6}} \cdot \alpha ,~~~
\bar{\pi} = (\Lambda^{(40)})^{-\frac{1}{6}} \cdot \pi ,
\end{eqnarray}
which preserves the commutation relation $[\bar{\alpha} , \bar{\pi}]
= i$, the Hamiltonian (\ref{H_b}) is written as
\begin{eqnarray}
H - E_0 = (\Lambda^{(40)})^{\frac{1}{3}} \cdot \Big(~ \frac{1}{4}
\bar{\alpha}^4 + \frac{1}{2} \bar{\pi}^2 + \frac{\Lambda^{(30)}
(\Lambda^{(40)})^{-\frac{5}{6}}}{3} \bar{\alpha}^3 \nonumber \\
+ \frac{\Lambda^{(12)} (\Lambda^{(40)})^{-\frac{1}{6}}}{4}
\{\bar{\alpha}, \bar{\pi}^2\} + \frac{\Lambda^{(22)}
(\Lambda^{(40)})^{-\frac{1}{3}}}{8} \{ \bar{\alpha}^2 , \bar{\pi}^2
\} \nonumber \\
+ \frac{\Lambda^{(04)} (\Lambda^{(40)})^{\frac{1}{3}}}{4}
\bar{\pi}^4 + \frac{\Lambda^{(60)}
(\Lambda^{(40)})^{-\frac{4}{3}}}{6} \bar{\alpha}^6  + \ldots  ~\Big)
.~~
\end{eqnarray}
Thus the term $\frac{\Lambda^{(40)}}{4} \alpha^4$ is dominant if
coefficients of other terms, e.g. $\Lambda^{(60)}
(\Lambda^{(40)})^{-\frac{4}{3}}$, are small. We consider their
dependence on the collectivity factor $\Omega$ in the factorizable
force model. Let the quadrupole operator $q$ have the property
specified in Appendix \ref{app_qq_coherent}. Eq. (\ref{norm_c_QQ})
gives $(\kappa Q^{(10)})^2 \sim \Omega^{-1}$. Eq. (\ref{L30_c_QQ})
gives $\Lambda^{(30)} \sim \Omega^{-\frac{1}{2}}$. Eq.
(\ref{L12_c_QQ}) gives $\Lambda^{(12)} + 2 \frac{Q^{(20)}}{Q^{(10)}}
\sim \Omega^{-\frac{1}{2}}$, and we assume $\Lambda^{(12)} \sim
\Omega^{-\frac{1}{2}}$, $\frac{Q^{(20)}}{Q^{(10)}} \sim
\Omega^{-\frac{1}{2}}$. Eq. (\ref{L40_c_QQ}) gives $\Lambda^{(40)}
\sim \Omega^{-1}$. A consistent estimation gives $\Lambda^{(22)}
\sim \Omega^{-1}$, $\Lambda^{(04)} \sim \Omega^{-1}$. In the
expression of $\Lambda^{(60)}$ there should be terms like $(\kappa
Q^{(10)})^6 \cdot$ [trace with six $q$'s], thus $\Lambda^{(60)} \sim
\Omega^{-2}$. In conclusion,
\begin{eqnarray}
\Lambda^{(30)} (\Lambda^{(40)})^{-\frac{5}{6}} \sim
\Omega^{\frac{1}{3}} ,~~~ \Lambda^{(12)}
(\Lambda^{(40)})^{-\frac{1}{6}} \sim \Omega^{-\frac{1}{3}}  ,    \nonumber \\
\Lambda^{(22)} (\Lambda^{(40)})^{-\frac{1}{3}} \sim
\Omega^{-\frac{2}{3}} ,~~~ \Lambda^{(04)}
(\Lambda^{(40)})^{\frac{1}{3}} \sim \Omega^{-\frac{4}{3}} ,\nonumber
\\
\Lambda^{(60)} (\Lambda^{(40)})^{-\frac{4}{3}} \sim
\Omega^{-\frac{2}{3}} .  \label{L_est}
\end{eqnarray}
The estimates (\ref{L_est}) are consistent with those in Ref.
\cite{Zele_estimate}. All terms except $\Lambda^{(30)}$ are
suppressed by powers of $1/\Omega$. The $\Lambda^{(30)}$ term is
given by three-body loops (\ref{L30_c_QQ}), which are usually
suppressed, because of cancelations due to the approximate
particle-hole symmetry near the Fermi surface, similarly to the
Furry theorem of QED. In the case of a spherical nucleus,
$\Lambda^{(30)}$ should be small.

\section{Details of Realistic Nuclear Application  \label{app_real_nucl}}

Here we supply the details for Sec. \ref{sec_real_nucl}. In eq.
(\ref{R_K_def_mu}) $R$ is Hermitian, $K$ is antisymmetric. The
Hermitian of $K$ is $(K^\dagger)_{12} = a_2^\dagger a_1^\dagger$.
$W\{R\}$ and $f\{R\}$ in eq. (\ref{W_def_mu}) are Hermitian,
$\Delta\{K\}$ in eq. (\ref{Delta_def_mu}) is antisymmetric. The
Hermitian of $\Delta\{K\}$ is
\begin{eqnarray}
\Delta^\dagger\{K\}_{12} \equiv (\Delta\{K\}_{21})^\dagger =
\frac{1}{2} \sum_{34} V_{4312} (K^\dagger)_{34} .
\end{eqnarray}
The expansion of the operator $R$ replacing eq. (\ref{R_exp}) is
\begin{eqnarray}
R = \rho + R^N = \rho + \sum_\mu r^{(10)}_\mu \alpha_{\mu}^\dagger +
\sum_\mu r^{(01)}_\mu \pi_{\mu}^\dagger \nonumber \\
+ \frac{1}{2}
\sum_{L = 0,2,4} \sum_\mu r^{(20)}_{L\mu} (\alpha^\dagger
\times \alpha^\dagger)^L_{\mu} \nonumber \\
+ \frac{1}{2} \sum_{L = 0,2,4} \sum_\mu r^{(02)}_{L\mu} (\pi^\dagger
\times \pi^\dagger)^L_{\mu} \nonumber \\
+ \frac{1}{2} \sum_{L =
0,1,2,3,4} \sum_\mu r^{(11)}_{L\mu} \{ \alpha^\dagger , \pi^\dagger
\}^L_{\mu}
\nonumber \\
+ \frac{1}{6} \sum_{L = 0,2,3,4,6} \sum_\mu r^{(30)}_{L\mu}
\{(\alpha^\dagger \times \alpha^\dagger)^{l_L} ,
\alpha^\dagger\}^L_\mu \nonumber \\
+ \frac{1}{6} \sum_{L =
0,2,3,4,6} \sum_\mu r^{(03)}_{L\mu} \{(\pi^\dagger \times
\pi^\dagger)^{l_L} , \pi^\dagger\}^L_\mu \nonumber
\\
+ \frac{1}{4} \sum_{L = 0,1,2,3,4,5,6} ~ \sum_{l = 0,2,4} \sum_\mu
r^{(21)}_{Ll\mu} \{(\alpha^\dagger
\times \alpha^\dagger)^l , \pi^\dagger\}^L_\mu  \nonumber \\
+ \frac{1}{4} \sum_{L = 0,1,2,3,4,5,6} ~ \sum_{l = 0,2,4} \sum_\mu
r^{(12)}_{Ll\mu} \{\alpha^\dagger , (\pi^\dagger \times
\pi^\dagger)^l \}^L_\mu + \ldots  ~~  \label{R_exp_mu}
\end{eqnarray}
Three identical $d$ bosons can couple to $L = 0,2,3,4,6$. In the
$\alpha^3$ and $\pi^3$ terms of eq. (\ref{R_exp_mu}) we choose the
intermediate quantum number for each $L$ to be $l_L$; this choice
does not influence the results. $R$ is Hermitian, time-even,
invariant under rotation and parity \cite{BZ2}. This implies that
the coefficient $r^{(mn)}_{L\mu}$ has the same symmetries as the
operator part $\{\alpha^m,\pi^n\}$: $(r^{(mn)}_{L\mu})^\dagger$ has
angular momentum $L$ and projection $\mu$, even parity, sign of
$(-)^n$ under time-reversal, $(r^{(mn)}_{L\mu})^\dagger =
(-)^{L-\mu} r^{(mn)}_{L-\mu}$. Similarly the expansion of the
operator $K$ is
\begin{eqnarray}
K = \kappa + K^N = \kappa + \sum_\mu k^{(10)}_\mu
\alpha_{\mu}^\dagger + \sum_\mu k^{(01)}_\mu \pi_{\mu}^\dagger
\nonumber \\
+ \frac{1}{2} \sum_{L = 0,2,4} \sum_\mu k^{(20)}_{L\mu}
(\alpha^\dagger \times \alpha^\dagger)^L_{\mu} + \ldots ~~~
\label{K_exp_mu}
\end{eqnarray}
$K$ is anti-symmetric, time-even, invariant under rotation and
parity. Thus $(k^{(mn)}_{L\mu})^\dagger$ has angular momentum $L$
and projection $\mu$, even parity, sign of $(-)^n$ under
time-reversal, $k^{(mn)}_{L\mu 12} = - k^{(mn)}_{L\mu 21}$. The
Hermitian of eq. (\ref{K_exp_mu}) is
\begin{eqnarray}
K^\dagger = \kappa^\dagger + (K^\dagger)^N = \kappa^\dagger +
\sum_\mu (\bar{k}^\dagger)^{(10)}_\mu \alpha_{\mu}^\dagger \nonumber
\\
+ \sum_\mu (\bar{k}^\dagger)^{(01)}_\mu \pi_{\mu}^\dagger +
\frac{1}{2} \sum_{L = 0,2,4} \sum_\mu
(\bar{k}^\dagger)^{(20)}_{L\mu} (\alpha^\dagger \times
\alpha^\dagger)^L_{\mu} + \ldots ~~ \label{Kd_exp_mu}
\end{eqnarray}
where
\begin{eqnarray}
\bar{k}_{L \mu} \equiv (-)^{L - \mu} k_{L -\mu}   ~~~\Rightarrow~~~
\bar{k}^\dagger_{L \mu} = (-)^{L - \mu} k^\dagger_{L -\mu} .
\end{eqnarray}
The expansion of $N[a_4^\dagger a_3^\dagger a_2 a_1]$ replacing eq.
(\ref{RR_exp}) is
\begin{eqnarray}
N[a_4^\dagger a_3^\dagger a_2 a_1] = \frac{1}{2} \sum_{L = 0,2,4}
\sum_\mu r^{(20)}_{L\mu 1234}
(\alpha^\dagger \times \alpha^\dagger)^L_{\mu} \nonumber \\
+ \frac{1}{2} \sum_{L = 0,2,4} \sum_\mu r^{(02)}_{L\mu 1234}
(\pi^\dagger \times \pi^\dagger)^L_{\mu} \nonumber \\
+ \frac{1}{2}
\sum_{L = 0,1,2,3,4} \sum_\mu r^{(11)}_{L\mu 1234} \{ \alpha^\dagger
,
\pi^\dagger \}^L_{\mu}   \nonumber \\
+ \frac{1}{6} \sum_{L = 0,2,3,4,6} \sum_{l = 0,2,4} \sum_\mu
r^{(30)}_{Ll\mu 1234} \{(\alpha^\dagger \times \alpha^\dagger)^l ,
\alpha^\dagger\}^L_\mu
\nonumber \\
+ \frac{1}{6} \sum_{L = 0,2,3,4,6} \sum_{l = 0,2,4} \sum_\mu
r^{(03)}_{Ll\mu 1234} \{(\pi^\dagger \times \pi^\dagger)^l ,
\pi^\dagger\}^L_\mu  \nonumber
\\
+ \frac{1}{4} \sum_{L} \sum_{l = 0,2,4} \sum_\mu r^{(21)}_{Ll\mu
1234} \{(\alpha^\dagger \times \alpha^\dagger)^l ,
\pi^\dagger\}^L_\mu \nonumber \\
+ \frac{1}{8} \sum_{L} \sum_{l =
0,1,2,3,4} \sum_\mu r^{(11 \times 10)}_{Ll\mu 1234}
\{\{\alpha^\dagger ,
\pi^\dagger\}^l , \alpha^\dagger\}^L_\mu  \nonumber \\
+ \frac{1}{4} \sum_{L} \sum_{l = 0,2,4} \sum_\mu r^{(12)}_{Ll\mu
1234} \{\alpha^\dagger , (\pi^\dagger \times \pi^\dagger)^l \}^L_\mu
\nonumber \\
+ \frac{1}{8} \sum_{L} \sum_{l = 0,1,2,3,4} \sum_\mu
r^{(11 \times 01)}_{Ll\mu
1234} \{ \{\alpha^\dagger , \pi^\dagger\}^l , \pi^\dagger  \}^L_\mu  \nonumber \\
+ \ldots ~~   \label{RR_exp_mu}
\end{eqnarray}
In eq. (\ref{RR_exp_mu}), the $\alpha^3$, $\alpha^2 \pi$, $\alpha
\pi^2$ and $\pi^3$ terms are over-complete. This form is convenient
for finding expressions of $r^{(mn)}_{1234}$ in terms of
$r^{(mn)}_{12}$ by the saturation principle, as explained in
Appendix \ref{app_sat_mu}. Similarly we need the expansions of
$N[a_4^\dagger a_3 a_2 a_1]$, $N[a_4 a_3 a_2 a_1]$, $N[a_4^\dagger
a_3^\dagger a_2^\dagger a_1]$ and $N[a_4^\dagger a_3^\dagger
a_2^\dagger a_1^\dagger]$.

\subsection{Exact Equations of Motion}

The Hamiltonian (\ref{H_f}) in the normal ordering form is
\begin{eqnarray}
H = \langle \Phi | H | \Phi \rangle + \sum_{12} {f\{\rho\}}_{12}
N[a_1^\dagger a_2] \nonumber \\
+ \frac{1}{2} \sum_{12} \Delta^\dagger\{\kappa\}_{12} N[a_1 a_2] +
\frac{1}{2} \sum_{12} \Delta\{\kappa\}_{12} N[a_1^\dagger
a_2^\dagger] \nonumber \\
+ \frac{1}{4} \sum_{1234} V_{1234}
N[a_1^\dagger a_2^\dagger a_3 a_4] ,
\end{eqnarray}
where
\begin{eqnarray}
\langle \Phi | H | \Phi \rangle = \sum_{12} (Z_{12} + \frac{1}{2}
W\{\rho\}_{12}) \rho_{21} \nonumber \\
+ \frac{1}{2} \sum_{12}
\Delta\{\kappa\}_{12} (\kappa^\dagger)_{21}
\end{eqnarray}
is the average energy on $|\Phi\rangle$. The exact e.o.m. in the
full space replacing eq. (\ref{e.o.m.}) is
\begin{eqnarray}
[ R_{12} , H ] = [ a_2^\dagger a_1 , H ] = \nonumber
\\
{[}f\{\rho\} , \rho]_{12} - (\kappa ~ \Delta^\dagger\{\kappa\})_{12}
+ (\Delta\{\kappa\} ~ \kappa^\dagger )_{12} \nonumber \\
+ [f\{\rho\} , R^N]_{12} + [W\{R^N\} , \rho]_{12} - (K^N
\Delta^\dagger\{\kappa\})_{12} \nonumber \\
- ( \kappa ~
\Delta^\dagger\{K^N\} )_{12}  +  (\Delta\{\kappa\} ~ K^{\dagger N}
)_{12} + ( \Delta\{K^N\} ~ \kappa^\dagger )_{12} \nonumber \\
+ \frac{1}{2} \sum_{345} (V_{1345} N[a_2^\dagger a_3^\dagger a_4
a_5] - V_{5432} N[a_5^\dagger a_4^\dagger a_3 a_1]) ,~
\label{eom_R_mu}
\end{eqnarray}
and
\begin{eqnarray}
[ K_{12} , H ] = [ a_2 a_1 , H ] = ( \kappa ~ {f^T\{\rho\}} )_{12} +
( {f\{\rho\}} ~ \kappa )_{12} \nonumber \\
+ \Delta\{\kappa\}_{12} -
(\Delta\{\kappa\} ~ \rho^T)_{12} -
(\rho ~ \Delta\{\kappa\})_{12} \nonumber \\
+ ( K^N ~ {f^T\{\rho\}} )_{12} + ( {f\{\rho\}} ~ K^N )_{12} - (
\Delta\{\kappa\} ~ (R^T)^N
)_{12} \nonumber \\
- (R^N ~ \Delta\{\kappa\})_{12} + (\kappa ~ W^T\{R^N\})_{12} +
(W\{R^N\} ~ \kappa)_{12} \nonumber \\
+ \Delta\{K^N\}_{12} - (\rho ~
\Delta\{K^N\})_{12} - (\Delta\{K^N\} ~
\rho^T)_{12} \nonumber \\
+ \frac{1}{2} \sum_{345} (V_{2543} N[a_5^\dagger a_4 a_3 a_1] -
V_{1543} N[a_5^\dagger a_4 a_3 a_2]) .~~~   \label{eom_K_mu}
\end{eqnarray}
Then in eqs. (\ref{eom_R_mu}) and (\ref{eom_K_mu}) we equate the
l.h.s. and r.h.s. coefficients of the same phonon structure: $1$,
$\alpha_\mu$, $\pi_\mu$, $\frac{1}{2} (\alpha \times \alpha)^L_\mu$
$\ldots$ , and obtain e.o.m. in the collective band
(\ref{HFB}-\ref{Quartic4_mu}) of Sec. \ref{sec_real_nucl_eom_band}.

\subsection{Hartree-Fock-Bogoliubov Equation}

The HFB equation (\ref{HFB}) says that $S^{(00)}$ and $D^{(00)}$ can
be diagonalized simultaneously,
\begin{eqnarray}
\left[~ \left( \begin{array}{cc}
E & 0 \\
0 & - E \\
\end{array} \right) ~,~ \left(\begin{array}{cc}
         n & 0 \\
         0 & I - n \\
       \end{array} \right) ~\right] = 0 ,   \label{HFB_diag}
\end{eqnarray}
where $E$ and $n$ are diagonal matrices. The chemical potential
$\mu$ (buried in $f$) is determined by $N = \sum_1 \rho_{11} =
{\rm{Tr}}\{\rho\}$. The unitary canonical transformation from the
original s.p. operators $a_1^\dagger$, $a_1$ to the new
quasiparticle operators $b_\lambda^\dagger$, $b_\lambda$ are
\begin{eqnarray}
b_\lambda = \sum_1 ( u_{1\lambda}^* a_1 - v_{1\lambda} a_1^\dagger )
,~ b_\lambda^\dagger = \sum_1 ( u_{1\lambda} a_1^\dagger -
v_{1\lambda}^* a_1 ) .~~~~   \label{QP_tran}
\end{eqnarray}
If $|\Phi\rangle$ is a ``quasiparticle determinant'', $|\Phi\rangle
= {\rm{norm}} \cdot \prod_{\lambda} b_\lambda |0\rangle$, then the
normal ordering with respect to $|\Phi\rangle$ is to put
$b^\dagger$'s to the left of $b$'s. Eq. (\ref{dm_def_mu}) gives:
\begin{eqnarray}
\rho = v v^\dagger  ,~~~  \kappa = - v u^T .  \label{rho_kappa_mu}
\end{eqnarray}
In this case $D^{(00)}$ is diagonalized by the canonical
transformation (\ref{QP_tran}):
\begin{eqnarray}
U = \left( \begin{array}{cc}
u^\dagger & - v^T \\
- v^\dagger & u^T \\
\end{array} \right) ,  ~~~
U D^{(00)} U^\dagger = \left(
                         \begin{array}{cc}
                           0 & 0 \\
                           0 & I \\
                         \end{array}
                       \right) ,
\end{eqnarray}
where the matrix $n$ in eq. (\ref{HFB_diag}) vanishes. The HFB
equation (\ref{HFB}) requires $S^{(00)}$ is diagonalized by $U$
simultaneously. In this article we assume that $|\Phi\rangle$ is a
``quasiparticle determinant''.

It is convenient to solve the e.o.m. (\ref{HFB}-\ref{Quartic4_mu})
in the quasiparticle basis (multiplying $U$ from left and
$U^\dagger$ from right). The density matrix operators in this basis
are
\begin{eqnarray}
R^b_{12} \equiv b_2^\dagger b_1 = N[b_2^\dagger b_1] ,~~ K^b_{12}
\equiv b_2 b_1 = N[b_2 b_1] ,\nonumber \\
(K^{b\dagger})_{12} = b_2^\dagger b_1^\dagger = N[b_2^\dagger
b_1^\dagger] .
\end{eqnarray}
$R^b$ is a mix of $R$, $K$, $K^\dagger$ of eqs. (\ref{R_K_def_mu});
so do $K^b$ and $K^{b \dagger}$. The expansions of them are defined
similarly to eqs. (\ref{R_exp_mu}-\ref{Kd_exp_mu}):
\begin{eqnarray}
R^b = \sum_\mu r^{b(10)}_\mu \alpha_{\mu}^\dagger + \ldots ,~~ K^b =
\sum_\mu k^{b(10)}_\mu \alpha_{\mu}^\dagger + \ldots ,\nonumber \\
K^{b\dagger} = \sum_\mu (\bar{k}^{b\dagger})^{(10)}_\mu
\alpha_{\mu}^\dagger + \ldots  ~~~ \label{RK_b_exp_mu}
\end{eqnarray}

The field matrices in the quasiparticle basis are
\begin{eqnarray}
D^U = U D U^\dagger ,~~~ S^U = U S U^\dagger .  \label{matb_mu}
\end{eqnarray}
We need to express them in terms of $R^b$, $K^b$ and $K^{b \dagger}$
of eq. (\ref{RK_b_exp_mu}). The result of $D^U$ is simple:
\begin{eqnarray}
D^U = U D U^\dagger = \left(\begin{array}{cc}
         R^b & K^b \\
         K^{b\dagger} & I - (R^b)^{T} \\
       \end{array} \right) .
\end{eqnarray}
The result of $S^U$ is long:
\begin{eqnarray}
S^U = U S U^\dagger = \left(
    \begin{array}{cc}
      S_{A} & S_{B} \\
      S_{C} & S_{D} \\
    \end{array}
  \right) ,   \label{S_ABCD}
\end{eqnarray}
where
\begin{eqnarray}
S_{A} = u^\dagger f\{R\} u - u^\dagger \Delta\{K\} v^* \nonumber \\
- v^T
\Delta^\dagger\{K\} u - v^T f^T\{R\} v^* \nonumber \\
= \Big(~ u^\dagger Z u - v^T Z^T v^* + u^\dagger W\{ v v^\dagger \}
u \nonumber \\
+ u^\dagger \Delta \{ v u^T \} v^* + v^T
\Delta^\dagger\{ v u^T \} u -
v^T W^T\{ v v^\dagger \} v^* ~\Big)  \nonumber \\
+ \Big(~ u^\dagger W\{ u_{\cdot\beta} (u^\dagger)_{\alpha\cdot} \} u
+ u^\dagger \Delta\{ u_{\cdot\beta} (v^T)_{\alpha\cdot} \} v^*
\nonumber \\
+ v^T \Delta^\dagger\{ u_{\cdot\alpha}
(v^T)_{\beta\cdot} \} u - v^T W^T\{
u_{\cdot\beta} (u^\dagger)_{\alpha\cdot} \} v^*  \nonumber \\
- u^\dagger W\{ v_{\cdot\alpha} (v^\dagger)_{\beta\cdot} \} u -
u^\dagger \Delta \{ v_{\cdot\alpha} (u^T)_{\beta\cdot} \} v^*
\nonumber \\
- v^T \Delta^\dagger\{ v_{\cdot\beta}
(u^T)_{\alpha\cdot} \} u + v^T W^T\{ v_{\cdot\alpha}
(v^\dagger)_{\beta\cdot} \} v^* ~\Big)
b_\alpha^\dagger b_\beta  \nonumber \\
+ \Big(~ - u^\dagger W\{ v_{\cdot\beta} (u^\dagger)_{\alpha\cdot} \}
u - u^\dagger \Delta \{ v_{\cdot\beta} (v^T)_{\alpha\cdot} \} v^*
\nonumber \\
- v^T \Delta^\dagger\{ u_{\cdot\alpha}
(u^T)_{\beta\cdot} \} u + v^T W^T\{ v_{\cdot\beta}
(u^\dagger)_{\alpha\cdot} \} v^* ~\Big)
b_{\alpha}^\dagger b_\beta^\dagger  \nonumber \\
+ \Big(~ - u^\dagger W\{ u_{\cdot\beta} (v^\dagger)_{\alpha\cdot} \}
u - u^\dagger \Delta\{ u_{\cdot\beta} (u^T)_{\alpha\cdot} \} v^*
\nonumber \\
- v^T \Delta^\dagger \{ v_{\cdot\alpha} (v^T)_{\beta\cdot} \} u +
v^T W^T\{ u_{\cdot\beta} (v^\dagger)_{\alpha\cdot} \} v^* ~\Big)
b_\alpha b_\beta ,~~   \label{S_A}
\end{eqnarray}
and
\begin{eqnarray}
S_{B} = - u^\dagger f\{R\} v + u^\dagger \Delta\{K\} u^* \nonumber
\\
+ v^T \Delta^\dagger\{K\} v + v^T f^T\{R\} u^*  \nonumber \\
= \Big(~ - u^\dagger Z v + v^T Z^T u^* - u^\dagger W\{ v v^\dagger
\} v \nonumber \\
- u^\dagger \Delta \{ v u^T \} u^* - v^T
\Delta^\dagger\{ v u^T \} v
+ v^T W^T\{ v v^\dagger \} u^* ~\Big)  \nonumber \\
+ \Big(~ - u^\dagger W\{ u_{\cdot\beta} (u^\dagger)_{\alpha\cdot} \}
v - u^\dagger \Delta\{ u_{\cdot\beta} (v^T)_{\alpha\cdot} \} u^*
\nonumber \\
- v^T \Delta^\dagger\{ u_{\cdot\alpha}
(v^T)_{\beta\cdot} \} v + v^T W^T\{
u_{\cdot\beta} (u^\dagger)_{\alpha\cdot} \} u^*  \nonumber \\
+ u^\dagger W\{ v_{\cdot\alpha} (v^\dagger)_{\beta\cdot} \} v +
u^\dagger \Delta \{ v_{\cdot\alpha} (u^T)_{\beta\cdot} \} u^*
\nonumber \\
+ v^T \Delta^\dagger\{ v_{\cdot\beta}
(u^T)_{\alpha\cdot} \} v - v^T W^T\{ v_{\cdot\alpha}
(v^\dagger)_{\beta\cdot} \} u^* ~\Big) b_\alpha^\dagger b_\beta  \nonumber \\
+ \Big(~ u^\dagger W\{ v_{\cdot\beta} (u^\dagger)_{\alpha\cdot} \} v
+ u^\dagger \Delta \{ v_{\cdot\beta} (v^T)_{\alpha\cdot} \} u^*
\nonumber \\
+ v^T \Delta^\dagger\{ u_{\cdot\alpha}
(u^T)_{\beta\cdot} \} v - v^T W^T\{ v_{\cdot\beta}
(u^\dagger)_{\alpha\cdot} \} u^* ~\Big)
b_{\alpha}^\dagger b_\beta^\dagger  \nonumber \\
+ \Big(~ u^\dagger W\{ u_{\cdot\beta} (v^\dagger)_{\alpha\cdot} \} v
+ u^\dagger \Delta\{ u_{\cdot\beta} (u^T)_{\alpha\cdot} \} u^*
\nonumber \\
+ v^T \Delta^\dagger \{ v_{\cdot\alpha} (v^T)_{\beta\cdot} \} v -
v^T W^T\{ u_{\cdot\beta} (v^\dagger)_{\alpha\cdot} \} u^* ~\Big)
b_\alpha b_\beta ,~~ \label{S_B}
\end{eqnarray}
and
\begin{eqnarray}
S_{C} = - v^\dagger f\{R\} u + v^\dagger \Delta\{K\} v^* \nonumber
\\
+ u^T \Delta^\dagger\{K\} u + u^T f^T\{R\} v^* = (S_{B})^\dagger ,
\label{S_C}
\end{eqnarray}
and
\begin{eqnarray}
S_{D} = v^\dagger f\{R\} v - v^\dagger \Delta\{K\} u^* \nonumber \\
- u^T \Delta^\dagger\{K\} v - u^T f^T\{R\} u^*  = - (S_{A})^T .
\label{S_D}
\end{eqnarray}
From now on we will always work in the quasiparticle basis unless
otherwise specified. For simplicity we will drop the superscripts
$^b$ in eq. (\ref{RK_b_exp_mu}) and $^U$ in eq. (\ref{matb_mu}).

\subsection{Quasi-particle Random Phase Approximation}

The QRPA equations (\ref{QRPA1}) and (\ref{QRPA2}) are (we have
dropped the superscript $^U$)
\begin{eqnarray}
i D^{(10)}_\mu = [S^{(00)} , D^{(01)}_\mu] +
[S^{(01)}_\mu , D^{(00)}] ,  \nonumber \\
- i \omega^2 D^{(01)}_\mu = [S^{(00)} , D^{(10)}_\mu] +
[S^{(10)}_\mu , D^{(00)}] .  \nonumber
\end{eqnarray}
Each of the above two equations has four components, only two of
them are independent. The upper-left component gives
\begin{eqnarray}
r^{(10)}_{\mu 12} = r^{(01)}_{\mu 12} = 0 .
\end{eqnarray}
The upper-right component gives
\begin{eqnarray}
i k^{(10)}_{\mu 12} = (S_B)^{(01)}_{\mu 12} + (E_1
+ E_2) k^{(01)}_{\mu 12} ,  \\
- i \omega^2 k^{(01)}_{\mu 12} = (S_B)^{(10)}_{\mu 12} + (E_1 + E_2)
k^{(10)}_{\mu 12} .
\end{eqnarray}
The formal solution is
\begin{eqnarray}
k^{(10)}_{\mu 12} = \frac{ i
\omega^2 (S_B)^{(01)}_{\mu 12} - (E_1 + E_2) (S_B)^{(10)}_{\mu 12} }{(E_1 + E_2)^2 - \omega^2}  ,  \label{k10_mu} \\
k^{(01)}_{\mu 12} = \frac{ - (E_1 + E_2) (S_B)^{(01)}_{\mu 12} - i
(S_B)^{(10)}_{\mu 12} }{(E_1 + E_2)^2 - \omega^2} . \label{k01_mu}
\end{eqnarray}
From eqs. (\ref{S_B}), (\ref{k10_mu}) and (\ref{k01_mu}) we obtain a
linear homogenous set of equations for $(S_B)^{(10)}$ and
$(S_B)^{(01)}$, a non-zero solution requires a zero determinant,
from which we solve for $\omega^2$.

Again to fix the normalization of $k^{(10/01)}_{\mu 12}$ we need the
saturation principle. Since now we are solving everything in the
quasiparticle basis, it is convenient to redo the saturation
principle in the quasiparticle basis. After that we obtain the
normalization condition (independent of $\mu$):
\begin{eqnarray}
i = [\alpha_\mu^\dagger , \pi_\mu] =^{{\rm{constant~terms}}}
\nonumber \\
- \frac{1}{2} ~ {\rm{Tr}}\{ (k^{(01)}_\mu)^\dagger ~ k^{(10)}_\mu \}
+ \frac{1}{2} ~ {\rm{Tr}}\{ \bar{k}^{(01)}_{\mu} ~
\bar{k}^{\dagger(10)}_{\mu} \} .
\end{eqnarray}

\subsection{Cubic Anharmonicity and Quartic Anharmonicity}

The second order e.o.m. are eqs. (\ref{Cubic1_mu}-\ref{Cubic3_mu}).
$D^{(11)}_{L=1,3;\mu}$ is determined from eq. (\ref{Cubic3_mu})
alone. $D^{(20/11/02)}_{L=0,4;\mu}$ is determined from eqs.
(\ref{Cubic1_mu}-\ref{Cubic3_mu}). They are expressed in terms of
lower order quantities. When $L=2$, $\Lambda^{(30)}$ and
$\Lambda^{(12)}$ enter eqs. (\ref{Cubic1_mu}-\ref{Cubic3_mu}), and
$D^{(20/11/02)}_{L=2;\mu}$ is determined in terms of
$\Lambda^{(30)}$ and $\Lambda^{(12)}$.

Similarly to the situation in Sec. \ref{sec_GDM_cubic_anharm}, the
$E_1 = E_{1'}$ matrix elements $r^{(20/02)}_{L=0,2,4;\mu 11'}$ and
$r^{(11)}_{L=1,3;\mu 11'}$ are fixed by the saturation principle
($E_1 = E_{1'}$):
\begin{eqnarray}
\underline{L = 0,2,4:} ~~~ r^{(20)}_{L\mu 11'} = 2
(k^{(10)} \times \bar{k}^{\dagger(10)})^L_{\mu 11'} , \label{r20_11p_mu} \\
\underline{L = 0,2,4:} ~~~ r^{(02)}_{L\mu 11'} = 2 (k^{(01)} \times
\bar{k}^{\dagger(01)})^L_{\mu 11'} ,  \label{r02_11p_mu} \\
\underline{L = 0,1,2,3,4:} ~~~ r^{(11)}_{L\mu 11'} = (k^{(10)}
\times \bar{k}^{\dagger(01)})^L_{\mu 11'} \nonumber \\
+ (-)^L (k^{(01)} \times (\bar{k}^\dagger)^{(10)} )^L_{\mu 11'} .
\label{r11_11p_mu}
\end{eqnarray}
Eqs. (\ref{r20_11p_mu}-\ref{r11_11p_mu}) are consistent with the
second order e.o.m. (\ref{Cubic1_mu}-\ref{Cubic3_mu}).

The third order e.o.m. are eqs.
(\ref{Quartic1_mu}-\ref{Quartic4_mu}). The $L \ne 2$ quantities
$(D/S)^{(30/21/12/03)}_{L\ne2,\mu}$ are solved in terms of lower
order quantities. When $L=2$, $\Lambda^{(40)}$, $\Lambda^{(22)}_l$
and $\Lambda^{(04)}$ enter into the equations and we have the
solvability condition as explained in eqs. (\ref{quar_sol1_mu}) and
(\ref{quar_sol2_mu}).

\section{Saturation Principle for Section \ref{sec_real_nucl}  \label{app_sat_mu}}

Keeping only one-body terms in $\alpha_\mu$ and $\pi_\mu$,
\begin{eqnarray}
\alpha_\mu = \sum_{12} ( x_{\mu 12} a_1^\dagger a_2 + z_{\mu 12}
a_1 a_2 +  \bar{z}^\dagger_{\mu12} a_1^\dagger a_2^\dagger ) ,  \label{a1b_exp_mu} \\
\pi_\mu = \sum_{12} ( p_{\mu12} a_1^\dagger a_2 + o_{\mu12} a_1 a_2
+ \bar{o}^\dagger_{\mu12} a_1^\dagger a_2^\dagger ) ,
\label{p1b_exp_mu}
\end{eqnarray}
where
\begin{eqnarray}
x_\mu^\dagger = (-)^\mu x_{-\mu} ,~~~ p_\mu^\dagger = (-)^\mu
p_{-\mu} ,\nonumber \\
z_{\mu12} = - z_{\mu21} ,~~~ o_{\mu12} = -
o_{\mu21} ,
\end{eqnarray}
we have the following identities in the full space:
\begin{eqnarray}
[ R_{12} , \alpha_\mu ] = [x_{\mu} , R]_{12} - 2
( K z_{\mu} )_{12} + 2 ( \bar{z}^\dagger_{\mu} K^\dagger )_{12}  ,   \label{sat_aR_mu} \\
{[} K_{12} , \alpha_\mu ] = ( x_{\mu} K )_{12} - (x_{\mu} K)^T_{12}
\nonumber \\
+ 2 \bar{z}^\dagger_{\mu12} + 2 ( R
\bar{z}^\dagger_{\mu} )^T_{12} - 2 ( R \bar{z}^\dagger_{\mu}
)_{12}  ,   \label{sat_aK_mu} \\
{[} (K^\dagger)_{12} , \alpha_\mu ] = - ( K^\dagger x_{\mu} )_{12} +
(K^\dagger x_{\mu})^T_{12} \nonumber \\
- 2 z_{\mu12} - 2 ( z_{\mu} R
)^T_{12} + 2 ( z_{\mu} R )_{12} ,     \label{sat_aKd_mu}
\end{eqnarray}
and
\begin{eqnarray}
[ R_{12} , \pi_\mu ] = [p_{\mu} , R]_{12} - 2
( K o_{\mu} )_{12} + 2 ( \bar{o}^\dagger_{\mu} K^\dagger )_{12}  ,    \label{sat_pR_mu} \\
{[} K_{12} , \pi_\mu ] = ( p_{\mu} K )_{12} - (p_{\mu} K)^T_{12}
\nonumber \\
+ 2 \bar{o}^\dagger_{\mu12} + 2 ( R
\bar{o}^\dagger_{\mu} )^T_{12} - 2 ( R \bar{o}^\dagger_{\mu} )_{12}
,    \label{sat_pK_mu} \\
{[} (K^\dagger)_{12} , \pi_\mu ] = - ( K^\dagger p_{\mu} )_{12} +
(K^\dagger p_{\mu})^T_{12} \nonumber \\
- 2 o_{\mu12} - 2 ( o_{\mu} R )^T_{12} + 2 ( o_{\mu} R )_{12} .
\label{sat_pKd_mu}
\end{eqnarray}
We obtain a set of equations by equating the l.h.s. and r.h.s.
coefficients of the same phonon structure: $1$, $\alpha_\mu$,
$\pi_\mu$, $\frac{1}{2} (\alpha \times \alpha)^L_\mu$ $\ldots$
Considering the length we do not list them here.

Similarly we calculate the commutators of $N[a_4^\dagger a_3^\dagger
a_2 a_1]$, $N[a_4^\dagger a_3 a_2 a_1]$ and $N[a_4 a_3 a_2 a_1]$
with $\alpha_\mu$ and $\pi_\mu$. We give only the result of $[
N[a_4^\dagger a_3^\dagger a_2 a_1] , \alpha_\mu ]$ as an example:
\begin{eqnarray}
[ N[a_4^\dagger a_3^\dagger a_2 a_1] , \alpha_\mu ] = \nonumber \\
i r^{(01)}_{\mu24} N[a_3^\dagger a_1] - i r^{(01)}_{\mu23}
N[a_4^\dagger a_1] - i r^{(01)}_{\mu14} N[a_3^\dagger a_2] \nonumber
\\
+ i r^{(01)}_{\mu13} N[a_4^\dagger a_2] - i
\bar{k}^{\dagger(01)}_{\mu34} N[a_2
a_1] - i k^{(01)}_{\mu12} N[a_4^\dagger a_3^\dagger] \nonumber \\
- \sum_{5} x_{\mu25} N[a_4^\dagger a_3^\dagger a_1 a_5] + \sum_{5}
x_{\mu15} N[a_4^\dagger a_3^\dagger a_2 a_5] \nonumber \\
- \sum_{5} x_{\mu54} N[a_5^\dagger a_3^\dagger a_2 a_1] + \sum_{5}
x_{\mu53} N[a_5^\dagger a_4^\dagger a_2 a_1] \nonumber
\\
+ 2 \sum_{5} z_{\mu54} N[a_3^\dagger a_2 a_1 a_5] + 2 \sum_{5}
z_{\mu35}
N[a_4^\dagger a_2 a_1 a_5] \nonumber \\
+ 2 \sum_{5} \bar{z}^\dagger_{\mu51} N[a_5^\dagger a_4^\dagger
a_3^\dagger a_2] + 2 \sum_{5} \bar{z}^\dagger_{\mu25} N[a_5^\dagger
a_4^\dagger a_3^\dagger a_1] ,~     \label{example1_mu}
\end{eqnarray}
where we have used the lowest order results from eqs.
(\ref{sat_aR_mu}-\ref{sat_pKd_mu}). Equating the l.h.s. and r.h.s.
coefficients of the same phonon structure we obtain a set of
equations. We give only the $(\alpha \times \alpha)^L_\mu$ terms as
an example. Using results from eqs.
(\ref{sat_aR_mu}-\ref{sat_pKd_mu}) we have
\begin{eqnarray}
- i \frac{1}{\sqrt{5}} \sum_{L_f = 0,2,4} \sum_{L}
\frac{\sqrt{2L+1}}{2} \cdot \Big( r^{(21)}_{L L_f 1234} \times
(\alpha \times \alpha)^{L_f}
\Big)^2_\mu  \nonumber \\
- i \frac{1}{2 \sqrt{5}} \sum_{L_f = 0,2,4} \sum_{L} \sum_{l} (-)^{l
+ L_f} \sqrt{(2L+1)(2l+1)(2L_f + 1)} \nonumber \\
\cdot \left\{
                                                    \begin{array}{ccc}
                                                      2 & 2 & l \\
                                                      2 & L & L_f \\
                                                    \end{array}
                                                  \right\}
\cdot \Big(r^{(11 \times
10)}_{Ll 1234} \times ( \alpha \times \alpha )^{L_f}
\Big)^2_\mu \nonumber \\
= - i \frac{1}{\sqrt{5}} \sum_{{L_f} = 0,2,4} \sum_L \frac{\sqrt{
2L+1 }}{2} \cdot \Big(~ [~ - ( r^{(20)}_{L_f 13} r^{(01)}_{24})^L
\nonumber \\
+ ( r^{(20)}_{L_f 14} r^{(01)}_{23})^L + ( r^{(20)}_{L_f 23}
r^{(01)}_{14})^L - (r^{(20)}_{L_f 24} r^{(01)}_{13} )^L \nonumber
\\
+ ( k^{(20)}_{{L_f} 12} \bar{k}^{\dagger(01)}_{34})^L + (
\bar{k}^{\dagger(20)}_{{L_f} 34} k^{(01)}_{12})^L ~]  \times (\alpha
\times \alpha)^{L_f} ~\Big)^2_\mu   \nonumber \\
- i
\frac{1}{\sqrt{5}} \sum_{{L_f} = 0,2,4} \sum_{L} \sum_l (-)^{l +
L_f} \sqrt{(2L+1)(2l+1)(2 L_f + 1)}  \nonumber \\
\cdot
                                           \left\{
                                                    \begin{array}{ccc}
                                                      2 & 2 & l \\
                                                      2 & L & L_f \\
                                                    \end{array}
                                                  \right\}
\cdot \Big(~ [~ - (r^{(11)}_{l 24} \times r^{(10)}_{13})^L \nonumber
\\
+ (r^{(11)}_{l 23} \times r^{(10)}_{14})^L + (k^{(11)}_{l 12}
\times \bar{k}^{\dagger(10)}_{34})^L + (r^{(11)}_{l 14} \times
r^{(10)}_{23})^L \nonumber \\
- (r^{(11)}_{l 13} \times r^{(10)}_{24})^L + (\bar{k}^{\dagger
(11)}_{l 34} \times k^{(10)}_{12})^L ~] \times (\alpha \times
\alpha)^{L_f} ~\Big)^2_\mu .~~~~     \label{example2_mu}
\end{eqnarray}
The l.h.s. and r.h.s. of eq. (\ref{example2_mu}) come from the
l.h.s. and r.h.s. of eq. (\ref{example1_mu}), respectively. The
following expressions satisfy eq. (\ref{example2_mu}):
\begin{eqnarray}
r^{(21)}_{L l 1234} ~|^{L=0,1,2,3,4,5,6}_{l = 0,2,4} = \nonumber \\
( r^{(20)}_{l 14} r^{(01)}_{23})^L + ( r^{(20)}_{l 23}
r^{(01)}_{14})^L  - ( r^{(20)}_{l 13}
r^{(01)}_{24})^L  \nonumber \\
- (r^{(20)}_{l 24}  r^{(01)}_{13} )^L + ( k^{(20)}_{l 12}
\bar{k}^{\dagger(01)}_{34})^L + ( \bar{k}^{\dagger(20)}_{l 34}
k^{(01)}_{12})^L   ,
\end{eqnarray}
\begin{eqnarray}
r^{(11 \times 10)}_{Ll 1234} ~|^{L=0,1,2,3,4,5,6}_{l=0,1,2,3,4} = 2
\cdot \{~ (r^{(11)}_{l 14} \times r^{(10)}_{23})^L \nonumber \\
+ (r^{(11)}_{l 23} \times r^{(10)}_{14})^L - (r^{(11)}_{l 13} \times
r^{(10)}_{24})^L - (r^{(11)}_{l 24} \times r^{(10)}_{13})^L
\nonumber \\
+ (k^{(11)}_{l 12} \times \bar{k}^{\dagger(10)}_{34})^L +
(\bar{k}^{\dagger (11)}_{l 34} \times k^{(10)}_{12})^L ~\}  .~~~
\end{eqnarray}

\section{Values of $g^L_{l,l'}$ and $\gamma^{L}_{l,l'}$  \label{app_g_gamma}}

The definition of $g^L_{l,l'}$ is given by eq. (\ref{g_def1}),
\begin{eqnarray}
\frac{1}{8} \{ \{\alpha , \pi\}^{l'} , \alpha \}^L_\mu =
\sum_{l=0,2,4} g^L_{l,l'} \cdot \frac{1}{4} \{ (\alpha \times
\alpha)^l , \pi\}^L_\mu ,   \nonumber
\end{eqnarray}
which implies
\begin{eqnarray}
g^{L}_{l,l'} = (-)^{l - l'} \sqrt{(2 l + 1)(2 l' + 1)} \left\{
                \begin{array}{ccc}
                  2 & 2 & l \\
                  2 & L & l' \\
                \end{array}
              \right\} = g^L_{l',l} .  \label{g_def2}
\end{eqnarray}
The definition of $\gamma^{L}_{l,l'}$ is given by eq.
(\ref{gamma_def}),
\begin{eqnarray}
\{ (\alpha \times \alpha)^l , \alpha \}^L_\mu = \gamma^L_{l,l'}
\cdot \{ (\alpha \times \alpha)^{l'} , \alpha \}^L_\mu .   \nonumber
\end{eqnarray}
Analytical expressions of $\gamma^{L}_{l,l'}$ can be obtained in the
following way. Assume $l$ and $l'$ are \emph{even}. We have the
identity
\begin{eqnarray}
[~ \{(\alpha^\dagger \times \alpha^\dagger)^{l'} ,
\alpha^\dagger\}^L_\mu ~,~  (\pi \times \pi)^0_0 ~]
~|^{L=0,2,3,4,6}_{l'=0,2,4} = \nonumber \\
i \frac{2}{\sqrt{5}} ~ \sum_{l=0,2,4} (~  \delta_{l,l'} + 2 \cdot
g^L_{l,l'}   ~) \cdot \{ (\alpha^\dagger \times \alpha^\dagger)^{l}
, \pi^\dagger \}^L_\mu .  \label{a3p2_com1}
\end{eqnarray}
Replacing $l'$ in eq. (\ref{a3p2_com1}) by $l''$ we obtain
\begin{eqnarray}
[~ \{(\alpha^\dagger \times \alpha^\dagger)^{l''} ,
\alpha^\dagger\}^L_\mu ~,~  (\pi \times \pi)^0_0 ~] ~|^{L=0,2,3,4,6}_{l''=0,2,4} =  \nonumber \\
i \frac{2}{\sqrt{5}} ~ \sum_{l=0,2,4} (~  \delta_{l,l''} + 2 \cdot
g^L_{l,l''}   ~) \cdot \{ (\alpha^\dagger \times \alpha^\dagger)^{l}
, \pi^\dagger \}^L_\mu .   \label{a3p2_com2}
\end{eqnarray}
Let us take the ratio of eq. (\ref{a3p2_com1})/eq.
(\ref{a3p2_com2}). The l.h.s. is $\gamma^L_{l',l''}$ by eq.
(\ref{gamma_def}). Since in the r.h.s. of eq. (\ref{a3p2_com1}) and
eq. (\ref{a3p2_com2}), $\{ (\alpha^\dagger \times
\alpha^\dagger)^{l} , \pi^\dagger \}^L_\mu$ with different $l$ are
linearly independent, we have
\begin{eqnarray}
\gamma^L_{l',l''} = \frac{\delta_{l,l'} + 2 \cdot
g^L_{l,l'}}{\delta_{l,l''} + 2 \cdot g^L_{l,l''}} ,~~~ (l = 0,2,4) .
\label{gamma_g_rel}
\end{eqnarray}
The ratio on the r.h.s. is independent of $l$. Since the matrix
$\delta_{l,l'} + 2 \cdot g^L_{l,l'}$ is symmetric (with respect to
$l$, $l'$), eq. (\ref{gamma_g_rel}) implies
\begin{eqnarray}
\delta_{l,l'} + 2 \cdot g^L_{l,l'} = f^L_l \cdot f^L_{l'} ,
\label{g_val}
\end{eqnarray}
where $f^L_l = \sqrt{ 1 + 2 \cdot g^L_{l,l} }$. Then from eq.
(\ref{gamma_g_rel}) we obtain
\begin{eqnarray}
\gamma^L_{l',l''} = \frac{f^L_{l'}}{f^L_{l''}} .  \label{gamma_val}
\end{eqnarray}

We will use only $L = 2$:
\begin{eqnarray}
f^{L=2}_{l=0} = \sqrt{\frac{7}{5}} ,~~~ f^{L=2}_{l=2} =
\frac{2}{\sqrt{7}} ,~~~ f^{L=2}_{l=4} = \frac{6}{\sqrt{35}} .
\label{f_val}
\end{eqnarray}
In the main text the superscript $^{L=2}$ on $f^{L=2}_{l}$ is
dropped for simplicity.

\section{Conventions  \label{app_convention}}

Our convention for the Wigner-Eckart theorem is:
\begin{eqnarray}
\langle n_1 j_1 m_1 | T^\lambda_\mu | n_2 j_2 m_2 \rangle = C_{j_2
m_2 , \lambda \mu}^{j_1 m_1} \cdot \langle n_1 j_1 \| T^\lambda \|
n_2 j_2 \rangle .
\end{eqnarray}

The reduced matrix element of $q^\dagger_{\lambda \mu}$
(\ref{q_def_mu}) is
\begin{small}
\begin{eqnarray}
\langle nlj \| q^\dagger_{\lambda} \| n'l'j' \rangle = ~~~~~~~~~~~~~~~~~~~~~~~~~~~~~~~~~~~~~~~~~~~~~~~~  \nonumber \\
\left\{
   \begin{array}{cc}
     i^{l' + \lambda - l} (-)^{j' + \lambda - j} \sqrt{\frac{(2
\lambda + 1)(2 j' + 1)}{4 \pi (2 j + 1)}}  \nonumber \\
\cdot C_{j' \frac{1}{2} , \lambda 0}^{j \frac{1}{2}} \int dr ~ r^2
f_\lambda(r)
R_{nlj}(r) R_{n'l'j'}(r) , & l' + \lambda - l ~ {\rm{is~even}} ,  \\
     0 , & l' + \lambda - l ~ {\rm{is~odd}} ,  \\
   \end{array}
 \right.  \nonumber
\end{eqnarray}
\end{small}
where the s.p. levels $|nljm\rangle$ are defined as
\begin{eqnarray}
\psi_{nljm} = R_{n l j}(r) \cdot \sum_{m_l m_s} C_{l m_l, s
m_s}^{jm} ~ i^l Y_{lm_l}(\theta,\phi) ~ \chi_{m_s} ,  \nonumber
\end{eqnarray}
in which spin $s = \frac{1}{2}$, $R_{n l j}(r)$ is a real function,
and a factor $i^l$ is included.

In this article we have used the matrix elements of the realistic
quadrupole moment operator in the harmonic oscillator s.p. basis. In
this case $\lambda = 2$, $f_\lambda(r) = r^2$, $R_{n l j}(r)$ is
independent of $j$, and $n = 2 n_r + l$ is the major-shell quantum
number. The non-vanishing matrix elements of $\langle nlj \|
q^\dagger \| n'l'j' \rangle$ have $n - n' = -2, 0, 2,$ and $l - l' =
-2, 0, 2$. For these combinations the symmetric radial integral
becomes
\begin{eqnarray}
\int dr ~ r^{4} R_{nl}(r) R_{n'l'}(r) = ~~~~~~~~~~~~~~~~~~~~~~~~ \nonumber \\
b^2 \cdot
\left\{
                                              \begin{array}{cc}
                                                n + \frac{3}{2} , & n' = n ,~ l' = l , \\
                                                - \sqrt{(n+l+3)(n-l)} , & n' = n ,~ l' = l + 2 , \\
                                                - \frac{1}{2} \sqrt{(n+l+3)(n-l+2)} , & n' = n + 2 ,~ l' = l , \\
                                                \frac{1}{2} \sqrt{(n+l+3)(n+l+5)} , & n' = n + 2 ,~ l' = l + 2 , \\
                                                \frac{1}{2} \sqrt{(n-l+2)(n-l+4)} , & n' = n + 2 ,~ l' = l - 2 , \\
                                              \end{array}
                                            \right.
\nonumber
\end{eqnarray}
where $b = \sqrt{\frac{\hbar}{m \Omega_0}}$ is the length parameter,
$\Omega_0$ is the harmonic oscillator frequency. As mentioned at the
beginning of Sec. \ref{sec_QQPP_example}, the factor $b^2$ will be
combined with $\kappa$ to make $q^\dagger_{\mu}$ dimensionless.

\section{Details of Quadrupole plus Pairing Model  \label{app_QQPP_mu}}

Here we supply the details for Sec. \ref{sec_QQPP}. In the
quadrupole plus pairing model the HFB equation becomes the BCS
equation. The canonical transformation (\ref{QP_tran}) becomes
\begin{eqnarray}
u_{12} = \delta_{12} u_1 ,~~~ v_{12} = - \delta_{1\tilde{2}} v_1
,~~~ (u_1)^2 + (v_1)^2 = 1 ,   \label{norm_uv_QP_mu}
\end{eqnarray}
where $u_{1} = u_{\tilde{1}}$, $v_{1} = v_{\tilde{1}}$ are real
numbers. The density matrices (\ref{rho_kappa_mu}) become
\begin{eqnarray}
\rho_{12} = \delta_{12} (v_1)^2 ,~~~ \kappa_{12} =
\delta_{1\tilde{2}} u_1 v_1 .
\end{eqnarray}
The $f\{\rho\}_{12}$ field (\ref{W_def_mu}) becomes
\begin{eqnarray}
f\{\rho\}_{12} = \delta_{12} [ \epsilon_1 - \mu - G (v_1)^2 ]
\nonumber \\
- \kappa \sum_{3} \sum_\mu (q^\dagger_{\mu 12} q_{\mu
33} - q^\dagger_{\mu 13} q_{\mu 32} ) (v_3)^2   \nonumber
\\
\approx \delta_{12} e_1 - \kappa (q^\dagger_{\mu = 0})_{12} Q^{(00)}
,    \label{f_QP_mu}
\end{eqnarray}
where $Q^{(00)} = \sum_{3} (q_{\mu = 0})_{33} (v_3)^2$, and $e_1
\equiv \epsilon_1 - \mu - G (v_1)^2 = e_{\tilde{1}}$, and we neglect
the incoherent sum. In the case of a spherical mean field, $Q^{(00)}
= 0$, thus only the $\delta_{12} e_1$ term survives. The
$\Delta\{\kappa\}_{12}$ field (\ref{Delta_def_mu}) becomes
\begin{eqnarray}
\Delta\{\kappa\}_{12} = - \frac{G}{2} ~ \delta_{1 \tilde{2}}
\sum_{3} u_3 v_3 - \kappa \sum_{3} \sum_\mu q^\dagger_{\mu
1\tilde{3}} q_{\mu 23} ~ u_3 v_3 \nonumber \\
\approx - \delta_{1\tilde{2}} ~ \Delta ,~~~    \label{Delta_QP_mu}
\end{eqnarray}
where the pairing energy $\Delta \equiv \frac{G}{2} \sum_{3} u_3
v_3$, and we neglect the quadrupole-force contribution to the
pairing potential. The HFB equation (\ref{HFB}) gives the BCS set of
equations (\ref{gap_eq}-\ref{mu_eq}).

Quadrupole moment in the quasiparticle basis is given by
\begin{eqnarray}
Q_\mu = \sum_{12} q_{\mu 21} a_2^\dagger a_1 = \nonumber \\
\sum_{1}
(v_1)^2 q_{\mu 11} + \sum_{12} (u_1 u_2 - v_1 v_2)
q_{\mu 21} b_2^\dagger b_1 \nonumber \\
- \frac{1}{2} \sum_{12}  (u_1 v_2 + u_2 v_1) q_{\mu 21}
(b_{\tilde{2}} b_1 + b_2^\dagger b_{\tilde{1}}^\dagger) .
\label{Qb_mu}
\end{eqnarray}
Substituting the expansions of $R^b$, $K^b$ and $K^{b \dagger}$
(\ref{RK_b_exp_mu}) into eq. (\ref{Qb_mu}) we have
\begin{eqnarray}
Q_\mu = Q^{(10)} \alpha_\mu + Q^{(20)} \frac{(\alpha \times
\alpha)^2_\mu}{2} + Q^{(02)} \frac{(\pi \times \pi)^2_\mu}{2}
\nonumber \\
+ Q^{(30)} \frac{\{(\alpha \times
\alpha)^{l(L=2)},\alpha\}^2_\mu}{6} \nonumber \\
+ \sum_{l = 0,2,4}
Q^{(12)}_l \frac{\{\alpha, (\pi \times \pi)^l\}^2_\mu}{4} + \ldots
\label{Q_exp_mu}
\end{eqnarray}
where $Q^{(mn)}$ is expressed in terms of $r$'s and $k$'s in eq.
(\ref{RK_b_exp_mu}). Note on the r.h.s. only terms with the same
symmetry as $Q_\mu$ survive. The Hermitian property
(\ref{Q_herm_mu}) implies that all $Q^{(mn)}$ are real. Pairing
operator is given by
\begin{eqnarray}
P = \frac{1}{2} \sum_1 a_{\tilde{1}} a_1 = - \frac{1}{2} \sum_1 u_1
v_1 \nonumber \\
+ \sum_1 u_1 v_1 b_1^\dagger b_1 + \frac{1}{2} \sum_1 [ (u_1)^2
b_{\tilde{1}} b_1 - (v_1)^2 b_1^\dagger b_{\tilde{1}}^\dagger ] .
\label{Pb_mu}
\end{eqnarray}
Substituting the expansions of $R^b$, $K^b$ and $K^{b \dagger}$
(\ref{RK_b_exp_mu}) into eq. (\ref{Pb_mu}) we have
\begin{eqnarray}
P = - \frac{1}{2} \sum_1 u_1 v_1 + P^{(20)} \frac{(\alpha \times
\alpha)^0_0}{2} \nonumber \\
+ P^{(11)} \frac{\{\alpha ,
\pi\}^0_0}{2} + P^{(02)} \frac{(\pi \times \pi)^0_0}{2} + \ldots
 \label{P_exp_mu}
\end{eqnarray}
$P + P^\dagger$ is Hermitian and time-even, $P - P^\dagger$ is
anti-Hermitian and time-odd. Thus $P^{(20)}$ and $P^{(02)}$ are
real, $P^{(11)}$ is pure imaginary. The $f\{R\}$ field
(\ref{W_def_mu}) becomes
\begin{eqnarray}
f\{R\}_{12} = \delta_{12} e_1 - G N[a_{\tilde{1}}^\dagger
a_{\tilde{2}}] - \kappa \sum_\mu q^\dagger_{\mu 12} Q_\mu \nonumber
\\
+ \kappa \sum_{34} \sum_\mu q^\dagger_{\mu 13} q_{\mu 42}
a_4^\dagger a_3 \approx \delta_{12} e_1 - \kappa \sum_\mu
q^\dagger_{\mu 12} Q_\mu . \label{fR_QP_mu}
\end{eqnarray}
Here again we neglect the incoherent sum and the pairing
contribution beyond the mean field. The pairing field
(\ref{Delta_def_mu}) becomes
\begin{eqnarray}
\Delta\{K\}_{12} = \delta_{1\tilde{2}} G P  - \kappa \sum_{34}
\sum_\mu q^\dagger_{\mu 14} q_{\mu 23} a_3 a_4 \approx
\delta_{1\tilde{2}} G P ,~~~~    \label{DeltaK_QP_mu}
\end{eqnarray}
again neglecting the quadrupole-force contribution. Finally, the
field $S_{A/B/C/D}$ (\ref{S_A}-\ref{S_D}) become
\begin{eqnarray}
(S_{A})_{12} = \delta_{12} [(u_1)^2 - (v_1)^2] e_1 \nonumber \\
- \kappa [u_1 u_2 - v_1 v_2] \sum_\mu q^\dagger_{\mu 12} Q_\mu - u_1
v_1 \delta_{12} G
(P + P^\dagger) ,~    \\
(S_{B})_{12} = 2 u_1 v_1 \delta_{1\tilde{2}} e_1 - \kappa [u_1 v_2 +
u_2 v_1] \sum_\mu q^\dagger_{\mu 1\tilde{2}} Q_\mu \nonumber \\
+ \delta_{1\tilde{2}} G [(u_1)^2 P - (v_1)^2 P^\dagger] ,~
\end{eqnarray}
and $S_{C} = (S_{B})^\dagger$, $S_{D} = - (S_{A})^T$. Substituting
eqs. (\ref{Q_exp_mu}) and (\ref{P_exp_mu}) into the above equations
we obtain the expansions of $S_{A/B/C/D}$.

The QRPA secular equation (\ref{omega_QP_mu}) in the form of reduced
matrix elements is
\begin{eqnarray}
1 = \kappa \sum_{n_1 j_1 n_2 j_2} (-)^{j_2 - j_1} \frac{\sqrt{(2 j_1
+ 1)(2 j_2 + 1)}}{5} \nonumber \\
\cdot \frac{ E_1 + E_2 }{(E_1 + E_2)^2 - \omega^2} \cdot
\xi^{\dagger}_{\| 12} \xi^{\dagger}_{\| 21} .
\end{eqnarray}
The normalization condition (\ref{norm_QP_mu}) in the form of
reduced matrix elements is
\begin{eqnarray}
1 = (\kappa Q^{(10)})^2 \sum_{n_1 j_1 n_2 j_2} (-)^{j_2 - j_1}
\frac{\sqrt{(2 j_1 + 1)(2 j_2 + 1)}}{5} \nonumber \\
\cdot \frac{ (E_1 + E_2) }{[(E_1 + E_2)^2 - \omega^2]^2} \cdot
\xi^{\dagger}_{\| 12} \xi^{\dagger}_{\| 21} .~~~
\end{eqnarray}

The cubic potential term (\ref{L30_c_red_QP_mu}) in the original
form is
\begin{eqnarray}
\Lambda^{(30)} \doteq - (\kappa Q^{(10)})^3 ~ \{ {\rm{Tr}} \Big[ (
\xi^{{(1)}} \times \xi^{(1)} )^{L=2}_{\mu} \eta^{ \dagger}_{\mu}
\Big] \nonumber \\
+ {\rm{Tr}} \Big[ \{ \eta , \xi^{{(1)}}
\}^{L=2}_{\mu} \xi^{{(1)} \dagger}_{\mu} \Big] \} ,
\label{L30_c_QP_mu}
\end{eqnarray}
where each term on the r.h.s. is real and independent of $\mu$.

The quartic potential term (\ref{L40_c_QP_mu}) can be written in the
form of reduced matrix elements. The result is long, we write down
them term by term. The term with $(\Lambda^{(30)})^2$ is
\begin{eqnarray}
- 2 ~ f_2 \cdot (\kappa Q^{(10)})^2 (\Lambda^{(30)})^2 ~ {\rm{Tr}}
\Big[ \xi_{\mu} \xi^{{(5)}
\dagger}_{\mu} \Big] \nonumber \\
= - 2 ~ f_2 \cdot (\kappa Q^{(10)})^2 (\Lambda^{(30)})^2 \nonumber
\\
\cdot \sum_{n_1 j_1 n_2 j_2} (-)^{j_2 - j_1} \frac{\sqrt{(2 j_1 +
1)(2 j_2 + 1)}}{5} \cdot \xi^{{(5)} \dagger}_{\| 12} \xi^{
\dagger}_{\| 21} .~~
\end{eqnarray}
The terms with $\Lambda^{(30)} (\kappa Q^{(10)})^3$ are
\begin{eqnarray}
- f_2 \cdot \Lambda^{(30)} (\kappa Q^{(10)})^3 \Big( ~~ {\rm{Tr}}
\Big[ \{ \xi^{(1)} , \xi^{(3)} \}^{L=2}_{\mu} \eta^{\dagger}_{\mu}
\Big] \nonumber \\
+ {\rm{Tr}} \Big[ \{ \eta , \xi^{(3)} \}^{L=2}_{\mu} \xi^{{(1)}
\dagger}_{\mu} \Big] + 2 ~ {\rm{Tr}} \Big[ \{ \eta , \xi^{(1)}
\}^{L=2}_{\mu} \xi^{{(3)} \dagger}_{\mu} \Big] \nonumber \\
+ {\rm{Tr}} \Big[ \{ \eta , \xi^{{(2)}} \}^{L=2}_{\mu} \xi^{{(2)}
\dagger}_{\mu} \Big] ~~ \Big)  \nonumber
\\
= 2 ~ \Lambda^{(30)} f_2 \cdot (\kappa Q^{(10)})^3 \sum_{n_1 j_1 n_2
j_2 n_3 j_3}
\nonumber \\
\sqrt{\frac{(2 j_1 + 1)(2 j_2 + 1)(2 j_3 + 1)}{5}} \cdot \left\{
                                          \begin{array}{ccc}
                                            2   & 2   & 2 \\
                                            j_1 & j_2 & j_3 \\
                                          \end{array}
                                        \right\}  \nonumber
\\
\cdot \Big[~ 4 ~ \xi^{{(1)} \dagger}_{\| 13} \xi^{{(3)}
\dagger}_{\| 32} \eta^{ \dagger}_{\| 21} + \xi^{{(2)} \dagger}_{\|
13} \xi^{{(2)} \dagger}_{\| 32} \eta^{ \dagger}_{\| 21} ~\Big] .~~
\end{eqnarray}
The two terms with $P^{(20)}$ are
\begin{eqnarray}
- f_0 \cdot G P^{(20)} (\kappa Q^{(10)})^2 \sum_{12} (u_1
v_1 + u_2 v_2) \xi_{\mu 12} \xi^{{(2)} \dagger}_{\mu 21}  \nonumber \\
= - f_0 \cdot G P^{(20)} (\kappa Q^{(10)})^2 \sum_{n_1 j_1 n_2 j_2}
(-)^{j_2 - j_1} \nonumber \\
\frac{\sqrt{(2 j_1 + 1)(2 j_2 + 1)}}{5} \cdot (u_1 v_1 + u_2 v_2)
\cdot \xi^{\dagger}_{\| 12} \xi^{{(2)} \dagger}_{\| 21} ,~
\end{eqnarray}
and
\begin{eqnarray}
+ f_0 \cdot G P^{(20)} (\kappa Q^{(10)})^2 \nonumber \\
\cdot \sum_{12} \{
\frac{(u_1)^2 - (v_1)^2}{2 E_1} + \frac{(u_2)^2 - (v_2)^2}{2 E_2} \} \xi^{(1)}_{\mu 12} \eta^{\dagger}_{\mu 21}  \nonumber \\
= f_0 \cdot G P^{(20)} (\kappa Q^{(10)})^2 \nonumber \\
\cdot \sum_{n_1 j_1 n_2 j_2} (-)^{j_2 - j_1}
\frac{\sqrt{(2 j_1 + 1)(2 j_2 + 1)}}{5} \nonumber \\
\cdot ~\{~ \frac{(u_1)^2 - (v_1)^2}{2 E_1} + \frac{(u_2)^2 -
(v_2)^2}{2 E_2} ~\} \cdot \eta^{\dagger}_{\| 12} \xi^{{(1)}
\dagger}_{\| 21}  ,
\end{eqnarray}
where $P^{(20)}$ is given in eq. (\ref{P20_c_red_mu}). The terms
with $(\kappa Q^{(10)})^4$ are
\begin{eqnarray}
+ (\kappa Q^{(10)})^4 \sum_{l=0,2,4} f_l \cdot \Big( ~ {\rm{Tr}}
\Big[ \{ \xi , ( \xi^{(1)} \times \xi^{(1)} )^{l} \}^{L=2}_{\mu}
\xi^{{(1)}
\dagger}_{\mu} \Big] \nonumber \\
- {\rm{Tr}} \Big[ \{ \eta , \{ \eta , \xi^{{(1)}} \}^{l,(1)}
\}^{L=2}_{\mu} \xi^{{(1)} \dagger}_{\mu} \Big] \nonumber \\
-
{\rm{Tr}} \Big[ \{ \xi^{(1)} , \{ \eta , \xi^{{(1)}} \}^{l,(1)}
\}^{L=2}_{\mu} \eta^{\dagger}_{\mu} \Big] ~\Big) \nonumber \\
= 2 (\kappa Q^{(10)})^4 \sum_{l=0,2,4} f_l \cdot
\sum_{n_1 j_1 n_2 j_2 n_3 j_3 n_4 j_4} (-)^{j_2 - j_3}  \nonumber \\
\cdot \sqrt{\frac{(2 j_1 + 1)(2 j_2 + 1)(2 j_3 + 1)(2 j_4 + 1)(2 l +
1)}{5}} \nonumber \\
\cdot \left\{
                                          \begin{array}{ccc}
                                            l   & 2   & 2   \\
                                            j_1 & j_2 & j_3 \\
                                          \end{array}
                                        \right\} \cdot
\left\{
                                          \begin{array}{ccc}
                                            2   & 2   & l   \\
                                            j_3 & j_2 & j_4 \\
                                          \end{array}
                                        \right\}
\nonumber \\
\cdot \Big[~ \xi^{\dagger}_{\| 13} \xi^{{(1)} \dagger}_{\| 34}
\xi^{{(1)} \dagger}_{\| 42} \xi^{{(1)} \dagger}_{\| 21} - 2
\eta^{\dagger}_{\| 13} \cdot \frac{1}{E_{3} + E_{2}} \cdot
\eta^{\dagger}_{\| 34} \xi^{{(1)} \dagger}_{\| 42} \xi^{{(1)}
\dagger}_{\| 21} \nonumber \\
- 2 \eta^{\dagger}_{\| 13} \cdot \frac{1}{E_{3} + E_{2}} \cdot
\xi^{{(1)} \dagger}_{\| 34} \eta^{\dagger}_{\| 42} \xi^{{(1)}
\dagger}_{\| 21}   ~\Big] .~~~~
\end{eqnarray}

\section{Mode Coupling  \label{app_mode_coupling}}

In many soft nuclei there exists a low-lying octupole ($3^-$) mode.
It can interact strongly with the quadrupole ($2^+$) mode, and both
of them should be kept in the collective subspace. For convenience
we still use ${\alpha}_\mu$, ${\pi}_\mu$ for the quadrupole mode;
and use $\hat{\alpha}_\mu$, $\hat{\pi}_\mu$ for the octupole mode.
The collective bosonic Hamiltonian replacing eq. (\ref{H_b}) is
\begin{eqnarray}
H = \frac{\omega^2}{2} \sqrt{5} (\alpha \times \alpha)^0_0 +
\frac{1}{2} \sqrt{5} (\pi \times \pi)^0_0 \nonumber \\
+ \frac{\hat{\omega}^2}{2} \sqrt{7} (\hat{\alpha} \times
\hat{\alpha})^0_0 + \frac{1}{2} \sqrt{7} (\hat{\pi}
\times \hat{\pi})^0_0  \nonumber \\
+ \frac{\Lambda^{(10|20)}}{2} \sqrt{7} (\alpha \times (\hat{\alpha}
\times \hat{\alpha})^2)^0_0 + \ldots
\end{eqnarray}
$\Lambda^{(10|20)}$ is the most important mode-coupling term in the
case of soft vibrations with large amplitudes. Following the
procedure of Sec. \ref{sec_GDM} and \ref{sec_Soft}, we are able to
determine the leading constant term of $\Lambda^{(10|20)}$ in a
Taylor expansion over both $\omega^2$ and $\hat{\omega}^2$ [see eq.
(\ref{Lmn_exp})]. Below we give the result in the quadrupole plus
pairing model. The microscopic Hamiltonian is:
\begin{eqnarray}
H = \sum_1 (\epsilon_1 - \mu) a_1^\dagger a_1 - \frac{G}{4}
\sum_{12} a_1^\dagger a_{\tilde{1}}^\dagger a_{\tilde{2}} a_2 \nonumber \\
+ \frac{1}{4} \sum_{1234} \sum_{\mu} ( - \kappa q^\dagger_{\mu 14}
q_{\mu 23} + \kappa q^\dagger_{\mu 13} q_{\mu 24} \nonumber \\
- \hat{\kappa} \hat{q}^\dagger_{\mu 14} \hat{q}_{\mu 23} +
\hat{\kappa} \hat{q}^\dagger_{\mu 13} \hat{q}_{\mu 24} ) a_1^\dagger
a_2^\dagger a_3 a_4  .
\end{eqnarray}
Approximately, this Hamiltonian can be written as $H \approx \sum_1
(\epsilon_1 - \mu) a_1^\dagger a_1 - G P^\dagger P - \frac{1}{2}
\kappa \sum_{\mu} Q_\mu^\dagger Q_\mu - \frac{1}{2} \hat{\kappa}
\sum_{\mu} \hat{Q}_\mu^\dagger \hat{Q}_\mu$, the difference is in a
one-body term originating from the $Q \cdot Q$ part. $\hat{\kappa}$
is the strength of the octupole force. The mean field is determined
by the HFB equation. In the harmonic order the two modes do not mix,
the octupole mode satisfies the same QRPA equation
(\ref{omega_QP_mu}) and normalization condition (\ref{norm_QP_mu})
as the quadrupole mode, with necessary changes. In the next order we
have the main result:
\begin{eqnarray}
\Lambda^{(10|20)} \doteq 2 \kappa Q^{(10)} (\hat{\kappa}
\hat{Q}^{(10)})^2 \nonumber \\
\cdot \sum_{n_1 j_1 n_2 j_2 n_3 j_3} \sqrt{\frac{(2 j_1
+ 1)(2 j_2 + 1)(2 j_3 + 1)}{7}} \nonumber \\
\cdot \left\{
                                          \begin{array}{ccc}
                                            3   & 2   & 3 \\
                                            j_1 & j_2 & j_3 \\
                                          \end{array}
                                        \right\}
\cdot \Big[~ 2 ~ \xi^{(1) \dagger}_{\| 13} ~ \hat{\xi}^{(1)
\dagger}_{\| 32} ~ \hat{\eta}^{\dagger}_{\| 21} + \eta^{\dagger}_{\|
13} ~ \hat{\xi}^{(1) \dagger}_{\| 32} ~ \hat{\xi}^{(1) \dagger}_{\|
21}  ~\Big] .~~  \label{L10_20}
\end{eqnarray}
The octupole operator $\hat{q}$ connects s.p. levels with opposite
parity, thus the intruder state becomes important. This may destroy
in eq. (\ref{L10_20}) symmetry with respect to the Fermi surface.
Three-body forces will contribute to the $\Lambda^{(10|20)}$ term
quite differently.

\begin{table*}
\caption{\label{Table_1}  The first excitation energy $E_1 - E_0$
for different $J$ in the Lipkin model. The first three lines are the
results of diagonalizing different bosonic Hamiltonians in different
phonon spaces; the last two lines are the results of diagonalizing
eq. (\ref{H_J}) directly in the $\{ |JM\rangle \}$ space. In the
space $| n \le 2J \rangle$, the matrix of e.g. $\alpha^4$ is
calculated by multiplications of the $\alpha$ matrices, which is
different from truncating the $\alpha^4$ matrix of the space $| n
\le + \infty \rangle$. Higher excited states from the GDM method are
also in good agreement with the exact results; please see the figure
in Ref. \cite{Zele_AIP}. }
\begin{tabular}{|c|c|c|c|c|c|c|}
  \hline
  $E_1 - E_0$ & $J = 2$ & $J = 3$ & $J = 4$ & $J = 6$ & $J = 10$ & $J = 50$ \\
  \hline
  $\frac{2}{J} \frac{\alpha^4}{4} + \frac{\pi^2}{2} - \frac{1}{8 J} \frac{\pi^4}{4}$ in $| n \le 2J \rangle$ & 1.011 & 0.904 & 0.835 & 0.740 & 0.630 & 0.371 \\
  $\frac{2}{J} \frac{\alpha^4}{4} + \frac{\pi^2}{2}$ in $| n \le 2J \rangle$ & 1.087 & 0.949 & 0.863 & 0.754 & 0.636 & 0.372 \\
  $\frac{2}{J} \frac{\alpha^4}{4} + \frac{\pi^2}{2}$ in $| n \le + \infty \rangle$ & 1.087 & 0.950 & 0.863 & 0.754 & 0.636 &  0.372 \\ \hline
  exact , $\kappa = \frac{1}{2J + 1}$ & 0.950 & 0.869 & 0.808 & 0.722 & 0.620 & 0.370 \\
  exact , $\kappa = \frac{1}{2J - 1}$ & 0.895 & 0.776 & 0.707 & 0.625 & 0.537 & 0.334  \nonumber \\
  \hline
\end{tabular}
\end{table*}

\newpage
\phantom{a}
\newpage

\begin{table*}
\caption{\label{Table_2} Results of the quadrupole plus pairing
model at different pairing strength $G$. All quantities are in unit
of MeV. $\Delta$ is the solution of eq. (\ref{gap_eq}). The chemical
potential $\mu$ is the solution of eq. (\ref{mu_eq}). $\kappa_c$ is
the critical $\kappa$ such that $\omega^2$ in eq.
(\ref{omega_QP_mu}) becomes zero. $\Lambda^{(30)}$ is given by eq.
(\ref{L30_c_red_QP_mu}). $\Lambda^{(40)}$ is given by eq.
(\ref{L40_c_QP_mu}) setting $\Lambda^{(12)} = 0$. ``GDM $E_{2^+}$''
is the excitation energy of the first $2^+$ state by diagonalizing
eq. (\ref{H_b_mu}) for $\omega^2 = \Lambda^{(12)} = \Lambda^{(04)} =
\Lambda^{(22)}_L = 0$. ``NuShellX $E_{2_1^+}$'' is the exact
excitation energy of the first $2^+$ state by diagonalizing eq.
(\ref{H_QP_mu}), in which $G$ and $\kappa$ are given by $G$ and
$\kappa_c$ in the table. Similarly ``NuShellX $E_{4_1^+}$'' is the
exact excitation energy of the first $4^+$ state. }

\begin{tabular}{|c|ccccccccccc|}
  \hline
  $G$                             & 0          & 0.03       & 0.06       & 0.09        & 0.11        & 0.12    & 0.15       & 0.18        & 0.21        & 0.25       & 0.30     \\  \hline
  $\Delta$                        & 0.0        & 0.0        & 0.0        & 0.0         & 0.0         & 0.066   & 0.453      & 0.672       & 0.862       & 1.096      & 1.374      \\
  $\mu$                           & 0.5        & 0.5        & 0.5        & 0.5         & 0.5         & 0.454   & 0.444      & 0.429       & 0.415       & 0.395      & 0.370    \\
  $\kappa_c$                      & 0.102      & 0.105      & 0.107      & 0.110       & 0.112       & 0.113   & 0.113      & 0.122       & 0.135       & 0.154      & 0.179     \\
  $\Lambda^{(30)}$                & -0.160     & -0.173     & -0.188     & -0.203      & -0.213      & -0.219  & -0.234     & -0.270      & -0.310      & -0.378     & -0.474     \\
  $\Lambda^{(40)}$                & 0.483      & 0.526      & 0.572      & 0.621       & 0.655       & 0.616   & 1.185      & 1.918       & 2.901       & 4.683      & 7.830     \\
  ${\rm{GDM}} ~ E_{2^+}$          & 0.882      & 0.908      & 0.933      & 0.959       & 0.976       & 0.955   & 1.194      & 1.405       & 1.614       & 1.894      & 2.249    \\
  \hline
  ${\rm{NuShellX}} ~ E_{2_1^+}$   & 0.855      & 0.892      & 0.944      & 1.023       & 1.106       & 1.158   & 1.353      & 1.552       & 1.764       & 2.059      & 2.438     \\
  ${\rm{NuShellX}} ~ E_{4_1^+}$   & 0.778      & 0.827      & 0.927      & 1.110       & 1.284       & 1.383   & 1.705      & 2.076       & 2.465       & 2.987      & 3.631
  \\ \hline
\end{tabular}
\end{table*}

\newpage
\phantom{a}
\newpage

%

\newpage
\phantom{a}
\newpage

\newpage
\phantom{a}
\newpage

\newpage
\phantom{a}
\newpage

~

\end{document}